\documentclass[11pt]{article}
\usepackage{latexsym,times}
\usepackage{amsmath,amssymb}
\usepackage{amsxtra}
\usepackage{amscd}
\usepackage{amsthm}
\usepackage{amsfonts}
\usepackage{bm}
\usepackage{comment} 
\usepackage{dsfont}
\usepackage{xcolor}
\usepackage{cancel}
\usepackage{float}
\usepackage{framed}

\usepackage[final]{graphicx}
\usepackage{subcaption}

\usepackage{amssymb}
\usepackage{amsmath}
\usepackage{graphicx}
\usepackage{esint}

\usepackage{enumitem}
\usepackage{float}

\usepackage{hyperref}
\usepackage[square,sort&compress,comma,numbers]{natbib}

\usepackage{pifont}
\usepackage{tocloft}


\makeatletter

\@addtoreset{equation}{section} \makeatother

\def\bri{\bar{\imath}}
\def\brj{\bar{\jmath}}
\newcommand\diag{\mbox{diag}}
\newcommand\rD{{\rm D}}
\newcommand{\half}{{{\textstyle\frac{1}{2}}}}
\newcommand{\quarter}{{{\textstyle\frac{1}{4}}}}
\newcommand{\be}{\begin{equation}}
\newcommand{\ee}{\end{equation} }
\newcommand{\beqa}{\begin{eqnarray} }
\newcommand{\eeqa}{\end{eqnarray} }
\newcommand{\ba}{\begin{array}}
\newcommand{\ea}{\end{array}}
\newcommand{\bpm}{\begin{pmatrix}}
\newcommand{\epm}{\end{pmatrix}}

\newcommand\fcL{{{\widetilde{\cal L}}}}

\newcommand{\SO}{\mathbf{SO}}
\newcommand{\Spin}{\mathbf{Spin}}

\newcommand{\rmd}{{\rm d}}

\newcommand{\ODD}{\mathbf{O}(D,D)}

\newcommand{\SpinD}{{\Spin(1,D{-1})}}
\newcommand{\oSpinD}{{{\Spin}(D{-1},1)}}

\newcommand{\ls}{l_{s}}

\newcommand{\DFT}{{\scriptscriptstyle{\rm{\bf{DFT}}}}}

\newcommand{\Matter}{\rm{\bf{\scriptscriptstyle M}atter}}

\newcommand\Tr{{\rm Tr}}

\newcommand\rd{{\rm d}}

\newcommand{\betappn}{\beta_{{{{\rm{PPN}}}}}}
\newcommand{\gammappn}{\gamma_{{{\rm{PPN}}}}}

\newcommand\cC{{\cal C}}
\newcommand\cD{{\cal D}}

\newcommand\cF{{\cal F}}

\newcommand\cH{{\cal H}}

\newcommand\cJ{{\cal J}}
\newcommand\cK{{\cal K}}
\newcommand\cL{{\cal L}}

\newcommand\cP{{\cal P}}

\newcommand\cR{{\cal R}}

\newcommand\brcP{{\bar{\cal{P}}}}

\newcommand\hcL{{\hat{\cal L}}}
\newcommand\hBox{{\widehat{\Box}}}

\newcommand\hR{{\hat{R}}}

\newcommand\rhop{{\rho^{\prime}}{}}
\newcommand\psip{\psi^{\prime}}
\newcommand\varepsilonp{\varepsilon^{\prime}{}}

\newcommand\alphap{{\alpha^{\prime}}}

\newcommand\dis{\displaystyle}

\def\tx{\tilde{x}}

\def\tpartial{\tilde{\partial}}

\def\bre{\bar{e}}

\def\breta{\bar{\eta}}
\def\bralpha{\bar{\alpha}}
\def\brbeta{\bar{\beta}}
\def\brgamma{\bar{\gamma}}
\def\brdelta{\bar{\delta}}

\def\brn{{\bar{n}}}
\def\brp{{\bar{p}}}
\def\brq{{\bar{q}}}
\def\brr{{\bar{r}}}
\def\brs{{\bar{s}}}

\def\bromega{{\bar{\omega}}}

\def\brPhi{{{\bar{\Phi}}}}
\def\brDelta{{{\bar{\Delta}}}}

\def\brR{\bar{R}}

\def\brV{{\bar{V}}}

\def\brX{{\bar{X}}}
\def\brY{{\bar{Y}}}
\def\brP{{\bar{P}}}

\newcommand\bfA{{\mathbf{A}}}
\newcommand\bfF{{\mathbf{F}}}

\newcommand{\DO}{\mathbf{\nabla}}

\newcommand{\na}{{\nabla}}
\newcommand{\trd}{{\bigtriangledown}}

\newcommand\So{S_{\scriptscriptstyle{{(0)}}}}

\newcommand\To{T_{\scriptscriptstyle{{(0)}}}}

\newcommand{\fa}{\mathfrak{a}}

\newcommand{\fT}{\mathfrak{T}}

\definecolor{darkbrown}{RGB}{100, 20, 10}

\newcommand{\hh}{\mathfrak{h}}

\newcommand{\fR}{\mathfrak{R}}

\newcommand{\bBox}{\scalebox{1.23}{$\boldsymbol{\Box}$}}
\newcommand{\Deltab}{\mathbf{\Delta}}
\newcommand{\brDeltab}{\mathbf{\brDelta}}

\newcommand\p\partial

\newcommand{\aim}[1]{\begin{quote}\textbf{{Aim.}} #1\end{quote}}
\newcommand{\summary}[1]{\begin{quote}\textbf{{Summary.}} #1\end{quote}}

\begin{document}

\begin{titlepage}
\title{\vspace{-27pt}
{\bf Gravitational  Core  of Double Field Theory}\\
\vspace{18pt}
{\sc---Lecture Notes---}\\\vspace{5pt}}
\author{{Jeong-Hyuck Park}}
\date{}
\maketitle \vspace{-12pt}
\centering Department of Physics, Sogang University, 35 Baekbeom-ro, Mapo-gu, Seoul 04107,  Korea\\\vspace{12pt}
\centering\texttt{ park@sogang.ac.kr  }\vspace{3pt}\vspace{23pt}
\begin{figure}[H]
\centering\includegraphics[scale=0.11]{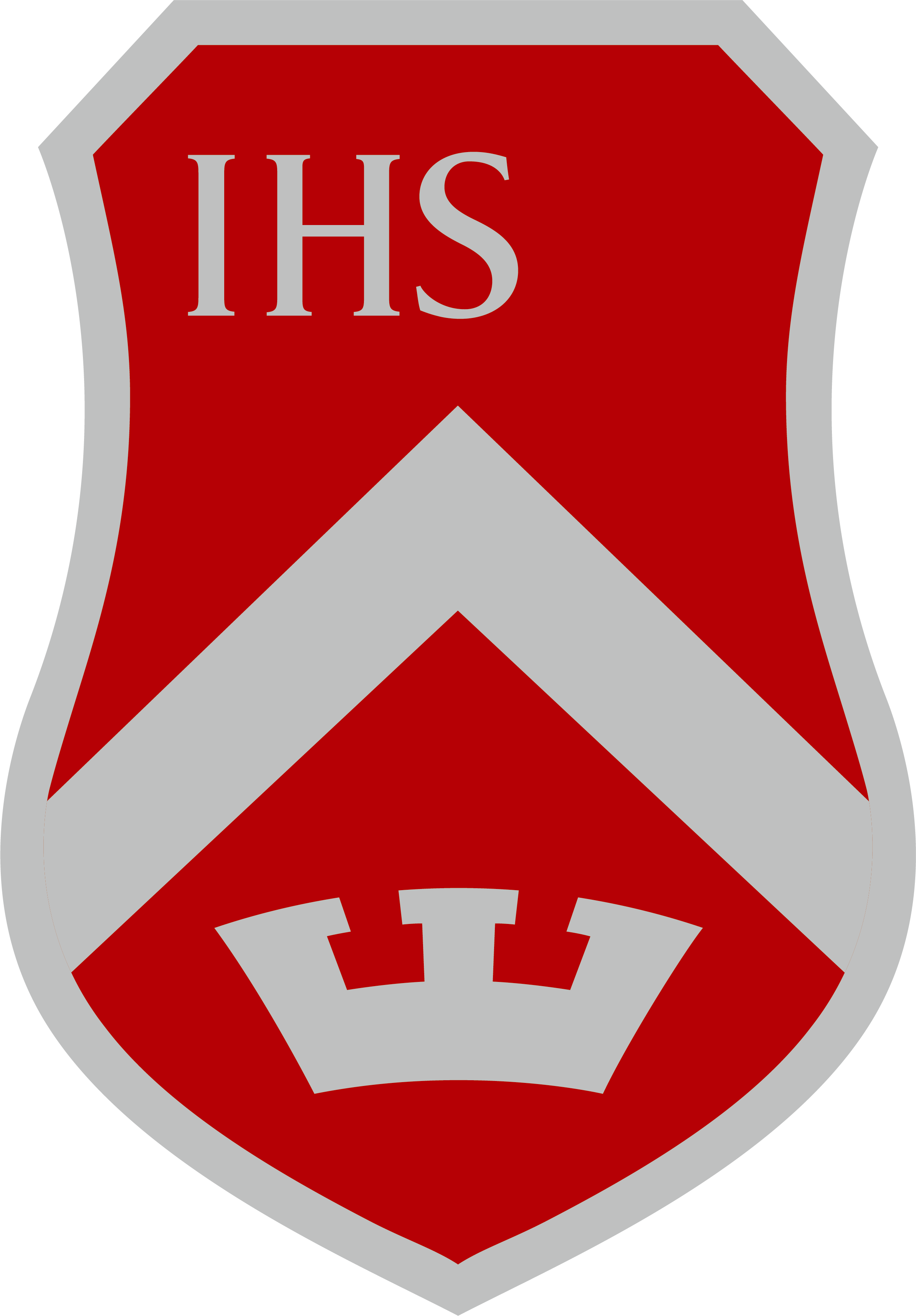}
\end{figure}
\vspace{10pt}
\begin{abstract}
\centering\begin{minipage}{\dimexpr\paperwidth-9.3cm}
\vskip0.1cm 
\noindent Double Field Theory (DFT) has emerged as a comprehensive framework for gravity, presenting a testable and robust alternative to General Relativity (GR), rooted in  the $\mathbf{O}(D,D)$  symmetry principle of string theory. These lecture notes aim to provide an accessible introduction to DFT, structured in a manner similar to traditional GR courses. Key topics  include   doubled-yet-gauged coordinates, Riemannian \textit{versus} non-Riemannian parametrisations of fundamental fields, covariant derivatives, curvatures, and the $\mathbf{O}(D,D)$-symmetric augmentation of the Einstein field equation, identified as the unified field equation for the closed string massless sector. By offering a novel perspective, DFT addresses unresolved questions in GR and enables the exploration of diverse physical phenomena, paving the way for significant future research.
\end{minipage}  
\end{abstract} 

\thispagestyle{empty}
\end{titlepage}
\newpage
~\\\vspace{-3.5cm}
\section*{Preface: A Physicist's Apology for Another Review\footnote{\textit{c.f.~}A Mathematician's Apology by G.~H. Hardy~\cite{Hardy}.}}
Double Field Theory (DFT) was initially conceived by Siegel in 1993~\cite{Siegel:1993xq,Siegel:1993th} and  independently developed, along with its name, by Hull and Zwiebach in 2009~\cite{Hull:2009mi,Hull:2009zb}, with further refinements introduced in 2010~\cite{Hohm:2010jy,Hohm:2010pp}.  The term ``double" reflects the use of doubled coordinates,  $x^{A} = (\tx_{\mu}, x^{\nu})$,  a concept originally introduced by Duff in 1990~\cite{Duff:1989tf},   to describe $D$-dimensional physics, in contrast to the conventional use of   $x^{\mu}$ alone.\\

\noindent Three influential review papers on DFT~\cite{Aldazabal:2013sca,Berman:2013eva,Hohm:2013bwa}, all written in 2013, provided timely insights into the field during its early development within the framework of string theory. Over the past decade, significant progress has been continuously made, including advancements, in particular, in cosmological, black hole-related, and phenomenological aspects of DFT~\textit{e.g.,}~\cite{Brandenberger:2018xwl,Brandenberger:2018bdc,Kang:2019owv,Lescano:2021nju,Liu:2021xfs,Liu:2021gjm,Lescano:2022nrb,Brandenberger:2023ver,Lescano:2023gge,Li:2023mhz,Arapoglu:2024umz,Ying:2024jjr,Angus:2024owg}, with more developments to be discussed later.\\

\noindent Furthermore, a unified field equation for DFT was proposed in~\cite{Angus:2018mep}, co-authored by the present author, which generalises the Einstein field equation of General Relativity by incorporating $\ODD$  symmetry.  Through subsequent developments,  DFT has, at least in the author's view, matured into a self-sustaining framework for gravity, characterised by its predictive and falsifiable nature.\\

\noindent These lecture notes are intended to provide an accessible introduction to the foundations of DFT, structured in a manner analogous to introductory General Relativity courses. The aim is to engage not only the \texttt{hep-th} community but also the \texttt{gr-qc} community in exploring and testing this remarkable and precise modified theory of gravity, as predicted by string theory.\\

\noindent These notes were developed in response to repeated requests from colleagues and participants at recent workshops held in 2025~\cite{YITP2025, Bangkok2025}. They are based on the author’s lectures from the HANGANG Gravity Workshop in 2024 and a graduate-level course at Sogang University in 2025.\\

\noindent Yet, this modest contribution traces its origins to the author's first encounter with DFT during Zwiebach's lecture at a summer school in M\"{u}nchen in 2010~\cite{Munchen2010}. That lecture sparked the author’s journey into the subject, which began in November 2010 with an exploration of the DFT analogue of the Christoffel symbols in collaboration with Imtak Jeon and Kanghoon Lee~\cite{Jeon:2010rw}.\\

\noindent Although most $\ODD$-manifest computations in DFT can be performed with pen and paper, deriving explicit expressions for $\ODD$-symmetric covariant derivatives and curvatures in terms of their $\ODD$-breaking component fields, such as $\{g_{\mu\nu}, B_{\mu\nu},  \phi\}$, often requires computational tools like \textit{Cadabra}~\cite{Peeters:2007wn}.

 \newpage
 
\tableofcontents 




~\\
~\\
\section{Introduction}
\noindent In electrodynamics, the electric field is conventionally represented by the capital letter $\mathbf{E}$ for obvious reasons. However, the magnetic field is represented by either $\mathbf{B}$ or $\mathbf{H}$ rather than $\mathbf{M}$. This convention arises because Maxwell did not use vector notations when formulating his equations in 1861. Instead, Maxwell used the letters $\{E, F, G\}$ to denote components of the electric field and neighboring letters, such as   $\{B, C, D\}$ or $\{H, I, J\}$, for  the magnetic field. It was Heaviside—or perhaps the rotational $\SO(3)$ symmetry—that reformulated Maxwell's equations into their modern vectorial form:
\be
\ba{llll}
\dis{\na\cdot\mathbf{E}=\rho\,,}\quad&\quad\dis{\na\times\mathbf{E}=-\frac{\partial\mathbf{B}}{\partial t}\,,}\quad&\quad\dis{
\na\cdot\mathbf{B}=0\,,}\quad&\quad\dis{ \na\times\mathbf{B}=\mathbf{J}+\frac{\partial\mathbf{E}}{\partial t}\,.}
\ea
\label{Maxwell1}
\ee
Minkowski, or the $\SO(1,3)$ Lorentz symmetry, then introduced further simplifications in 1908, unifying space and time into a four-dimensional spacetime framework and recasting Maxwell's equations in a more compact and elegant form:
\be
\ba{ll}
\partial_{\lambda}F^{\lambda\mu}=J^{\mu}\,,\qquad&\qquad
\partial_{\lambda}F_{\mu\nu}+\partial_{\mu}F_{\nu\lambda}+\partial_{\nu}F_{\lambda\mu}=0\,.
\ea
\label{Maxwell2}
\ee

In string theory, a similar unification has been accomplished through the $\ODD$ symmetry principle. The vanishing of the three beta-functions on a closed string worldsheet, 
\be
\ba{r}
R_{\mu\nu}+2\trd_{\mu}(\partial_{\nu}\phi)-\quarter H_{\mu\rho\sigma}H_{\nu}{}^{\rho\sigma}
=0\,,\\
\half e^{2\phi}\trd^{\rho}\!\left(e^{-2\phi}H_{\rho\mu\nu}\right)=0\,,\\
R+4\Box\phi-4\partial_{\mu}\phi\partial^{\mu}\phi-\textstyle{\frac{1}{12}}H_{\lambda\mu\nu}H^{\lambda\mu\nu}=0\,,
\ea
\label{3beta}
\ee
is  unified into a single formula characterized by the vanishing of the $\ODD$-symmetric augmentation of the Einstein curvature tensor~\cite{Park:2015bza},
\be
{G}_{AB}=0\,,
\label{G0}
\ee
which corresponds to the vacuum case of the more general unified field equation in Double Field Theory (DFT), dubbed the \textit{Einstein Double Field Equation} (EDFE)~\cite{Angus:2018mep}:
\be
G_{AB}=T_{AB}\,.
\label{iEDFE}
\ee
Hereafter, the capital letters $A, B, \cdots$ denote $\ODD$ vector indices, which span the doubled spacetime dimensions, $D+D$. As reviewed below, the DFT Einstein curvature, $G_{AB}$, on the left-hand side of the equation,  can be---but not necessarily---constructed from the trio $\{g_{\mu\nu},B_{\mu\nu},\phi\}$. 
This trio represents the traditional closed string massless sector, known as the Neveu–Schwarz Neveu–Schwarz (NSNS) sector, and constitutes the stringy gravitational degrees of freedom. On the right-hand side, the DFT energy-momentum tensor, $T_{AB}$, accounts for other sectors (or ``matter''), including the Ramond--Ramond (RR) sector and NSR fermions. Like General Relativity,   these quantities are governed by conservation laws derived from the action principle:
\be
\ba{ll}
\nabla_{A}G^{A}{}_{B}=0~~~:~~~\mbox{off-shell}\,,\qquad&\qquad
\nabla_{A}T^{A}{}_{B}=0~~~:~~~\mbox{on-shell}\,.
\ea
\label{CONSERV}
\ee
This conservation principle highlights the fundamental symmetries and consistency of the theory, mirroring  General Relativity (GR). With a greater number of components in the energy-momentum tensor, the gravitational physics in DFT becomes inherently richer compared to GR.

\summary{In analogy with Lorentz symmetry unifying Maxwell’s equations, $\ODD$ symmetry unifies the closed string beta-function equations. The EDFE~(\ref{iEDFE}) is a generalised and testable extension of Einstein’s field equation, providing a richer framework for gravity.}


\section{Double Field Theory Minimum}
In this main section, we delve into the geometric foundations of Double Field Theory (DFT). After establishing the necessary notation, we introduce the concept of double-yet-gauged coordinates and the fundamental field variables, followed by the formulation of semi-covariant derivatives and curvatures. The notion of 'semi-covariance' acts as an intermediate step in the formalism, leading to fully covariant derivatives and curvatures. The overarching goal is to present the action principle and derive the unified field equation, the EDFE~(\ref{iEDFE}), along with the  conservation law~(\ref{CONSERV}).

\subsection{Symmetry \& Notation}
\aim{To establish index conventions and identify the  symmetry group $\ODD$ as well as  the twofold Lorentz symmetries.}
\noindent DFT can be viewed as a top-down extension of GR, guided by the $\ODD$ symmetry principle, necessitating the introduction of notation for its fundamental symmetry groups: $\ODD$ and $\Spin(1, D-1) \times \Spin(D-1, 1)$. As summarised in Table~\ref{TABindices}, capital letters are used for $\ODD$ vector indices, while small letters are reserved for the twofold local Lorentz symmetries. To distinguish between the indices of the two distinct spin groups, barred and unbarred notations are employed: unbarred indices correspond to $\Spin(1, D-1)$, while barred indices correspond to $\Spin(D-1, 1)$. These conventions prepare the ground for consistently formulating doubled geometry.\\
{\begin{table}[H]
\begin{center}
\begin{tabular}{c|c|c}
\hline
~~~Symmetry~~~&~~Index~~&~~``Metric'' (raising/lowering indices)~~\\
\hline
~$\ODD$~&~$A,B,\cdots,M,N,\cdots$~  & $\cJ_{AB}={\scriptscriptstyle{\bf\left(\ba{cc}0&1\\1&0\ea\right)}}$\\
~$\SpinD\,$~&~$p,q,\cdots$~ &$\eta_{pq}=\mbox{diag}(-++\cdots+)$ \\
~$\oSpinD\,$~&~$\brp,\brq,\cdots$~  &$\breta_{\brp\brq}=\mbox{diag}(+--\cdots-)$ \\
\hline
\end{tabular}
\caption{Vectorial indices and ``metrics''  for  
$\ODD$, $\Spin(1,D-1)$, and $\Spin(D-1,1)$:\\
\noindent\textit{i)} $~\cJ_{AB}$ denotes the invariant metric for the $\ODD$ group.\\
\noindent\textit{ii)}  $~\eta_{pq}$ and $\breta_{\brp\brq}$ are flat $D$-dimensional Minkowskian metrics associated with $\Spin(1, D-1)$ and\\
$\mbox{~\,\quad}\Spin(D-1, 1)$, respectively. These metrics exhibit opposite signatures. }
\label{TABindices}
\end{center}
\end{table}}


\subsection{Doubled-yet-Gauged Coordinates \& Diffeomorphisms}
\aim{To introduce doubled coordinates, the section condition, and relevant diffeomorphisms.}
\noindent The $\ODD$-invariant metric $\cJ_{AB}$  presented  in Table~\ref{TABindices}  plays a crucial role in splitting the doubled coordinates into two parts, $x^{A} = (\tx_{\mu}, x^{\nu})$, where $\mu, \nu = 0, 1, 2, \cdots, D{-1}$ denote $D$-dimensional spacetime indices. A fundamental  requirement in DFT is that all functions in the theory, collectively denoted as $\{\Phi, \hat{\Phi}, \tilde{\Phi}, \cdots\}$---including physical fields and local gauge parameters---must satisfy the  \textit{section condition}~\cite{Berman:2010is}. This condition  restricts their dependence on the doubled coordinates via the constraint:
\be
\partial_{A}\partial^{A}=\partial_{\mu}\tpartial^{\mu}+\tpartial^{\mu}\partial_{\mu}=0\,,
\label{SC}
\ee
which ensures that any pairwise contraction of  (doubled) partial derivatives  vanishes:
\be
 \ba{lll}
 \partial_{A}\partial^{A}\Phi=0\,,\qquad&\quad
 \partial_{A}\Phi\partial^{A}\hat{\Phi}=0\,,\qquad&\quad
 \partial_{A}\partial^{A}(\Phi\hat{\Phi})=0\,.
 \ea
 \ee
In practice, the section condition is often solved by setting $\tpartial^{\mu} \equiv 0$, thereby  eliminating dependence on the dual coordinates $\tx^{\mu}$. The general solution, or ``section'', is obtained by applying a global $\ODD$ rotation to this specific choice. The $\ODD$ symmetry   is spontaneously broken by selecting the $D$-dimensional section, such as  $\tpartial^{\mu} \equiv 0$. However, in cases where a configuration admits an isometric direction,  \textit{e.g.~}$\partial_{1}\equiv0$, the $\mathbf{O}(1,1)$ rotation of the $(\tx_{1},x^{1})$ plane   generates new configurations, a procedure known as Buscher's rule~\cite{Buscher:1987qj,Buscher:1987sk}.
 
 The section condition is equivalent to a certain translational invariance generated by \textit{derivative-index-valued} shift parameters $\Delta^{A}$~\cite{Park:2013mpa, Lee:2013hma}:
\be
\ba{llll}
\Phi(x)=\Phi(x+\Delta)\,&\quad\mbox{where}\quad&
\Delta^{A}=\hat{\Phi}\partial^{A}\tilde{\Phi}\,,\quad&\quad\Delta^{A}\partial_{A}=0\,.
\ea
\label{trinv}
\ee
The geometric meaning of the section condition, as suggested in~\cite{Park:2013mpa}, is that the doubled coordinates are \textit{gauged} by the derivative-index-valued vectors:
\be
\ba{ll}
x^{A}~\sim~x^{A}+\Delta^{A}\,,\qquad&\qquad\Delta^{A}\partial_{A}=0\,.
\ea
\label{xgauge}
\ee

In this framework, each gauge orbit corresponds to a single spacetime point. The concept of "derivative-index-valued" vectors is made possible by the $\ODD$-invariant metric $\cJ_{AB}$, which raises the vector index of the partial derivative: $\partial^{A}=\cJ^{AB}\partial_{B}$.

In DFT, diffeomorphisms are governed by a generalised Lie derivative, denoted as $\hcL_{\xi}$~\cite{Siegel:1993th,Hull:2009zb}:
\be
\hcL_{\xi}T_{A_{1}\cdots A_{n}}:=\xi^{B}\partial_{B}T_{A_{1}\cdots A_{n}}+\omega\partial_{B}\xi^{B}T_{A_{1}\cdots A_{n}}+\sum_{i=1}^{n}(\partial_{A_{i}}\xi_{B}-\partial_{B}\xi_{A_{i}})T_{A_{1}\cdots A_{i-1}}{}^{B}{}_{A_{i+1}\cdots  A_{n}}\,,
\label{hcL}
\ee
where $\omega$ is the weight of the tensor or tensor density, and each tensor index $A_{i}$ is rotated by an  infinitesimal $\mathbf{so}(D,D)$ element, $\partial_{A_{i}}\xi_{B}-\partial_{B}\xi_{A_{i}}$.    For consistency,  generalised Lie derivatives are closed under commutation, provided the section condition holds:
\be
\left[\hcL_{\zeta},\hcL_{\xi}\right]=\hcL_{\left[\zeta,\xi\right]_{\rm{C}}}\,,
\label{closeda}
\ee
where the commutator is given by the C-bracket:
\be
\left[\zeta,\xi\right]^{A}_{\rm{C}}=\half\!\left(\hcL_{\zeta}\xi^{A}-\hcL_{\xi}\zeta^{A}\right)= \zeta^{B}\partial_{B}\xi^{A}-\xi^{B}\partial_{B}\zeta^{A}+\half \xi^{B}\partial^{A}\zeta_{B}-\half \zeta^{B}\partial^{A}\xi_{B}\,.
\label{Cbracket}
\ee
It is noteworthy that the $\ODD$-metric is invariant under generalised diffeomorphisms, ${\hcL_{\xi}\cJ_{AB}=0}$, and the generalised Lie derivative itself is covariant~\cite{Angus:2018mep}:
\be
\delta_{\xi}(\hcL_{\zeta}T_{A_{1}\cdots A_{n}})=
\hcL_{\zeta}(\delta_{\xi}T_{A_{1}\cdots A_{n}})+\hcL_{\delta_{\xi}\zeta}T_{A_{1}\cdots A_{n}}=\hcL_{\xi}(\hcL_{\zeta}T_{A_{1}\cdots A_{n}})\,.
\label{covhcL}
\ee

\subsubsection{Explicit Gauging of the Doubled  Coordinates}
\noindent The section condition encapsulates the concept of \textit{doubled-yet-gauged coordinates}~\cite{Park:2013mpa} in DFT, where coordinates serve only as labels for the dynamical fields. This contrasts with the particle worldline or string worldsheet actions, where the target spacetime coordinates themselves become dynamical fields and must be explicitly gauged~\cite{Lee:2013hma, Park:2016sbw, Ko:2016dxa,Basile:2019pic}.

The usual coordinate basis of differential one-forms, $\rd x^{A}$, is not $\ODD$-symmetric under infinitesimal passive coordinate transformations,\footnote{Active transformations are set  by the generalised Lie derivative~(\ref{hcL}).} $x^{A} \rightarrow x^{A} + \xi^{A}$, as shown below:
\be
\delta(\rd x^{A})=\rd\xi^{A}=\rd x^{B}\partial_{B}\xi^{A}\neq\rd x^{B}(\partial_{B}\xi^{A}-\partial^{A}\xi_{B})\,.
\ee
To restore $\ODD$ symmetry, a derivative-index-valued gauge potential is introduced to explicitly gauge the doubled coordinates. This defines a new gauged differential:
\be
\ba{ll}
\rD x^{A}:=\rd x^{A}-\fa^{A}\,,\qquad&\quad \fa^{A}\partial_{A}=0\,.
\ea
\label{rDx}
\ee
The  passive transformations for the coordinates $x^{A}$ and the gauge potential $\fa^{A}$ are given by:
\be
\ba{lll}
\delta x^{A}=\xi^{A}\,,\quad&\quad\delta \fa^{A}=\rD x^{B}\partial^{A}\xi_{B}\,,\quad&\quad\delta\fa^{A}\partial_{A}=0\,.
\ea
\ee
This leads to an $\ODD$-symmetric transformation for the gauged differential:
\be
\delta (\rD x^{A})=\rD x^{B}(\partial_{B}\xi^{A}-\partial^{A}\xi_{B})\,.
\ee

To measure the distance between gauge orbits, a proper length can be defined through a path integral~\cite{Morand:2017fnv}:
\be
L=-\ln\left[\int\mathfrak{D}\fa~\exp\left(-\int\sqrt{\rD x^{A}\rD x^{B}\cH_{AB}}~\right)\right]\,,
\label{PL}
\ee 
where $\rD x^{A}$ is the gauged coordinate differential defined in~\eqref{rDx}, 
$\fa$ is the auxiliary gauge potential to be integrated out, and 
$\cH_{AB}$ is the  DFT-metric, or the generalised metric, which will be detailed later.

This definition of proper length naturally leads to a doubled  worldline action for a particle~\cite{Ko:2016dxa}:
\be\dis{
S=\frac{1}{\,\ls}
\int\rd\tau~\frac{1}{2} e^{-1\,}\rD_{\tau}x^{A}\rD_{\tau}x^{B}\cH_{AB}(x)-\frac{1}{2} (m\,\ls)^{2}e\,,}
\label{particleaction}
\ee
which further generalises to a completely covariant, doubled  worldsheet action for a string~\cite{Hull:2006va,Lee:2013hma},
\be
S=\frac{1}{4\pi\alphap}\displaystyle{\int}\rmd^{2}\sigma~-\half\sqrt{-h}h^{\alpha\beta}\rD_{\alpha}x^{M}\rD_{\beta}x^{N}\cH_{MN}
-\epsilon^{\alpha\beta}\rD_{\alpha}x^{M}\fa_{\beta M}\,.
\label{stringaction}
\ee
This framework can be extended to a $\kappa$-symmetric Green–Schwarz superstring, as discussed in~\cite{Park:2016sbw}.

\summary{The doubled-yet-gauged construction preserves $\ODD$ symmetry while constraining  dynamics to consistent $D$-dimensional sections. The  gauging  procedure leads to $\ODD$-symmetric particle and string actions, extendable to supersymmetric cases.}

\subsection{Fundamental  Field Variables: Non-Riemannian Geometry}
\aim{To define the generalised metric and dilaton, introduce projection operators and vielbeins, and set up the non-Riemannian sector.}
\noindent The fundamental fields of DFT  are the generalised metric, $\cH_{AB}$, and the dilaton, $d$, which together define the geometric and gravitational structure of the theory. These fields encapsulate the stringy nature of spacetime geometry and play a central role in the formulation of DFT.  

By definition, the generalised metric $\cH_{AB}$ is a symmetric $\ODD$ element:
\be
\ba{ll}
\cH_{AB}=\cH_{BA}\,,\qquad&\qquad\cH_{A}{}^{C}\cH_{B}{}^{D}\cJ_{CD}=\cJ_{AB}\,,
\ea
\label{cHdef}
\ee
and the  exponentiated form of the dilaton, $e^{-2d}$, serves as the integral measure in DFT: it is a scalar density of unit weight.  In words, the generalised metric  squares  to the identity: $\cH_{A}{}^{B}\cH_{B}{}^{C}=\delta_{A}{}^{C}$. Their generalised Lie derivatives~(\ref{hcL}) are explicitly given by
\be
\ba{ll}
\hcL_{\xi}\cH_{AB}=\xi^{C}\partial_{C}\cH_{AB}+2\partial_{[A}\xi_{C]}\cH^{C}{}_{B}+2\partial_{[B}\xi_{C]}\cH_{A}{}^{C}\,,~~~\,&~~~\hcL_{\xi}d=-\frac{1}{2}e^{2d}\hcL_{\xi}\left(e^{-2d}\right)=\xi^{A}\partial_{A}d-\frac{1}{2}\partial_{A}\xi^{A}\,.
\ea
\ee

Combining the invariant metric $\cJ_{AB}$ and the generalised metric $\cH_{AB}$, we define symmetric projection matrices~\cite{Jeon:2010rw}:
\be
\ba{ll}
P_{AB}=P_{BA}=\half(\cJ_{AB}+\cH_{AB})\,,\quad&\quad P_{A}{}^{B}P_{B}{}^{C}=P_{A}{}^{C}\,,\\
\brP_{AB}=\brP_{BA}=\half(\cJ_{AB}-\cH_{AB})\,,\quad&\quad \brP_{A}{}^{B}\brP_{B}{}^{C}=\brP_{A}{}^{C}\,,
\ea
\label{P1}
\ee
which are orthogonal and complete:
\be
\ba{ll}
 P_{A}{}^{B}\brP_{B}{}^{C}=0\,,\qquad&\qquad
P_{A}{}^{B}+\brP_{A}{}^{B}=\delta_{A}{}^{B}\,.
\ea
\label{P2}
\ee

Taking the ``square roots" of the  projection matrices, we introduce a pair of vielbeins, $V_{Ap}$ and $\brV_{B\brq\,}$:
\be
\ba{ll}
P_{AB}=V_{A}{}^{p}V_{B}{}^{q}\eta_{pq}\,,\qquad&\qquad
\brP_{AB}=\brV_{A}{}^{\brp}\brV_{B}{}^{\brq}\breta_{\brp\brq}\,.
\ea
\label{PPVV}
\ee
The vielbeins satisfy the following defining property:
\be
V_{A}{}^{p}V_{B}{}^{q}\eta_{pq}+\brV_{A}{}^{\brp}\brV_{B}{}^{\brq}\breta_{\brp\brq}=\cJ_{AB}\,.
\label{VVdef1}
\ee
Essentially,  viewed as a $({D+D}) \times ({D+D})$ matrix, the pair $\big(V_{A}{}^{p}, \brV_{A}{}^{\brp}\big)$ simultaneously diagonalises $\cJ_{AB}$ and $\cH_{AB}$ into $\diag(\eta, +\breta)$ and $\diag(\eta, -\breta)$, respectively. Furthermore, since the left inverse coincides with the right inverse, this property~\eqref{VVdef1} is equivalent to:
\be
\ba{lll}
V_{Ap}V^{A}{}_{q}=\eta_{pq}\,,\qquad&\qquad
\brV_{A\brp}\brV^{A}{}_{\brq}=\breta_{\brp\brq}\,,\qquad&\qquad
V_{Ap}\brV^{A}{}_{\brq}=0\,.
\ea
\label{VVdef2}
\ee
It follows that
\be
\ba{llll}
P_{A}{}^{B}V_{Bp}=V_{Ap}\,,\qquad&\quad
\brP_{A}{}^{B}\brV_{B\brp}=\brV_{A\brp}\,,\qquad&\quad
P_{A}{}^{B}\brV_{B\brp}=0=\brP_{A}{}^{B}V_{Bp}\,.
\ea
\ee

The presence of twofold vielbeins, $V_{A}{}^{p}$ and $\brV_{A}{}^{\brp}$, and spin groups, $\SpinD \times \oSpinD$, is a distinctly string-theoretic feature. This reflects the existence of two separate locally inertial frames: one associated with the left-moving sector, and
the other associated with the right-moving sector of closed strings~\cite{Duff:1986ne}.  Naturally, 
there are also two types of spinorial fermions,  as seen in supersymmetric DFTs~\cite{Jeon:2011sq,Jeon:2012hp} and as conjectured  in the Standard Model of particle physics coupled to DFT~\cite{Choi:2015bga}.  This dual structure could offer testable predictions for DFT and string theory.  Moreover, it facilitates the unification of type IIA and IIB supergravities in the ${D=10}$ maximally supersymmetric DFT~\cite{Jeon:2012hp} that has been explicitly constructed to the complete, quartic order in fermions.

The generalised metric and  the vielbeins are constrained by the properties~(\ref{cHdef}), (\ref{VVdef1}), and (\ref{VVdef2}). Consequently, their infinitesimal variations satisfy the following relations:\footnote{Observe the implicit index contractions in $(P\delta\cH\brP)_{AB}$, where the expression expands as $P_{A}{}^{C}\delta\cH_{CD}\brP^{D}{}_{B}$.}
\be
\delta\cH_{AB}=(P\delta\cH\brP)_{AB}+(\brP\delta\cH P)_{AB}\,,
\label{deltacH}
\ee
and
\be
\ba{ll}
\delta V_{Ap}=\brV_{A\brq}\brV^{B\brq}\delta V_{Bp}+V_{A}{}^{q}V_{B[q}\delta V^{B}{}_{p]}\,,\quad&\quad
\delta \brV_{A\brp}=V_{Aq}V^{Bq}\delta \brV_{B\brp}+\brV_{A}{}^{\brq}\brV_{B[\brq}\delta \brV^{B}{}_{\brp]}\,,
\ea
\label{deltaVVVV}
\ee
where  the second terms on the right-hand sides correspond to local Lorentz rotations.

\subsubsection{Parametrisations: Riemannian  \textit{\,vs.}  non-Riemannian}
\aim{To present explicit parametrisations of the generalised metric in both Riemannian and non-Riemannian forms.}
\noindent DFT and its supersymmetric extensions offer a unified framework to describe stringy spacetime dynamics via the fundamental fields, $\{\cH_{AB},d\}$ and $\{V_{Ap},\brV_{B\brq},d\}$. These fields satisfy key defining relations, such as~\eqref{cHdef} and~\eqref{VVdef1}, and allow for the classification of DFT geometries through the introduction of two non-negative integers, $(n,\brn)$, whose sum is bounded: $0\leq n+\brn\leq D$.   

With  $1\leq i,j\leq n$ and $1\leq \bri,\brj\leq\brn$,    the DFT metric satisfying (\ref{cHdef})  takes the most general form~\cite{Morand:2017fnv},
\be
\cH_{AB}=\left(\ba{cc}\cH^{\mu\nu}&
-\cH^{\mu\sigma}B_{\sigma\lambda}+Y_{i}^{\mu}X^{i}_{\lambda}-
\brY_{\bri}^{\mu}\brX^{\bri}_{\lambda}\\
B_{\kappa\rho}\cH^{\rho\nu}+X^{i}_{\kappa}Y_{i}^{\nu}
-\brX^{\bri}_{\kappa}\brY_{\bri}^{\nu}\quad&~~
~~\cK_{\kappa\lambda}-B_{\kappa\rho}\cH^{\rho\sigma}B_{\sigma\lambda}
+2X^{i}_{(\kappa}B_{\lambda)\rho}Y_{i}^{\rho}
-2\brX^{\bri}_{(\kappa}B_{\lambda)\rho}\brY_{\bri}^{\rho}
\ea\right)\,.
\label{cHFINAL}
\ee
Here, $\cH$ and $\cK$ are symmetric,    $B$ is skew-symmetric,
\be
\ba{lll}
\cH^{\mu\nu}=\cH^{\nu\mu}\,,\quad&\qquad \cK_{\mu\nu}=\cK_{\nu\mu}\,,\quad&\qquad
B_{\mu\nu}=-B_{\nu\mu}\,; 
\ea
\ee
$\cH$ and $\cK$ admit kernels, 
\be
\ba{ll} 
\cH^{\mu\nu}X^{i}_{\nu}=0=\cH^{\mu\nu}\brX^{\bri}_{\nu}\,,\quad&\qquad
\cK_{\mu\nu}Y_{j}^{\nu}=0=\cK_{\mu\nu}\brY_{\brj}^{\nu}\,;
\ea
\label{HXX}
\ee
and  a completeness relation must hold,
\be
\cH^{\mu\rho}\cK_{\rho\nu}
+Y_{i}^{\mu}X^{i}_{\nu}+\brY_{\bri}^{\mu}\brX^{\bri}_{\nu}
=\delta^{\mu}{}_{\nu}\,.
\label{COMPL}
\ee
\noindent The linear independence of the kernel’s zero-eigenvectors implies that
\be
\ba{lllll}
Y^{\mu}_{i}X_{\mu}^{j}=\delta_{i}{}^{j}\,,&\,\,
\brY^{\mu}_{\bri}\brX_{\mu}^{\brj}=\delta_{\bri}{}^{\brj}\,,&\,\,
Y^{\mu}_{i}\brX_{\mu}^{\brj}=0=
\brY^{\mu}_{\bri}X_{\mu}^{j}\,,&\,\,
\cH^{\rho\mu}\cK_{\mu\nu}\cH^{\nu\sigma}=\cH^{\rho\sigma}\,,&\,\,
\cK_{\rho\mu}\cH^{\mu\nu}\cK_{\nu\sigma}=\cK_{\rho\sigma}\,,
\ea
\ee
and  hence the $\ODD$-invariant trace of the DFT metric amounts to $\cH_{A}{}^{A}=2(n-\brn)$.

The precise  expression  of the $(n,\brn)$ DFT  metric~(\ref{cHFINAL}) and  the  fundamental algebraic relations (\ref{HXX}), (\ref{COMPL}) are all   invariant under  $\mathbf{GL}(n)\times\mathbf{GL}(\brn)$ local rotations,
\be
\ba{ccc}
\left(X^{i}_{\mu}\,,\,Y_{i}^{\mu}\,,\,\brX^{\bri}_{\mu}\,,\,\brY_{\bri}^{\nu}\right)\quad&\longmapsto&\quad
\left(X^{j}_{\mu}\,R_{j}{}^{i}\,,\,R^{-1}{}_{i}{}^{j}\,Y_{j}^{\nu}\,,\,\brX^{\brj}_{\mu}\,\brR_{\brj}{}^{\bri}\,,\,\brR^{-1}{}_{\bri}{}^{\brj}\,\brY_{\brj}^{\nu}\right)\,,
\ea
\label{GL2}
\ee
and, with  arbitrary local parameters, $V_{\mu i},\brV_{\mu\bri}$, under the following transformations,
\be
\ba{cll}
Y_{i}^{\mu}~~&\longmapsto&~~ Y_{i}^{\mu}+\cH^{\mu\nu}V_{\nu i}\,,\\
\brY_{\bri}^{\mu}~~&\longmapsto&~~\brY_{\bri}^{\mu}+\cH^{\mu\nu}\brV_{\nu\bri}\,,\\
\cK_{\mu\nu}~~&\longmapsto&~~ \cK_{\mu\nu}-2X^{i}_{(\mu}\cK_{\nu)\rho}\cH^{\rho\sigma}V_{\sigma i}-2\brX^{\bri}_{(\mu}\cK_{\nu)\rho}\cH^{\rho\sigma}\brV_{\sigma\bri}+(X_{\mu}^{i}V_{\rho i}+\brX_{\mu}^{\bri}\brV_{\rho\bri})\cH^{\rho\sigma}(X_{\nu}^{j}V_{\sigma j}+\brX_{\nu}^{\brj}\brV_{\sigma\brj})\,,\\
B_{\mu\nu}~~&\longmapsto&~~ B_{\mu\nu}
-2X^{i}_{[\mu}V_{\nu]i}+2\brX^{\bri}_{[\mu}\brV_{\nu]\bri}
+2X^{i}_{[\mu}\brX^{\bri}_{\nu]}\left(Y_{i}^{\rho}\brV_{\rho\bri}
+\brY_{\bri}^{\rho}V_{\rho i}+V_{\rho i}\cH^{\rho\sigma}\brV_{\sigma\bri}\right)\,.
\ea
\label{Milne2}
\ee
In fact, these two symmetries~(\ref{GL2}), (\ref{Milne2}) are  identifiable   as the twofold local Lorentz symmetries.

On the one hand,  in the case $(n,\brn)=(0,0)$,     $\cK_{\mu\nu}$ and  $\cH^{\mu\nu}$  coincide with the usual (invertible) Riemannian metric and its inverse. In this Riemannian limit, the $(0,0)$      DFT metric  takes the familiar  form~\cite{Giveon:1988tt},
\be
\cH_{AB}=\left(\ba{cc}
\quad g^{\mu\nu}\quad&\quad -g^{\mu\lambda}B_{\lambda\tau}\quad\\
\quad B_{\sigma\kappa}g^{\kappa\nu}\quad&\quad
g_{\sigma\tau}-B_{\sigma\kappa}g^{\kappa\lambda}B_{\lambda\tau}\quad
\ea\right)\,.
\label{00H}
\ee
The $\ODD$-symmetric proper length~(\ref{PL}), as well as the doubled particle and string actions~(\ref{particleaction}), (\ref{stringaction}), all reduce to their conventional (undoubled) forms after integrating out the auxiliary variables $\fa^{A}$.

On the other  hand, cases with $(n,\brn)
\neq (0,0)$ are inherently non-Riemannian,  as they lack an invertible Riemannian metric. Specifically, in the cases $(n,\brn) = (D,0)$ or $(0,D)$, $\cK_{\mu\nu}$ and $\cH^{\mu\nu}$ vanish, the relations  $Y_{i}^{\mu}X^{i}_{\nu}=\delta^{\mu}{}_{\nu}$ or $\brY_{\bri}^{\mu}\brX^{\bri}_{\nu}=\delta^{\mu}{}_{\nu}$ hold respectively, and the $B$-field becomes irrelevant.  In these  extremely non-Riemannian limits, the DFT metric coincides with the $\ODD$-invariant metric up to  sign, 
\be
\cH_{AB}=\pm\cJ_{AB}=\left(\ba{cc}
~ 0\quad&\quad\pm 1~\\
~ \pm 1\quad&\quad0~
\ea\right)\,.
\label{D0H}
\ee  
These cases represent two perfectly $\ODD$-symmetric vacua in bosonic DFT that are maximally non-Riemannian, characterized by the saturation, ${n+\brn=D}$. Remarkably, these vacua exhibit infinitely many isometries~\cite{Blair:2020gng}, as any local parameter $\xi^{A}$ automatically satisfies the Killing equation, defined by the condition ${\hcL_{\xi}\cH_{AB}=0}$. 

In this scenario,  the ordinary Riemannian spacetime~(\ref{00H}) emerges after spontaneous symmetry breaking of the fully $\ODD$-symmetric vacua~(\ref{D0H}), while the Riemannian metric and the $B$-field are identified as  Nambu–Goldstone bosons~\cite{Berman:2019izh,Park:2020ixf}.

On a generic $(n, \brn)$ non-Riemannian background, a particle described by the doubled action~(\ref{particleaction}) freezes in $n+\brn$ untilde directions, 
\be
\ba{ll}
\dis{X^{i}_{\mu}\frac{\rd x^{\mu}}{\rd\tau}=0}\,,\qquad&\qquad
\dis{\brX^{\bri}_{\mu}\frac{\rd x^{\mu}}{\rd\tau}=0}\,,
\ea
\label{FROZEN}
\ee
while the string in (\ref{stringaction}) becomes chiral and anti-chiral in the $n$ and $\brn$ untilde  directions, respectively:
\be
\ba{ll}
X^{i}_{\mu}\left(\partial_{\alpha}x^{\mu}+\textstyle{\frac{1}{\sqrt{-h}}}\epsilon_{\alpha}{}^{\beta}\partial_{\beta}x^{\mu}\right)=0\,,
\qquad&~~\quad
\brX^{\bri}_{\mu}\left(\partial_{\alpha}x^{\mu}-\textstyle{\frac{1}{\sqrt{-h}}}\epsilon_{\alpha}{}^{\beta}\partial_{\beta}x^{\mu}\right)=0\,.
\ea
\ee
This behaviour arises as the components of the auxiliary gauge potential $\fa^{A}$ act as Lagrange multipliers~\cite{Morand:2017fnv}.

Nonetheless, the $(n,\brn)$-classification of the generalised metric~(\ref{cHFINAL}) described above remains valid for bosonic DFT geometries. In the presence of fermions, particularly in supersymmetric DFTs~\cite{Jeon:2011sq,Jeon:2012hp}, the twofold spin groups are \textit{a priori} fixed, and the permissible values of $(n,\brn)$ are constrained by their dimensions and signatures. Specifically, the Euclidean spin group, $\Spin(D)\times\Spin(D)$, does not permit any non-Riemannian geometries, whereas the Minkowskian case, $\Spin(1,{D-1})\times\Spin({D-1},1)$, as assumed in Table~\ref{TABindices}, allows $(1,1)$ non-Riemannian geometry~\cite{Lee:2013hma,Ko:2015rha}, which corresponds to non-relativistic~\cite{Gomis:2000bd,Danielsson:2000gi,Gomis:2005pg} or Newton–Cartan string theories~\cite{Christensen:2013lma,Hartong:2015zia,Harmark:2017rpg,Harmark:2018cdl,Bergshoeff:2018yvt,Bergshoeff:2019pij,Harmark:2019upf,Bergshoeff:2021bmc,Oling:2022fft,Hartong:2022lsy}. Furthermore, certain known singularities in supergravity theories can be appropriately identified as regular $(1,1)$ non-Riemannian geometries~\cite{Morand:2021xeq}; see section~\ref{SphericalSOL} for an example.

 While we refer to \cite{Morand:2017fnv} for the general $(n,\brn)$ cases, here we   merely spell out the $(0,0)$ Riemannian parametrisation of the  DFT vielbeins,
\be
\ba{ll}
V_{Ap}=\frac{1}{\sqrt{2}}\left(\ba{c}e_{p}{}^{\mu}\\
e_{\nu}{}^{q}\eta_{qp}+B_{\nu\sigma}e_{p}{}^{\sigma}\ea\right)\,,\quad&\quad
\brV_{A\brp}=\frac{1}{\sqrt{2}}\left(\ba{c}\bre_{\brp}{}^{\mu}\\
\bre_{\nu}{}^{\brq}\breta_{\brq\brp}+B_{\nu\sigma}\bre_{\brp}{}^{\sigma}\ea\right)\,.
\ea
\label{00V}
\ee
Here $e_{\mu}{}^{p}$ and $\bre_{\mu}{}^{\brp}$ are a pair of (Riemannian) vielbeins which square to  the same  metric, 
\be
e_{\mu}{}^{p}e_{\nu p}=-\bre_{\mu}{}^{\brp}\bre_{\nu\brp}=g_{\mu\nu}\,.
\label{eeg}
\ee
Their dual presence allows  $\ODD$  to safely rotate the capital-letter indices of the vielbeins exclusively~\cite{Jeon:2011cn}.

Lastly, the DFT dilaton is  parametrised    by the Riemannian metric and the (weightless) string dilaton:
\be
e^{-2d}=\sqrt{-g}\,e^{-2\phi}\,.
\label{00d}
\ee

\summary{DFT geometries are classified into Riemannian and non-Riemannian cases, unifying stringy spacetime structures. Ordinary spacetime appears as the $(0,0)$ case, while maximally $\ODD$-symmetric vacua are  non-Riemannian and  chiral.}

\subsection{Christoffel Symbols \& Spin Connections}
\aim{To build a master derivative that unifies diffeomorphisms and local Lorentz symmetries.}
\noindent  The master derivative, denoted as $\cD_{A}$, is introduced to unify and covariantly describe all the  local symmetries: the doubled-yet-gauged diffeomorphisms and the twofold local Lorentz symmetries, $\Spin(1,D{-1}) \times \Spin(D{-1},1)$.  More explicitly, the master derivative is postulated as:
\be
\cD_{A}=\partial_{A}+\Gamma_{A}+\Phi_{A}+\brPhi_{A}\,.
\label{MastercD}
\ee
Here, $\Gamma_{A}$ refers to the Christoffel connection for the doubled-yet-gauged diffeomorphisms,  introduced in~\cite{Jeon:2011cn} and based  on  earlier work~\cite{Jeon:2010rw}.\footnote{One key lesson from \cite{Jeon:2010rw} is that the generalised metric alone is insufficient for fully constructing covariant derivatives and curvatures; the inclusion of the dilaton $d$ is essential.}\footnote{An alternative formulation was proposed in \cite{Siegel:1993th} and developed in \cite{Hohm:2010xe}, where curved $\ODD$ indices are mapped to flat Lorentz indices via the vielbeins~(\ref{VVdef1}). Although this appears to bypass the Christoffel connection, the spin connection necessarily reintroduces its content, as in (\ref{PhibrPhi}), so that the framework ultimately preserves parallels with conventional differential geometry.} The connection is defined as:
\be
\ba{ll}
\Gamma_{CAB}=&2\left(P\partial_{C}P\brP\right)_{[AB]}
+2\left({{\brP}_{[A}{}^{D}{\brP}_{B]}{}^{E}}-{P_{[A}{}^{D}P_{B]}{}^{E}}\right)\partial_{D}P_{EC}\\
{}&-4\left(\textstyle{\frac{1}{P_{M}{}^{M}-1}}P_{C[A}P_{B]}{}^{D}+\textstyle{\frac{1}{\brP_{M}{}^{M}-1}}\brP_{C[A}\brP_{B]}{}^{D}\right)\!\left(\partial_{D}d+(P\partial^{E}P\brP)_{[ED]}\right)\,.
\ea
\label{Gamma}
\ee
Additionally, $\Phi_{A}$ and $\brPhi_{A}$ are the spin connections corresponding to the twofold local Lorentz symmetries \cite{Jeon:2011vx},  given as:
\be
\ba{ll}
\Phi_{Apq}=\Phi_{A[pq]}=V^{B}{}_{p}\na_{A}V_{Bq}\,,\quad&\quad
\brPhi_{A\brp\brq}=\brPhi_{A[\brp\brq]}=\brV^{B}{}_{\brp}\na_{A}\brV_{B\brq}\,.
\ea
\label{PhibrPhi}
\ee
In our notation, $\na_{A}$ is   the  diffeomorphism-covariant derivative that involves the Christoffel symbols only,
\be
\na_{A}:=\partial_{A}+\Gamma_{A}\,,
\label{napG}
\ee
and,  ignoring  any  local Lorentz  indices,  acts   on a tensor density with weight $\omega$ as:
\be
\na_{C}T_{A_{1}A_{2}\cdots A_{n}}
:=\partial_{C}T_{A_{1}A_{2}\cdots A_{n}}-\omega_{{\scriptscriptstyle{T\,}}}\Gamma^{B}{}_{BC}T_{A_{1}A_{2}\cdots A_{n}}+
\sum_{i=1}^{n}\,\Gamma_{CA_{i}}{}^{B}T_{A_{1}\cdots A_{i-1}BA_{i+1}\cdots A_{n}}\,.
\label{asemicov}
\ee
In particular, in (\ref{PhibrPhi}),
\be
\ba{ll}
\na_{A}V_{Bq}=\partial_{A}V_{Bq}+\Gamma_{AB}{}^{C}V_{Cq}\,,\quad&\quad
\na_{A}\brV_{B\brq}=\partial_{A}\brV_{B\brq}+\Gamma_{AB}{}^{C}\brV_{C\brq}\,.
\ea
\label{naVV}
\ee

 The Christoffel connection~(\ref{Gamma}) is uniquely fixed by requiring the following three properties:
\begin{itemize}
\item[\textit{i)}] Full  compatibility with all  fundamental fields, 
\be
\ba{ll}
\cD_{A}P_{BC}=\na_{A}P_{BC}=0\,,\quad&\quad\cD_{A}\brP_{BC}=\na_{A}\brP_{BC}=0\,,\\
\multicolumn{2}{r}{
\cD_{A}d\,=\na_{A}d\,=-\half e^{2d}\na_{A}(e^{-2d})=\partial_{A}d+\half\Gamma^{B}{}_{BA}=0\,,}
\ea
\label{Gcomp}
\ee
which implies
\be
\ba{ll}
\cD_{A}\cJ_{BC}=\na_{A}\cJ_{BC}=0\,,\quad&\qquad\cD_{A}\cH_{BC}=\na_{A}\cH_{BC}=0\,,
\ea
\ee
such that $\Gamma_{A}$ is, as expected,  $\mathbf{so}(D,D)$-valued:
\be
\Gamma_{ABC}=-\Gamma_{ACB}\,.
\label{Gskew}
\ee
\item[\textit{ii)}] Torsionless cyclic property,\footnote{In full-order supersymmetric DFTs~\cite{Jeon:2011sq,Jeon:2012hp}, the torsions are quadratic in fermionic fields and naturally implement the 1.5 formalism characteristic of supergravity theories.}
\be
\Gamma_{ABC}+\Gamma_{BCA}+\Gamma_{CAB}=0\,,
\label{GABC}
\ee
which makes  $\na_{A}$    compatible  with  the generalised Lie derivative~(\ref{hcL}) and the C-bracket~(\ref{Cbracket})  so that ordinary derivatives in these expressions can be freely replaced by  $\na_{A}$:
\be
\hcL_{\xi}T_{A_{1}\cdots A_{n}}=\xi^{B}\na_{B}T_{A_{1}\cdots A_{n}}+\omega\na_{B}\xi^{B}T_{A_{1}\cdots A_{n}}+\sum_{i=1}^{n}~(\na_{A_{i}}\xi_{B}-\na_{B}\xi_{A_{i}})T_{A_{1}\cdots A_{i-1}}{}^{B}{}_{A_{i+1}\cdots  A_{n}}\,.
\label{hcLna}
\ee
\item[\textit{iii)}] Projection constraints,
\be
\ba{ll}
\cP_{ABC}{}^{DEF}\Gamma_{DEF}=0\,,\quad&\quad\bar{\cP}_{ABC}{}^{DEF}\Gamma_{DEF}=0\,.
\label{kernel}
\ea
\ee
These constraints involve   six-index  projectors,
\be
\ba{l}
\cP_{ABC}{}^{DEF}:=P_{A}{}^{D}P_{[B}{}^{[E}P_{C]}{}^{F]}+\textstyle{\frac{2}{P_{M}{}^{M}-1}}P_{A[B}P_{C]}{}^{[E}P^{F]D}\,,\\
\bar{\cP}_{ABC}{}^{DEF}:=\brP_{A}{}^{D}\brP_{[B}{}^{[E}\brP_{C]}{}^{F]}+\textstyle{\frac{2}{\brP_{M}{}^{M}-1}}\brP_{A[B}\brP_{C]}{}^{[E}\brP^{F]D}\,,
\ea
\label{P6}
\ee
which satisfy  the following projection properties,
\be
\ba{ll}
\cP_{ABC}{}^{DEF}\cP_{DEF}{}^{GHI}=\cP_{ABC}{}^{GHI}\,,\qquad&\qquad
\brcP_{ABC}{}^{DEF}\brcP_{DEF}{}^{GHI}=\brcP_{ABC}{}^{GHI}\,,
\ea
\ee
various symmetric  relations~\cite{Cho:2015lha},
\be
\ba{lll}
\cP_{ABCDEF}=\cP_{DEFABC}\,,\quad&\quad\cP_{ABCDEF}=\cP_{A[BC]D[EF]}\,,
\quad&\quad \cP_{[AB]CDEF}=\cP_{CAB[EF]D}\,,\\
\brcP_{ABCDEF}=\brcP_{DEFABC}\,,\quad&\quad\brcP_{ABCDEF}=\brcP_{A[BC]D[EF]}\,,
\quad&\quad \brcP_{[AB]CDEF}=\brcP_{CAB[EF]D}\,,
\ea
\label{symP6}
\ee
and traceless conditions,
\be
\ba{ll}
\quad P^{AB}\cP_{ABCDEF}=0\,,\qquad&\quad
\quad \brP^{AB}\brcP_{ABCDEF}=0\,.
\ea
\label{symP6s}
\ee
\end{itemize}
~\\

Unlike the  GR Christoffel symbols, there exist  no normal coordinates where the DFT Christoffel symbols would vanish pointwise. The Equivalence Principle holds   for  point particles but not for  extended objects such as strings~\cite{Choi:2015bga}.  Parallel  to the variation of the  Christoffel symbols in GR:
\[\delta\gamma^{\lambda}_{\mu\nu}=\frac{1}{2}(\trd_{\mu}\delta g^{\lambda}{}_{\nu}+\trd_{\nu}\delta g_{\mu}{}^{\lambda}-\trd^{\lambda}\delta g_{\mu\nu})\,,
\]  
the DFT Christoffel symbols vary infinitesimally by the  DFT metric and dilaton: with $\delta P_{AB}=\frac{1}{2}\delta\cH_{AB}$,
\be
\ba{rll}
{\delta\Gamma}_{CAB}&=&2P_{[A}^{~D}\brP_{B]}^{~E}\DO_{C}\delta P_{DE}+2(\brP_{[A}^{~D}\brP_{B]}^{~E}-P_{[A}^{~D}P_{B]}^{~E})\DO_{D}\delta P_{EC}\\
{}&{}&-\textstyle{\frac{4}{D-1}}(\brP_{C[A}\brP_{B]}^{~D}+P_{C[A}P_{B]}^{~D})(\partial_{D}\delta d+P_{E[G}\DO^{G}\delta P^{E}_{~D]})-\Gamma_{FDE\,}\delta(\cP+\brcP)_{CAB}{}^{FDE}\,.
\ea
\label{deltaG}
\ee
Once the DFT Christoffel symbols are fixed, the spin connections~(\ref{PhibrPhi}) follow  naturally  from the compatibility of the master derivative with the DFT vielbeins: from (\ref{naVV}),
\be
\ba{ll}
\cD_{A}V_{Bp}=\na_{A}V_{Bp}+\Phi_{Ap}{}^{q}V_{Bq}=0\,,\qquad&\quad
\cD_{A}\brV_{B\brp}=\na_{A}\brV_{B\brp}+\brPhi_{A\brp}{}^{\brq}\brV_{B\brq}=0\,.
\ea
\label{cDVV}
\ee
The master derivative is also compatible with the metrics and gamma matrices of the twofold spin groups,
\be
\ba{llll}
\cD_{A}\eta_{pq}=0\,,\quad&\quad
\cD_{A}\breta_{\brp\brq}=0\,,\quad&\quad\cD_{A}(\gamma^{p})^{\alpha}{}_{\beta}=0\,,\quad&\quad
\cD_{A}(\brgamma^{\brp})^{\bralpha}{}_{\brbeta}=0\,,
\ea
\ee
and thus, analogous to GR,
\be
\ba{llll}
\Phi_{Apq}=-\Phi_{Aqp}\,,\quad&\quad
\brPhi_{A\brp\brq}=-\brPhi_{A\brq\brp}\,,\quad&\quad
\Phi_{A}{}^{\alpha}{}_{\beta}=\quarter\Phi_{Apq}(\gamma^{pq})^{\alpha}{}_{\beta}\,,\quad&\quad 
\brPhi_{A}{}^{\bralpha}{}_{\brbeta}=\quarter\brPhi_{A\brp\brq}(\brgamma^{\brp\brq})^{\bralpha}{}_{\brbeta}\,.
\ea
\label{Phisv}
\ee
It is worthwhile to note a formula that relates the three connections~\cite{Cho:2015lha}:
\be
\Gamma_{CAB}=
V_{A}{}^{p}\partial_{C}V_{Bp}+\brV_{A}{}^{\brp}\partial_{C}\brV_{B\brp}+
V_{A}{}^{p}V_{B}{}^{q}\Phi_{Cpq}+
\brV_{A}{}^{\brp}\brV_{B}{}^{\brq}\brPhi_{C\brp\brq}\,,
\label{GPhibrPhi}
\ee
ensuring that Cartan's structure equations in GR hold analogously in DFT~\cite{Cho:2015lha}.

Due to the section condition,    the Christoffel symbols satisfy
\be
P_{A}{}^{C}\brP_{B}{}^{D}\Gamma^{E}{}_{CD}\partial_{E}=0\,.
\label{Gps}
\ee
\summary{The  Christoffel and spin connections in DFT are uniquely determined by compatibility and extra (torsionless)  conditions, extending GR into the doubled framework.}

\subsection{From Semi-Covariance to Full Covariance}
\aim{To refine the semi-covariant derivative into a fully covariant derivative through projections.}
\noindent Explicitly, the master  derivative~(\ref{MastercD}) acts as
\be
\ba{l}
\cD_{A}T_{Bp}{}^{\alpha}{}_{\brp}{}^{\bralpha}=
\na_{A}T_{Bp}{}^{\alpha}{}_{\brp}{}^{\bralpha}+
\Phi_{Ap}{}^{q}
T_{Bq}{}^{\alpha}{}_{\brp}{}^{\bralpha}+\Phi_{A}{}^{\alpha}{}_{\beta}T_{Bp}{}^{\beta}{}_{\brp}{}^{\bralpha}+\brPhi_{A\brp}{}^{\brq}
T_{Bp}{}^{\alpha}{}_{\brq}{}^{\bralpha}+\brPhi_{A}{}^{\bralpha}{}_{\brbeta}T_{Bp}{}^{\alpha}{}_{\brp}{}^{\brbeta}\\
\scalebox{0.97}{$=\partial_{A}T_{Bp}{}^{\alpha}{}_{\brp}{}^{\bralpha}-\omega
\Gamma^{C}{}_{CA}T_{Bp}{}^{\alpha}{}_{\brp}{}^{\bralpha}
{+\Gamma_{AB}{}^{C}}T_{Cp}{}^{\alpha}{}_{\brp}{}^{\bralpha}
{+\Phi_{Ap}{}^{q}}
T_{Bq}{}^{\alpha}{}_{\brp}{}^{\bralpha}{+\Phi_{A}{}^{\alpha}{}_{\beta}}T_{Bp}{}^{\beta}{}_{\brp}{}^{\bralpha}{+\brPhi_{A\brp}{}^{\brq}}
T_{Bp}{}^{\alpha}{}_{\brq}{}^{\bralpha}{+\brPhi_{A}{}^{\bralpha}{}_{\brbeta}}T_{Bp}{}^{\alpha}{}_{\brp}{}^{\brbeta}$}\,.
\ea
\label{MastercDexplicit}
\ee
The master derivative is fully covariant with respect to the twofold local Lorentz symmetries but is, \textit{a priori}, only \textit{semi-covariant} under the doubled-yet-gauged diffeomorphisms. To achieve complete covariance under these diffeomorphisms, an additional step of projection  is required, which we explain.

\subsubsection{Tensorial Covariant Derivatives} 
\aim{To obtain a fully covariant derivative for tensors.}
\noindent Under diffeomorphisms, the DFT Christoffel symbols transform as
\be
\delta_{\xi}\Gamma_{CAB}=\hcL_{\xi}\Gamma_{CAB}
-2\partial_{C}\partial_{[A}\xi_{B]}+2(\cP+\brcP)_{CAB}{}^{DEF}\partial_{D}\partial_{[E}\xi_{F]}\,,
\label{GPbrP}
\ee
such that $\na_{A}$, and consequently $\cD_{A}$, are not inherently covariant under the diffeomorphisms:
\be
\delta_{\xi}\big(\na_{C}T_{A_{1}\cdots A_{n}}\big)=\hcL_{\xi}\big(\na_{C}T_{A_{1}\cdots A_{n}}\big)+
\dis{\sum_{i=1}^{n}2(\cP{+\brcP})_{CA_{i}}{}^{BDEF}
\partial_{D}\partial_{E}\xi_{F}\,T_{A_{1}\cdots A_{i-1} BA_{i+1}\cdots A_{n}}\,.}
\label{diffeoanormalous}
\ee
However, the anomalous terms consistently appear through the six-index projectors~(\ref{P6}), making them straightforward to project out. For instance, in~(\ref{diffeoanormalous}), projecting the derivative index and the tensor indices in opposite ways allows the construction of fully covariant derivatives~\cite{Jeon:2011cn}:
\be
\ba{ll}
P_{C}{}^{D}{\brP}_{A_{1}}{}^{B_{1}}\cdots{\brP}_{A_{n}}{}^{B_{n}}
\DO_{D}T_{B_{1}\cdots B_{n}}\,,\quad&\qquad
{\brP}_{C}{}^{D}P_{A_{1}}{}^{B_{1}}\cdots P_{A_{n}}{}^{B_{n}}
\DO_{D}T_{B_{1}\cdots B_{n}}\,.
\ea
\label{covD}
\ee
Similarly, using  (\ref{symP6}) and  (\ref{symP6s}),  we obtain fully covariant \textit{divergences}, 
\be
\ba{ll}
P^{AB}{\brP}_{C_{1}}{}^{D_{1}}\cdots{\brP}_{C_{n}}{}^{D_{n}}\DO_{A}T_{BD_{1}\cdots D_{n}}\,,\quad&\qquad
\brP^{AB}{P}_{C_{1}}{}^{D_{1}}\cdots{P}_{C_{n}}{}^{D_{n}}\DO_{A}T_{BD_{1}\cdots D_{n}}\,.
\ea
\label{covdiv}
\ee
For instance, the  divergence of  a weightless vector, $J^{A}$, reads
\be
\na_{A}\!\left(e^{-2d}J^{A}\right)=\partial_{A}\!\left(e^{-2d}J^{A}\right)=e^{-2d}\na_{A}J^{A}=e^{-2d}\left(P^{AB}+\brP^{AB}\right)\na_{A}J_{B}\,.
\label{divJ}
\ee

Nevertheless, when acting on the two-index projectors,  $P_{AB}$ and $\brP_{AB}$, the anomalous terms in (\ref{diffeoanormalous}) are automatically eliminated due to the properties of the six-index projectors, (\ref{symP6}) and (\ref{symP6s}). This guarantees the full covariance of the compatibility condition, $\na_{C}P_{AB}=0=\na_{C}\brP_{AB}$, as postulated in (\ref{Gcomp}).

In accordance with (\ref{deltacH}), the partial derivative of the generalised metric satisfies a projection property,
\be
\partial_C\cH_{AB}=(P\partial_C\cH\brP)_{AB}+(\brP \partial_C\cH P)_{AB}\,,
\ee
where the two free indices are projected in opposite manners. This property essentially precludes the possibility of canceling the anomalous terms in (\ref{GPbrP}) by adding any derivatives of the generalised metric or dilaton to $\Gamma_{CAB}$ in (\ref{Gamma}). While introducing additional complementary fields could, in principle, eliminate the anomalous terms, such fields remain "undetermined" or are not naturally identifiable within string theory.

 Notably, the additional projection step~(\ref{covD}) highlights the more refined and rigid structure of DFT compared to GR. For instance, while DFT provides two available "metrics," namely $\cJ_{AB}$ and $\cH_{AB}$, the pairwise contraction of $\ODD$ indices is significantly restricted. Specifically, when constructing a scalar from the squared derivative $
\na_{A}T_{B_{1}B_{2}\cdots B_{n}}$ for a kinetic term, from (\ref{covD}) there are only two viable ways to "square" it, rather than the $2^{n+1}$ possibilities that might otherwise exist.

As a further illustration, consider a doubled Yang--Mills potential $\bfA_{A}$. The fully covariant skew-symmetric field strength is expressed as~\cite{Jeon:2011kp,Choi:2015bga}
\be
P_{A}{}^{C}\brP_{B}{}^{D}\bfF_{CD}=
P_{A}{}^{C}\brP_{B}{}^{D}\big(\na_{C}\bfA_{D}-\na_{D}\bfA_{C}-i\left[\bfA_{C},\bfA_{D}\right]\big)\,,
\label{YMF0}
\ee
and there is  only one way to construct the doubled Yang--Mills   action:
\be
\dis{\int_{\Sigma_D}e^{-2d}P^{AC}\brP^{BD}\Tr(\bfF_{AB}\bfF_{CD})\,,}
\label{YMaction}
\ee
where  $\Sigma_{D}$ is a $D$-dimensional section over which the integral is taken with the measure, $e^{-2d}$.  

The doubled Yang--Mills potential  decomposes into a  displacement vector  and an ordinary one-form: $\bfA_{A} = (\varphi^{\mu}, A_{\nu})$. This structure enables the doubled Yang--Mills action to provide a unified description of gluons/photons and (non-Abelian) phonons~\cite{Angus:2021jvm}. However, it is always possible to eliminate the displacement phonon vector by imposing an $\ODD$-symmetric constraint, $\bfA^{A} \partial_{A} = 0$, analogous to the section condition~(\ref{SC})~\cite{Choi:2015bga}.

 We proceed to  construct fully   covariant second-order differential operators, for which  we need to identify the relevant  anomalous terms.  Making use of  (\ref{diffeoanormalous}) repeatedly with care,   we get
\be
\ba{rll}
\big(\delta_{\xi}-\hcL_{\xi}\big)\big(\na_{B}\na_{C}T_{A_{1}\cdots A_{n}}\big)&=&2 (\cP{+\brcP})_{BC}{}^{DEFG}
\partial_{E}\partial_{F}\xi_{G}\,\na_{D}T_{A_{1}\cdots A_{n}}\\
{}&{}&+
\dis{\sum_{i=1}^{n}}\left[\ba{l}
2(\cP{+\brcP})_{CA_{i}}{}^{DEFG}T_{A_{1}\cdots A_{i-1} DA_{i+1}\cdots A_{n}}\na_{B}\left(
\partial_{E}\partial_{F}\xi_{G}\right)\\
+
2 (\cP{+\brcP})_{CA_{i}}{}^{DEFG}
\partial_{E}\partial_{F}\xi_{G}\,\na_{B}T_{A_{1}\cdots A_{i-1} DA_{i+1}\cdots A_{n}}\\
+
2 (\cP{+\brcP})_{BA_{i}}{}^{DEFG}
\partial_{E}\partial_{F}\xi_{G}\,\na_{C}T_{A_{1}\cdots A_{i-1} DA_{i+1}\cdots A_{n}}\ea
\right]\,.
\ea
\label{diffeoanormalous2}
\ee
Again from the symmetric and traceless relations of the six-index projectors~(\ref{symP6}),  (\ref{symP6s}),  it is straightforward to obtain   fully covariant \textit{d'Alembertians}:
\be
\ba{ll}
P^{AB}{\brP}_{C_{1}}{}^{D_{1}}\cdots{\brP}_{C_{n}}{}^{D_{n}}
\DO_{A}\DO_{B}T_{D_{1}\cdots D_{n}}\,,\quad&\qquad
{\brP}^{AB}P_{C_{1}}{}^{D_{1}}\cdots P_{C_{n}}{}^{D_{n}}
\DO_{A}\DO_{B}T_{D_{1}\cdots D_{n}}\,.
\ea
\label{coco1}
\ee
Later, through  equations (\ref{DELTA}), (\ref{brDELTA}), (\ref{covseccon}), and (\ref{bBox}), we will encounter a pair of general forms of the d'Alembertian,  which yield    a box operator.  These operators are crafted to act on arbitrary multi-index tensor densities while \textit{a priori} incorporating four-index (Riemann) curvature.

\summary{Projections remove anomalous terms and enforce strict covariance, showing that DFT is structurally more rigid than GR.}

\subsubsection{Spinorial Covariant Derivatives} 
\aim{To extend covariance to spinors and the  RR sector by constructing \textit{Dirac operators}.}
\noindent It is possible to freely replace $\na_A$ in the fully covariant derivatives~(\ref{covD}), (\ref{covdiv}), and (\ref{coco1}) with the master derivative $\cD_A$, while simultaneously contracting their projected, otherwise unconstrained $\ODD$ vector indices using the DFT vielbeins, $V_{Ap}$ and $\brV_{A\brq}$. Due to the compatibility of the DFT vielbeins with the master derivative~(\ref{cDVV}), this replacement reduces the fully covariant derivatives to more streamlined forms:
\be
\ba{llllll}
\cD_{p}T_{\brq_{1}\cdots \brq_{n}}\,,\quad&
\cD_{\brp}T_{q_{1}\cdots q_{n}}\,,\quad&
\cD_{p}T^{p}{}_{\brq_{1}\cdots\brq_{n}}\,,\quad&
\cD_{\brp}T^{\brp}{}_{q_{1}\cdots q_{n}}\,,\quad&
\cD_{p}\cD^{p}T_{\brq_{1}\cdots \brq_{n}}\,,\quad&
\cD_{\brp}\cD^{\brp}T_{q_{1}\cdots q_{n}}\,,
\ea
\label{coco2}
\ee
where $\cD_{p}=V^{A}{}_{p}\cD_{A}$ and  $\cD_{\brp}=\brV^{A}{}_{\brp}\cD_{A\,}$.

The fully covariant doubled   Yang--Mills field strength~(\ref{YMF0})  also reads
\be
\bfF_{p\brq}:=V^{A}{}_{p}\brV^{B}{}_{\brq}\big(\na_{A}\bfA_{B}-\na_{B}\bfA_{A}-i\left[\bfA_{A},\bfA_{B}\right]\big)=\cD_{p}\bfA_{\brq}-\cD_{\brq}\bfA_{p}-i\left[\bfA_{p},\bfA_{\brq}\right]\,.
\label{YMF}
\ee
The full covariance of (\ref{coco2}) and (\ref{YMF}) can also be directly confirmed by observing  that, from
\be
\ba{ll}
\delta_{\xi}\Phi_{Apq}=\hcL_{\xi}\Phi_{Apq}+2\cP_{Apq}{}^{DEF}\partial_{D}\partial_{[E}\xi_{F]}\,,\quad&\qquad
\delta_{\xi}\brPhi_{A\brp\brq}=\hcL_{\xi}\brPhi_{A\brp\brq}+2\brcP_{A\brp\brq}{}^{DEF}\partial_{D}\partial_{[E}\xi_{F]}\,,
\ea
\ee 
the following projected components of the spin connections are fully covariant under doubled-yet-gauged  diffeomorphisms:\footnote{The quantities in (\ref{covPhi}) are essentially the ``generalised fluxes''  considered in \cite{Aldazabal:2011nj,Grana:2012rr,Geissbuhler:2013uka} as the building blocks of DFT.}
\be
\ba{llllll}
\Phi_{\brr pq}\,,\quad&\quad\brPhi_{r\brp\brq}\,,\quad&\quad\Phi_{[pqr]}\,,\quad&~
\brPhi_{[\brp\brq\brr]}\,,\quad&\quad\Phi^{p}{}_{pq}\,,\quad&\quad
\brPhi^{\brp}{}_{\brp\brq}\,,
\ea
\label{covPhi}
\ee 
where  we have set $\Phi_{\brr pq}=\brV^{A}{}_{\brr}\Phi_{A pq}\,$ and  $\,\Phi_{rpq}=V^{A}{}_{r}\Phi_{A pq}$, \textit{etc.}

Consequently, when acting on $\SpinD$ spinors, such as $\rho^{\alpha}$ or $\psi_{\brp}^{\alpha}$, or on $\oSpinD$ spinors, such as $\rho^{\prime\bralpha}$ or $\psi^{\prime\bralpha}_{p}$, the   fully  covariant \textit{Dirac operators} are~\cite{Jeon:2011vx,Jeon:2011sq}
\be
\ba{llllllll}
\gamma^{p}\cD_{p}\rho\,,\quad&~
\gamma^{p}\cD_{p}\psi_{\brp}\,,\quad&~
\cD_{\brp}\rho\,,\quad&~
\cD_{\brp}\psi^{\brp}\,,\quad&~
\brgamma^{\brp}\cD_{\brp}\rhop\,,\quad&~
\brgamma^{\brp}\cD_{\brp}\psi^{\prime}_{p}\,,\quad&~
\cD_{p}\rhop\,,\quad&\quad
\cD_{p}\psip{}^{p}\,.
\ea
\label{covDirac}
\ee

Further, in the maximally supersymmetric type II DFT~\cite{Jeon:2012hp}, as well as in the pure spinor formalism~\cite{Berkovits:2001ue}, the Ramond--Ramond (RR) sector is characterized by a $\SpinD \times \oSpinD$ bi-fundamental spinorial potential: $\cC^{\alpha}{}_{\bralpha}$  (\textit{c.f.~}\cite{Rocen:2010bk,Hohm:2011zr,Hatsuda:2014aza,Cederwall:2016ukd,Butter:2022gbc,Butter:2022sfh}). Subsequently, a pair of fully covariant and  nilpotent derivatives, $\cD_{+}$ and $\cD_{-}$, are introduced~\cite{Jeon:2012kd}:
\be
\ba{ll}
\cD_{\pm}\cC:=\gamma^{p}\cD_{p}\cC\pm\gamma^{(D+1)}\cD_{\brp}\cC\brgamma^{\brp}\,,\quad&\qquad \cD_{\pm}^{2}\cC=0\,,
\ea
\label{Dpm}
\ee
where, with (\ref{Phisv}),  $\cD_{A}\cC=\partial_{A}\cC+\Phi_{A}\cC-\cC\brPhi_{A}$.  In particular, the RR field strength, $\cF^{\alpha}{}_{\bralpha}$, is given by  one of these  operators and remains invariant under RR gauge transformations,
\be
\ba{ll}
\cF:=\cD_{+}\cC\,,\qquad&\qquad \delta \cC=\cD_{+}\lambda\quad\longrightarrow\quad
\delta\cF=0\,.
\ea
\label{RRfs}
\ee

\subsubsection{Generalised Kosmann Derivative,  aka the Further-Generalised Lie Derivative}
\aim{To adapt the Lie derivative so that it respects local Lorentz symmetry.}
\noindent The generalised Lie derivative is compatible with the semi-covariant derivative,  $\hcL_{\xi}(\partial_{A}) = \hcL_{\xi}(\na_{A})$, as demonstrated in (\ref{hcLna}), and is in fact fully covariant under the doubled-yet-gauged diffeomorphisms~(\ref{covhcL}). However, when acting on spinorial tensors, it fails to preserve the local Lorentz symmetries. To achieve full covariance under both diffeomorphisms and local Lorentz rotations, the generalised Lie derivative must be further extended to incorporate spin connections. This approach was originally introduced in the context of GR by Kosmann in 1971~\cite{Kosmann}.\footnote{See \cite{Kim:2024ewt} for a recent application of the Kosmann derivative.} The DFT counterpart to the Kosmann derivative, also referred to as the further-generalised Lie derivative, is defined in~\cite{Angus:2018mep} as
\be
\ba{lll}
\fcL_{\xi}T_{Ap\brp}{}^{\alpha}{}_{\beta}{}^{\bralpha}{}_{\brbeta}&:=&\xi^{B}\cD_{B}T_{Ap\brp}{}^{\alpha}{}_{\beta}{}^{\bralpha}{}_{\brbeta}+\omega\cD_{B}\xi^{B}T_{Ap\brp}{}^{\alpha}{}_{\beta}{}^{\bralpha}{}_{\brbeta}+2\cD_{[A}\xi_{B]}T^{B}{}_{p\brp}{}^{\alpha}{}_{\beta}{}^{\bralpha}{}_{\brbeta}\\
{}&{}&+2\cD_{[p}\xi_{q]}T_{A}{}^{q}{}_{\brp}{}^{\alpha}{}_{\beta}{}^{\bralpha}{}_{\brbeta}+
\half\cD_{[r}\xi_{s]}(\gamma^{rs})^{\alpha}{}_{\delta}T_{Ap\brp}{}^{\delta}{}_{\beta}{}^{\bralpha}{}_{\brbeta}- \half\cD_{[r}\xi_{s]}(\gamma^{rs})^{\delta}{}_{\beta}T_{Ap\brp}{}^{\alpha}{}_{\delta}{}^{\bralpha}{}_{\brbeta}\\
{}&{}&+2\cD_{[\brp}\xi_{\brq]}T_{Ap}{}^{\brq\alpha}{}_{\beta}{}^{\bralpha}{}_{\brbeta}+
\half\cD_{[\brr}\xi_{\brs]}(\brgamma^{\brr\brs})^{\bralpha}{}_{\brdelta}T_{Ap\brp}{}^{\alpha}{}_{\beta}{}^{\brdelta}{}_{\brbeta}-
\half\cD_{[\brr}\xi_{\brs]}(\brgamma^{\brr\brs})^{\brdelta}{}_{\brbeta}T_{Ap\brp}{}^{\alpha}{}_{\beta}{}^{\bralpha}{}_{\brdelta}\,,
\ea
\label{Kosmann}
\ee
which consists of the generalised Lie derivative combined with infinitesimal local Lorentz rotations characterized by
\be
\ba{ll}
\xi^{A}\Phi_{Apq}+2\cD_{[p}\xi_{q]}={2\partial_{[p}\xi_{q]}+\Phi_{\brr pq}\xi^{\brr}+3\Phi_{[pqr]}\xi^{r}}\,,~~&~~
\xi^{A}\brPhi_{A\brp\brq}+2\cD_{[\brp}\xi_{\brq]}={2\partial_{[\brp}\xi_{\brq]}+\brPhi_{r\brp\brq}\xi^{r}
+3\brPhi_{[\brp\brq\brr]}\xi^{\brr}}\,.
\ea
\label{additional}
\ee
As listed in (\ref{covPhi}),  all of these are fully covariant under the doubled-yet-gauged diffeomorphisms.

\subsubsection{Riemannian Parametrisation of the Covariant Derivatives}
\aim{To show how covariant derivatives reduce under the $(0,0)$ Riemannian parametrisation.}
\noindent Under the $(0,0)$ Riemannian parametrisation, (\ref{00V}) and (\ref{00d}),    the  projected  components of the spin connections~(\ref{covPhi}) reduce explicitly to
\be
\ba{ll}
\Phi_{\brp pq}=\frac{1}{\sqrt{2}}\bre_{\brp}{}^{\mu}\left(\omega_{\mu pq}+\half H_{\mu pq}\right)\,,\quad&\quad
\brPhi_{p\brp\brq}=\frac{1}{\sqrt{2}}e_{p}{}^{\mu}\left(\bromega_{\mu\brp\brq}+\half H_{\mu \brp\brq}\right)\,,\\

\Phi_{[pqr]}=\frac{1}{\sqrt{2}}\left(\omega_{[pqr]}+\textstyle{\frac{1}{6}} H_{pqr}\right)\,,\quad&\quad
\brPhi_{[\brp\brq\brr]}=\frac{1}{\sqrt{2}}\left(\bromega_{[\brp\brq\brr]}+\textstyle{\frac{1}{6}} H_{\brp\brq\brr}\right)\,,\\
\Phi^{p}{}_{pq}=\frac{1}{\sqrt{2}}\left(e^{p\mu}\omega_{\mu pq}-2e_{q}{}^{\mu}\partial_{\mu}\phi\right)\,,\quad&\quad
\brPhi^{\brp}{}_{\brp\brq}=\frac{1}{\sqrt{2}}\left(\bre^{\brp\mu}\bromega_{\mu\brp\brq}-2\bre_{\brq}{}^{\mu}\partial_{\mu}\phi\right)\,,
\ea
\label{covPhi2}
\ee 
where we have a pair of  undoubled spin connections, 
\be
\ba{ll}
\omega_{\mu pq}=e_{p}{}^{\nu}(\partial_{\mu}e_{\nu q}-\gamma_{\mu}^{\lambda}{}_{\nu}e_{\lambda q})\,,\qquad&\quad\bromega_{\mu \brp\brq}=\bre_{\brp}{}^{\nu}(\partial_{\mu}\bre_{\nu\brq}-\gamma_{\mu}^{\lambda}{}_{\nu}\bre_{\lambda\brq})\,,
\ea
\ee
which along with the Christoffel connection $\gamma^{\lambda}_{\mu\nu}$ imply an   undoubled master derivative:
\be
\ba{llllll}
\trd_{\mu}:=\partial_{\mu}+\gamma_{\mu}+\omega_{\mu}+\bromega_{\mu}\,,&~
\trd_{\mu} e_{\nu}{}^{p}=0\,,&~
\trd_{\mu}\eta_{pq}=0\,,&~
\trd_{\mu}\bre_{\nu}{}^{\brq}=0,&~
\trd_{\mu}\breta_{\brp\brq}=0\,,&~
\trd_{\lambda} g_{\mu\nu}=0\,.
\ea
\label{trdmaster}
\ee
For example, from \cite{Jeon:2011vx}  we obtain for a $\SpinD$ tensor,
\be
\cD_{\brp}T_{q_1q_2\cdots q_{n}}=
\frac{1}{\sqrt{2}}\bre_{\brp}{}^{\mu}\left[\partial_{\mu}T_{q_1q_2\cdots q_{n}}+\sum_{i=1}^{n}~(\omega_{\mu q_{i}}{}^{r}+\half H_{\mu q_{i}}{}^{r})
\,T_{q_{1}\cdots q_{i-1}rq_{i+1}\cdots q_{n}}\right]\,,
\label{cDT}
\ee
and for a weightless spinor, we produce  known expressions from generalised geometry~\cite{Coimbra:2011nw}, 
\be
\ba{ll}
\cD_{\brp}\rho=\frac{1}{\sqrt{2}}\!\left(\partial_{\brp} \rho{\, + \frac{1}{4}} \omega_{\brp q r} \gamma^{q r} \rho {\,+ \frac{1}{8}}H_{\brp q r} \gamma^{q r} \rho \right),
~&
\gamma^{p}\cD_{p}\rho=\frac{1}{\sqrt{2}} \gamma^{\mu}\!\left( \partial_{\mu} \rho{\, + \frac{1}{4}} \omega_{\mu  qr} \gamma^{qr} \rho {\,+ \frac{1}{24}} H_{\mu qr} \gamma^{qr} \rho - \partial_{\mu} \phi \rho \right).
\label{cDrho}
\ea
\ee

Furthermore, choosing the gauge $e_{\mu}{}^{p} \equiv \bre_{\mu}{}^{\brp}$ reduces the twofold spin groups to a diagonal subgroup, which can be identified with the single spin group of GR. This gauge condition forces the spinors and the RR fields to transform under $\ODD$ rotations~\cite{Jeon:2012hp}, and it allows the single RR potential field to be expanded using gamma matrices:
\be
\cC^{\alpha}{}_{\beta}=\sum_{p}~\frac{1}{p!}C_{\mu_{1}\mu_2\cdots\mu_{p}}(\gamma^{\mu_1\mu_2\cdots\mu_p})^{\alpha}{}_{\beta}\,,
\ee
such that the conventional (even or odd) RR form fields appear, and the pair of nilpotent differential  operators~(\ref{Dpm}) reduce to a twisted exterior derivative and its Hodge dual~\cite{Jeon:2012kd}:
\be
\ba{ll}
\cD_{+}\quad\!\longrightarrow\!\quad\rd+(H-\rd\phi)\,\wedge~\,,\qquad&\qquad
\cD_{-}\quad\!\longrightarrow\!\quad\star\,\big[\,\rd+(H-\rd\phi)\,\wedge~\big]\star~\,.
\ea
\ee
 For the non-Riemannian parametrisations of the fully covariant derivatives~(\ref{covD}) and  doubled Yang--Mills theory~(\ref{YMaction}), we refer to \cite{Cho:2019ofr} (section 4.3 therein) and \cite{Angus:2021jvm}, respectively.

\subsection{Curvatures\label{SECCURVATURE}}
\noindent Now we turn to curvatures. The semi-covariant four-index Riemann curvature is defined as~\cite{Jeon:2011cn}:
\be
S_{ABCD}:=\half\left(\fR_{ABCD}+\fR_{CDAB}-\Gamma^{E}{}_{AB}\Gamma_{ECD}\right)\,.
\label{RiemannS}
\ee
Here $\Gamma_{ABC}$ is the DFT  Christoffel connection~(\ref{Gamma}) and  $\fR_{ABCD}$ represents  its  `field strength':\footnote{In accordance with the decomposition of an $\ODD$ index into  $D$-dimensional upper and (dual) lower  indices, the field strength $\fR_{ABCD}$ includes $\fR^{\kappa}{}_{\lambda\mu\nu}$ which should not be confused with the ordinary (undoubled) Riemann curvature  $R^{\kappa}{}_{\lambda\mu\nu}$.  Notably,  $S$ follows $R$ alphabetically, signifying its intended connection within the semi-covariant formalism. }
\be
\fR_{CDAB}=\partial_{A}\Gamma_{BCD}-\partial_{B}\Gamma_{ACD}+\Gamma_{AC}{}^{E}\Gamma_{BED}-\Gamma_{BC}{}^{E}\Gamma_{AED}\,,
\label{FSR}
\ee 
which arises in the commutator of the semi-covariant derivatives, 
\be
\left[\na_{A\,},\na_{B}\right]T_{C_1C_2\cdots C_n}=-\Gamma^{D}{}_{AB}\na_{D}T_{C_1C_2\cdots C_{n}}+\sum_{i=1}^{n}~\fR_{C_{i}}{}^{D}{}_{AB}T_{C_1\cdots C_{i-1}DC_{i+1}\cdots C_{n}}\,,
\label{commna}
\ee
satisfies  symmetric and projective properties,
\be
\fR_{ABCD}=\fR_{[AB][CD]}=(P_{A}{}^{E}P_{B}{}^{F}+\brP_{A}{}^{E}\brP_{B}{}^{F})\fR_{EFCD}\,,
\ee
and, under arbitrary variation of the Christoffel symbols~(\ref{deltaG}), transforms infinitesimally as
\be
\delta \fR_{ABCD}=\na_{C}\delta\Gamma_{DAB}-\na_{D}\delta\Gamma_{CAB}+\Gamma^{E}{}_{CD}\delta\Gamma_{EAB}\,.
\label{deltafR}
\ee
In particular, under  doubled-yet-gauged diffeomorphisms,  it varies as
\be
\ba{rll}
\delta_{\xi}\fR_{ABCD}&=&\hcL_{\xi}\fR_{ABCD}
-2\Gamma^{E}{}_{AB}\partial_{E}\partial_{[C}\xi_{D]}
+2\Gamma^{E}{}_{CD}(\cP{+\brcP})_{EAB}{}^{FGH}\partial_{F}\partial_{[G}\xi_{H]}\\
{}&{}&+4\na_{[C}\Big((\cP{+\brcP})_{D]AB}{}^{FGH}\partial_{F}\partial_{[G}\xi_{H]}\Big)\,.
\ea
\label{dfRhcL}
\ee
Further, the Jacobi identity of the commutators~(\ref{commna}) implies the following two sets of  identities~\cite{Jeon:2010rw}:
\be
\ba{ll}
\fR_{[A}{}^{D}{}_{BC]}+\na_{[A}\Gamma^{D}{}_{BC]}+\Gamma^{D}{}_{E[A}\Gamma^{E}{}_{BC]}=0\,,\qquad&\quad
\na_{[A}{}\fR^{DE}{}_{BC]}+\fR^{DE}{}_{F[A}\Gamma^{F}{}_{BC]}=0\,.
\ea
\ee
Consequently,  the semi-covariant Riemann curvature~(\ref{RiemannS})  appears through a projected commutator,
\be
\left[\cD_{p\,},\cD_{\brq\,}\right]T_{C_1C_2\cdots C_n}=\sum_{i=1}^{n}~2S_{p\brq C_{i}}{}^{D}T_{C_1\cdots C_{i-1}DC_{i+1}\cdots C_{n}}\,,
\label{cDcDS}
\ee
satisfies symmetric properties, including an algebraic Bianchi  identity,
\be
\ba{ll}
S_{ABCD}=S_{CDAB}=S_{[AB][CD]}\,,\qquad&\quad
S_{A[BCD]}=0\,,
\ea
\label{aB}
\ee
and transforms as 	total (covariant) derivatives under  the arbitrary variation of the   Christoffel symbols, 
\be
\delta S_{ABCD}=\na_{[A}\delta\Gamma_{B]CD}+\na_{[C}\delta\Gamma_{D]AB}\,.
\label{deltaS}
\ee
In particular, it is semi-covariant under  doubled-yet-gauged diffeomorphisms:
\be
\delta_{\xi}S_{ABCD}=\hcL_{\xi}S_{ABCD}+
2\na_{[A}\Big(\!(\cP{+\brcP})_{B][CD]}{}^{EFG}\partial_{E}\partial_{F}\xi_{G}\Big)
+2\na_{[C}\Big(\!(\cP{+\brcP})_{D][AB]}{}^{EFG}\partial_{E}\partial_{F}\xi_{G}\Big)\,.
\label{anomalyS}
\ee
Obvious methods of eliminating the anomalous terms  via projection  end up producing only identically vanishing trivial quantities,
\be
\ba{lll}
S_{pq\brp\brq}=0\,,\qquad&\qquad S_{\brp\brq pq}=0\,,\qquad&\qquad 
S_{p\brp q\brq}=0\,,
\ea
\label{RSp}
\ee
and there appears to be no fully covariant four-index curvature in DFT~\cite{Jeon:2011cn,Hohm:2011si}.   Fully  covariant  curvature tensors are  then  obtained by contracting the indices:  with $S_{AB}=S_{BA}=S^{C}{}_{ACB}$, we have  the fully covariant two-index or  Ricci curvature in DFT,
\be
S_{p\brq}=V^{A}{}_{p}\brV^{B}{}_{\brq}S_{AB}\,,
\label{Ricci}
\ee
and  further the scalar curvature,\footnote{The subscript $\scriptstyle{(0)}$  indicates the scalar nature of $\So$ and serves to distinguish $\So$ from  the notation used for an action, \textit{e.g.~}(\ref{pureDFT}). }  
\be
\So:=\left(P^{AC}P^{BD}-\brP^{AC}\brP^{BD}\right)S_{ABCD}=S_{pq}{}^{pq}-S_{\brp\brq}{}^{\brp\brq}\,.
\label{Sc}
\ee
These  contain  both $\cH_{AB}$ and $d$ through the Christoffel symbols~(\ref{Gamma}): the generalised  metric alone cannot generate the  covariant curvature~\cite{Jeon:2010rw}.      Explicitly, we recover the original expression in \cite{Hohm:2010pp},
\be
\!\So=\cH^{AB}\!\left(\textstyle{\frac{1}{8}}\partial_{A}\cH_{CD}\partial_{B}\cH^{CD}+\textstyle{\frac{1}{2}}\partial_{C}\cH_{A}{}^{D}\partial_{D}\cH_{B}{}^{C}-4\partial_{A}d\partial_{B}d+4\partial_{A}\partial_{B}d\right)-\partial_{A}\partial_{B}\cH^{AB}+4\partial_{A}\cH^{AB}\partial_{B}d\,.
\ee

Like (\ref{RSp}), the Ricci curvature satisfies the identities~\cite{Jeon:2011cn,Cho:2015lha}:
\be
S_{pr\brq}{}^{r}=S_{p\brr\brq}{}^{\brr}=\half S_{p\brq}\,,
\label{SSS}
\ee
and it arises through the commutators of the covariant differential operators~\cite{Coimbra:2011nw,Cho:2015lha}: using (\ref{cDcDS}) and (\ref{SSS}),
\be
\ba{llll}
\!\!\!\big[\cD_{p},\cD_{\brq}\big]T^{p}=S_{p\brq}T^{p}\,,~&\,
\big[\cD_{\brq},\cD_{p}\big]T^{\brq}=S_{p\brq}T^{\brq}\,,~&\,
{}\big[\gamma^{p}\cD_{p},\cD_{\brq}\big]\varepsilon=\half S_{p\brq}\gamma^{p}\varepsilon\,,~&\,
{}\big[\brgamma^{\brq}\cD_{\brq},\cD_{p}\big]\varepsilon^{\prime}=\half S_{p\brq}\gamma^{\brq}\varepsilonp\,.
\ea
\label{usefulS}
\ee
Lastly, the scalar curvature satisfies
\be
\ba{ll}
S_{pq}{}^{pq}+S_{\brp\brq}{}^{\brp\brq}=0\,,\qquad&\qquad
\So=2S_{pq}{}^{pq}=-2S_{\brp\brq}{}^{\brp\brq}\,,
\ea
\ee
and manifests itself through  successive application of the Dirac operators~(\ref{covDirac}),
\be
\ba{ll}
(\gamma^{p}\cD_{p})^{2}\varepsilon
+\cD_{\brp}\cD^{\brp}\varepsilon=-\quarter S_{pq}{}^{pq}\varepsilon=-\textstyle{\frac{1}{8}}\So\varepsilon\,,\quad&\quad
(\brgamma^{\brp}\cD_{\brp})^{2}\varepsilonp
+\cD_{p}\cD^{p}\varepsilonp=-\quarter S_{\brp\brq}{}^{\brp\brq}\varepsilonp=\textstyle{\frac{1}{8}}\So\varepsilonp\,.
\ea
\ee
For the curvatures of the twofold spin connections and their relation to $\fR_{ABCD}$ and $S_{ABCD}$, we refer to  \cite{Cho:2015lha} (section 2 therein).

 Restricting to the  Riemannian parametrisation~(\ref{00V}), (\ref{00d}),  we have explicitly, 
\be
S_{p\brq}=\half e_{p}{}^{\mu}\bre_{\brq}{}^{\nu}\Big[
R_{\mu\nu}+2\trd_{\mu}(\partial_{\nu}\phi)-\quarter H_{\mu\rho\sigma}H_{\nu}{}^{\rho\sigma}
+\half e^{2\phi}\trd^{\rho}\big(e^{-2\phi}H_{\rho\mu\nu}\big)
\Big]\,,
\label{RiemannSpbrq}
\ee
and
\be
\So=R+4\Box\phi-4\partial_{\mu}\phi\partial^{\mu}\phi-\textstyle{\frac{1}{12}}H_{\lambda\mu\nu}H^{\lambda\mu\nu}\,.
\label{RiemannSo}
\ee
For  non-Riemannian parametrisations, we refer to  \cite{Cho:2019ofr} (section 4.3 therein).

\summary{While a fully covariant four-index curvature is absent, Ricci and scalar curvatures exist and underpin the DFT action.}

\subsection{Doubled Einstein--Hilbert Action \& Gravitational Wave\label{SECDEH}}
\aim{To formulate the DFT action and examine its variation, conserved currents, and wave equations.}
\noindent The `pure' DFT action, or the doubled Einstein--Hilbert action,  is naturally given by the scalar curvature $\So$  multiplied by the $\ODD$-symmetric integral measure $e^{-2d}$, integrated over a section $\Sigma_{D}$:
\be
S_{\DFT}=\int_{\Sigma_{D}}~e^{-2d}\So\,.
\label{pureDFT}
\ee
From the compatibility  condition of the semi-covariant derivative~(\ref{Gcomp}) and the nice properties of the semi-covariant four-index curvature~(\ref{aB}), (\ref{deltaS}),  it is straightforward to vary the pure DFT Lagrangian:
\be
\delta \big(e^{-2d}\So\big)=4e^{-2d}\big(\brV_{A}{}^{\brq}\delta V^{Ap}S_{p\brq}-\half\delta d\,\So\big)+\partial_{A}\big(e^{-2d}\Theta^{A}\big)\,,
\label{deltaEH}
\ee
where in the total derivative we have set~\cite{Park:2015bza,Blair:2015eba}
\be
\Theta^{A}=2\big(P^{AC}P^{BD}-\brP^{AC}\brP^{BD}\big)\delta \Gamma_{BCD}
=4\cH^{AB}\partial_{B}\delta d-\na_{B}\delta\cH^{AB}\,.
\label{Theta}
\ee
It is worthwhile to note  different ways of rewriting  the variation of the vielbeins  in (\ref{deltaEH}):
\be
\brV_{A}{}^{\brq}\delta V^{Ap}=- V^{Ap}\delta\brV_{A}{}^{\brq} =\half\big(\brV_{A}{}^{\brq}\delta V^{Ap}- V^{Ap}\delta\brV_{A}{}^{\brq}\big)=\half V^{Ap}\brV^{B\brq}\delta \cH_{AB}\,.
\label{deltaVcH}
\ee
In particular, when the variation is generated  by the Kosmann derivative~(\ref{Kosmann}), \textit{i.e.~}$\delta_{\xi}=\fcL_{\xi}$, we have 
\be
\ba{ll} 
\brV_{A}{}^{\brq}\delta_{\xi} V^{Ap}=\brV_{A}{}^{\brq}\fcL_{\xi} V^{Ap}
=2\cD^{[\brq}\xi^{p]}\,,\qquad&\quad \delta_{\xi}d=\fcL_{\xi}d=\hcL_{\xi}d=-\frac{1}{2}\na_{A}\xi^{A}\,,\\
\multicolumn{2}{c}{
\delta_{\xi}\cH_{AB}=\fcL_{\xi}\cH_{AB}=\hcL_{\xi}\cH_{AB}=
8\brP_{(A}{}^{C}P_{B)}{}^{D}\na_{[C}\xi_{D]}=8\brV_{(A}{}^{\brq}V_{B)}{}^{p}\cD_{[\brq}\xi_{p]}\,.}
\ea
\label{deltaxi2}
\ee
Applying these to (\ref{deltaEH}), \textit{i.e.}~the action principle, we can identify the off-shell conserved Einstein curvature in DFT, which satisfies a differential Bianchi identity~\cite{Park:2015bza}:
\be
\ba{ll}
G_{AB}=4(PS\brP)_{[AB]}-\half\cJ_{AB}\So=4V_{[A}{}^{p}\brV_{B]}{}^{\brq}S_{p\brq}-\half\cJ_{AB}\So\,,\quad&\quad\na_{A}G^{AB}=0\,.
\ea
\label{EinsteinC}
\ee
The complete equations of motion of the pure DFT action—derived from the action principle~(\ref{deltaEH})—are, \textit{a priori}, expressed by the independent vanishing of the Ricci curvature and the scalar curvature~\cite{Siegel:1993xq, Hohm:2010pp}. However, by utilising the relations, $G_{A}{}^{A}=-D\So$ and $(PG\brP)_{AB}=2(PS\brP)_{AB}$, the equations of motion can be seen, in a unified manner, as equivalent to the vanishing of the Einstein curvature~\cite{Park:2015bza},
\be
\ba{lll}
{S_{p\brq}=0}\quad\&\quad{\So=0}\quad&\Longleftrightarrow&\quad
G_{AB}=0\,.
\ea
\label{vacuumEDFE}
\ee

Although the Einstein curvature is not symmetric,   $G_{AB}\neq G_{BA}$,  and does not satisfy $\na_{B}G^{AB}\neq 0\,$,  it is possible to symmetrise the  curvature  by multiplying the generalised metric from the right. Specifically, 
\be
\ba{ll}
(G\cH)_{AB}=(G\cH)_{BA}=G_{AC}\cH^{C}{}_{B}=-4 V_{(A}{}^{p}\brV_{B)}{}^{\brq}S_{p\brq}-\half\cH_{AB}\So\,,
\quad&\quad
\na_{A}(G\cH)^{AB}=0\,.
\ea
\label{GH}
\ee

Under the $(0,0)$ Riemannian parametrisation, from (\ref{RiemannSo}), the pure DFT action~(\ref{pureDFT}) coincides with the NSNS gravity action up to a total derivative (denoted by $\simeq$):
\be
S_{\DFT}=\int_{\Sigma_{D}}~e^{-2d}\So~\simeq\,\int\rd^D x~\sqrt{-g}e^{-2\phi}\Big(R+4\partial_{\mu}\phi\partial^{\mu}\phi-\textstyle{\frac{1}{12}}H_{\lambda\mu\nu}H^{\lambda\mu\nu}\Big)\,,
\label{DFT00}
\ee
and  the upper left $D{\times D}$  block of $(G\cH)_{AB}$  contains the undoubled  Einstein curvature:
\be
(G\cH)^{\mu\nu}=R^{\mu\nu}-\half g^{\mu\nu}R+2\trd^{\mu}(\partial^{\nu}\phi)
-2g^{\mu\nu}(\Box\phi-\partial_{\sigma}\phi\partial^{\sigma}\phi)-\quarter H^{\mu\rho\sigma}H^{\nu}{}_{\rho\sigma}
+\textstyle{\frac{1}{24}} g^{\mu\nu}H_{\rho\sigma\tau}H^{\rho\sigma\tau}\,.
\label{GcHuu}
\ee
Moreover,  the differential Bianchi identity~(\ref{EinsteinC}) decomposes into  two separate identities:
\be
\ba{ll}
\trd_{\mu}\big(R^{\mu\nu}-\half g^{\mu\nu}R\big)=0\,,\qquad&\quad
\trd_{\mu}\trd_{\nu}\big(e^{-2\phi}H^{\lambda\mu\nu}\big)=0\,.
\ea
\label{dBI00}
\ee

\subsubsection{Gamma Squared  Action  \& Noether Currents}
\aim{To rewrite the action in $\Gamma^2$ form and identify associated symmetries.}
\noindent The doubled Einstein--Hilbert action~(\ref{pureDFT}) contains  terms with two derivatives of the fundamental fields which can be eliminated by a partial integral. Subtracting a specific  total derivative~\cite{Park:2015bza}, 
\be
\ba{ll}
\cL_{\Gamma^{2}}=e^{-2d}\So-\partial_{A}\big(e^{-2d}B^{A}\big)\,,\quad&~~~
B^{A}=2(P^{AC}P^{BD}-\brP^{AC}\brP^{BD})
\Gamma_{BCD}  =4\cH^{AB}\partial_{B}d -\partial_{B}\cH^{AB}\,,
\ea
\label{defB}
\ee
we acquire a doubled $\Gamma^{2}$-action, free of any two-derivative terms (\textit{c.f.~}\cite{Dyer:2008hb}, \cite{Berman:2011kg}),
\be
S_{\Gamma^{2}}=\int_{\Sigma_{D}}~\cL_{\Gamma^{2}}=\int_{\Sigma_{D}}~e^{-2d}
\big(P^{AC}P^{BD}-\brP^{AC}\brP^{BD}\big)
\big(\Gamma_{AC}{}^{E}\Gamma_{BDE}-\Gamma_{AB}{}^{E}\Gamma_{DCE}+\half\Gamma^{E}{}_{AB}\Gamma_{ECD}\big)\,,
\label{GammaDFT}
\ee
which still admits  diffeomorphisms as a Noether symmetry. We note generically from (\ref{deltaEH}) and (\ref{defB}),
\be
\delta\cH_{BC}\frac{\partial\cL_{\Gamma^{2}}}{\partial(\partial_{A}\cH_{BC})}+ \delta d\frac{\partial\cL_{\Gamma^{2}}}{\partial(\partial_{A}d)}=e^{-2d}\Theta^{A}-\delta\big(e^{-2d}B^{A}\big)\,,
\ee
and especially under diffeomorphisms,
\be
\delta_{\xi}\cL_{\Gamma^{2}}=\partial_{A}\left[\xi^{A}e^{-2d}\So-\delta_{\xi}\big(e^{-2d}B^{A}\big)\right]\,.
\ee
Thus,  with the explicit expression  of $\Theta^{A}$~(\ref{Theta}) and the commutator relations~(\ref{usefulS}), the   on-shell conserved Noether current for the doubled-yet-gauged  diffeomorphisms    is~\cite{Park:2015bza}
\be
J_{{\rm{on-shell}}}^{A}=e^{-2d}\big(4\cH^{AB}\partial_{B}\delta_{\xi}d-\na_{B}\delta_{\xi}\cH^{AB}-\xi^{A}\So\big)
=\partial_{B}\big(e^{-2d}{K}^{[AB]}\big)+2e^{2d}G^{A}{}_{B}\xi^{B}+\mathfrak{j}^{A}\,,
\ee
where  ${K}^{AB}$ is a skew-symmetric Noether potential, constituting an off-shell conserved Noether current,
\be
\ba{ll}
K^{AB}=4\brV^{[A}{}_{\brp}V^{B]}{}_{q}\big(\cD^{\brp}\xi^{q}+\cD^{q}\xi^{\brp}\big)\,,\quad&\qquad
J_{{\rm{off-shell}}}^{A}=\partial_{B}\big(e^{-2d}{K}^{[AB]}\big)\,,
\ea
\label{Kpotential}
\ee
and $\mathfrak{j}^{A}$ denotes a harmless derivative-index-valued vector,
\be
\ba{ll}
\mathfrak{j}^{A}= 2e^{-2d}\Big[V_{B}{}^{p}\brV_{C}{}^{\brq}\cD_{(p}\xi_{\brq)}\partial^{A}\cH^{BC}
-\partial^{A}\big(\cH^{BC}
\na_{B}\xi_{C}\big)\Big]\,,\quad&\quad \mathfrak{j}^{A}\partial_{A}=0\,,
\ea
\ee
which does not contribute to any Noether charge.

\subsubsection{Box Operator That Unveils The  Riemann Curvature Tensor: Gravitational Wave Equations}
\aim{To build a covariant box operator that encodes curvature and governs wave equations.}
\noindent By combining the  results from equations such as  (\ref{diffeoanormalous2}),  (\ref{dfRhcL}), and (\ref{anomalyS}), it is possible to construct a pair of fully covariant d'Alembertians that act on an arbitrary tensor density $T_{A_{1}A_{2}\cdots A_{n}}$~(\ref{hcL})  and  incorporate  $S_{ABCD}$~(\ref{RiemannS}) and $\fR_{ABCD}$~(\ref{FSR})~\cite{Lee:2025fme}:
\be
\ba{rll}
\Deltab T_{A_{1}A_{2}\cdots A_{n}}&:=&
P^{BC}\na_{B}\na_{C}T_{A_{1}A_{2}\cdots A_{n}}\\
{}&{}&
+\dis{\sum_{i=1}^{n}~2P_{A_{i}}{}^{C}P_{B}{}^{D}\Big(\fR_{[CD]}-\half\Gamma^{EF}{}_{C}\Gamma_{EFD}-\Gamma^{E}{}_{CD}\na_{E}\Big)\,T_{A_{1}\cdots A_{i-1}}{}^{B}{}_{A_{i+1}\cdots A_{n}}}\\
{}&{}&+\dis{\sum_{i<j}~2\left(\ba{l}
P_{A_{i}}{}^{D}P_{B}{}^{E}\fR_{A_{j}CDE}\\
+P_{A_{j}}{}^{D}P_{C}{}^{E}\fR_{A_{i}BDE}\\
-2P_{A_{i}}{}^{D}P_{B}{}^{E}P_{A_{j}}{}^{F}P_{C}{}^{G}S_{DEFG}\ea\right)T_{A_{1}\cdots A_{i-1}}{}^{B}{}_{A_{i+1}\cdots A_{j-1}}{}^{C}{}_{A_{j+1}\cdots A_{n}}}\,,
\ea
\label{DELTA}
\ee
and with the replacement of every explicit occurrence of the projector $P_{A}{}^{B}$ with the opposite projector $\brP_{A}{}^{B}$, 
\be
\ba{rll}
\brDeltab T_{A_{1}A_{2}\cdots A_{n}}&:=&
\brP^{BC}\na_{B}\na_{C}T_{A_{1}A_{2}\cdots A_{n}}\\
{}&{}&
+\dis{\sum_{i=1}^{n}~2\brP_{A_{i}}{}^{C}\brP_{B}{}^{D}\Big(\fR_{[CD]}-\half\Gamma^{EF}{}_{C}\Gamma_{EFD}-\Gamma^{E}{}_{CD}\na_{E}\Big)\,T_{A_{1}\cdots A_{i-1}}{}^{B}{}_{A_{i+1}\cdots A_{n}}}\\
{}&{}&+\dis{\sum_{i<j}~2\left(\ba{l}
\brP_{A_{i}}{}^{D}\brP_{B}{}^{E}\fR_{A_{j}CDE}\\
+\brP_{A_{j}}{}^{D}\brP_{C}{}^{E}\fR_{A_{i}BDE}\\
-2\brP_{A_{i}}{}^{D}\brP_{B}{}^{E}\brP_{A_{j}}{}^{F}\brP_{C}{}^{G}S_{DEFG}\ea\right)T_{A_{1}\cdots A_{i-1}}{}^{B}{}_{A_{i+1}\cdots A_{j-1}}{}^{C}{}_{A_{j+1}\cdots A_{n}}}\,.
\ea
\label{brDELTA}
\ee
While these two mirroring operators annihilate each other identically due to the section condition:
\be
\left(\Deltab+\brDeltab\right)T_{A_{1}A_{2}\cdots A_{n}}=0\,,
\label{covseccon}
\ee
their difference defines the fully covariant \textit{box operator}:
\be
\bBox T_{A_{1}A_{2}\cdots A_{n}}:=\left(\Deltab-\brDeltab\right)T_{A_{1}A_{2}\cdots A_{n}}=\cH^{BC}\na_{B}\na_{C}T_{A_{1}A_{2}\cdots A_{n}}+\sum_{i}~\cdots~+\sum_{i<j}~\cdots\,.
\label{bBox}
\ee

In particular, acting on projected tensors such as ${\brP}_{C_{1}}{}^{D_{1}}\cdots{\brP}_{C_{n}}{}^{D_{n}}T_{D_{1}\cdots D_{n}}$, $P_{C_{1}}{}^{D_{1}}\cdots P_{C_{n}}{}^{D_{n}}T_{D_{1}\cdots D_{n}}$, and  $(PT\brP)_{AB}=P_{A}{}^{C}\brP_{B}{}^{D}T_{CD}$, the pair of d'Alembertians and the box operator produce the two expressions in (\ref{coco1}) and the following result: 
\be
\ba{rrl}
\bBox (PT\brP)_{AB}&=&
\cH^{CD}\na_{C}\na_{D}(PT\brP)_{AB}-2P_{A}{}^{C}\brP_{B}{}^{D}(\fR_{CEDF}-\fR_{DFCE})(PT\brP)^{EF}\\
{}&{}&+2P_{A}{}^{C}(\fR_{[CD]}-\frac{1}{2}\Gamma^{EF}{}_{C}\Gamma_{EFD}-\Gamma^{E}{}_{CD}\na_{E})(PT\brP)^{D}{}_{B}\\
{}&{}&
-2\brP_{B}{}^{C}(\fR_{[CD]}-\frac{1}{2}\Gamma^{EF}{}_{C}\Gamma_{EFD}-\Gamma^{E}{}_{CD}\na_{E})(PT\brP)_{A}{}{}^{D}\,.
\ea
\label{BXMAIN}
\ee
Equivalent yet   more compact expressions are available through  contraction with the vielbeins, $V_{Ap},\brV_{B\brq}$:
\be
\ba{l}
\Deltab T_{p\brq}=
\cD_{r}\cD^{r}T_{p\brq}+2\fR_{\brq \brs pr}T^{r\brs}
+2\big(\fR_{[pr]}-\half\Gamma^{AB}{}_{p}\Gamma_{ABr}-\Gamma^{C}{}_{pr}\cD_{C}\big)T^{r}{}_{\brq}\,,\\
\brDeltab T_{p\brq}=
\cD_{\brr}\cD^{\brr}T_{p\brq}+2\fR_{pr\brq\brs}T^{r\brs}
+2\big(\fR_{[\brq\brs]}-\half\Gamma^{AB}{}_{\brq}\Gamma_{AB\brs}-\Gamma^{C}{}_{\brq\brs}\cD_{C}\big)T_{p}{}^{\brs}\,.
\ea
\ee

Under the $(0,0)$ Riemannian parametrisation~(\ref{00V}), (\ref{00d}),  we set for the two-index tensor,
\be
\ba{lll}
T_{p\brq}=
V^{A}{}_{p}\brV^{B}{}_{\brq}T_{AB}=\fT^{\mu\nu}e_{\mu p\,}\bre_{\nu\brq}\quad&~\Longleftrightarrow~&\quad
\fT_{\mu\nu}=-e_{\mu}{}^{p}\bre_{\nu}{}^{\brq} T_{p\brq}\,,
\ea
\ee
and  the box operator reduces to
\be
\hBox\fT_{\mu\nu}=-e_{\mu}{}^{p}\bre_{\nu}{}^{\brq}\bBox T_{p\brq}= e^{2\phi}\trd^{\rho}\big(e^{-2\phi}\trd_{\rho}\fT_{\mu\nu}\big)-
 H_{\rho\sigma\mu} \trd^{\rho}\fT^{\sigma}{}_{\nu}+
H_{\rho\sigma\nu}\trd^{\rho}\fT_{\mu}{}^{\sigma}+2\hR_{\mu}{}^{\rho}{}_{\nu}{}^{\sigma}\fT_{\rho\sigma}\,,
\label{hBox}
\ee
where $\hR_{\mu}{}^{\rho}{}_{\nu}{}^{\sigma}$ contains the Riemann curvature  tensor and the $H$-flux,
\be
\ba{rll}
\hR_{\mu}{}^{\rho}{}_{\nu}{}^{\sigma}&=&R_{\mu}{}^{\rho}{}_{\nu}{}^{\sigma}
-\half H_{(\mu}{}^{\rho\kappa}H_{\nu)}{}^{\sigma}{}_{\kappa}-\quarter  H_{\mu\nu\kappa}H^{\rho\sigma\kappa}
+\half\trd_{(\mu}H_{\nu)}{}^{\rho\sigma}+\half\trd^{(\rho}H^{\sigma)}{}_{\mu\nu}\\
{}&{}&
+\frac{1}{4}\delta_{\mu}{}^{\rho}\Big[e^{2\phi}\trd_{\kappa}\big(e^{-2\phi}H^{\kappa\sigma}{}_{\nu}\big)-\frac{1}{2}H_{\nu\kappa\lambda}H^{\sigma\kappa\lambda}\Big]
-\frac{1}{4}\Big[e^{2\phi}\trd_{\kappa}\big(e^{-2\phi}H^{\kappa\rho}{}_{\mu}\big)+\frac{1}{2}H_{\mu\kappa\lambda}H^{\rho\kappa\lambda}\Big]\delta_{\nu}{}^{\sigma}
\,.
\ea
\label{hR4}
\ee

 In the context of applications, the imposition of an $\ODD$-symmetric harmonic gauge,
\be
e^{2d}\na_{A}\delta\big(e^{-2d}\cH^{AB}\big)=\na_{A}\delta\cH^{AB}-2\cH^{AB}\na_{A}\delta d=0\,,
\label{harmonic}
\ee
leads to a simplification of the linearised equations of motion in pure DFT~\cite{Ko:2015rha} (see also  \cite{Hohm:2015ugy, Cho:2019npq}). The resulting $\ODD$-symmetric gravitational wave equations are, with the box operator~(\ref{BXMAIN})~\cite{Lee:2025fme}:
\be
\ba{ll}
\bBox(P\delta\cH\brP)_{AB}=0\,,~\qquad&\quad\cH^{AB}\na_{A}\partial_{B}\delta d=0\,.
\ea
\label{WAVEEQ}
\ee
When applied to a Riemannian background, the $\ODD$-symmetric harmonic gauge  condition~(\ref{harmonic}) decomposes into the following components:
\be
\ba{ll}
e^{2\phi}\trd^{\rho}\!\left(e^{-2\phi}\delta g_{\rho\mu}\right)
-\frac{1}{2}H_{\mu}{}^{\rho\sigma}\delta B_{\rho\sigma}+2\partial_{\mu}\delta d=0\,,\quad&\qquad
e^{2\phi}\trd^{\rho}\!\left(e^{-2\phi}\delta B_{\rho\mu}\right)=0\,,
\ea
\ee
where $\delta d=\delta\phi-\frac{1}{4}g^{\rho\sigma}\delta g_{\rho\sigma\,}$, and the wave equations~(\ref{WAVEEQ})  reduce further, with (\ref{hBox}) and (\ref{hR4}),   to
\be
\ba{ll}
\hBox(\delta g_{\mu\nu}-\delta B_{\mu\nu})=0\,,\qquad&\qquad 
e^{2\phi}\trd^{\rho}\big(e^{-2\phi}\partial_{\rho}\delta d\big)=0\,.
\ea
\ee

\subsection{Einstein Double Field Equation: Unified Field Equation}
\aim{To couple matter fields to DFT consistently and derive the unified field equation.}
\noindent  By employing the fully covariant derivatives~(\ref{covD}), (\ref{covdiv}), (\ref{coco2}), (\ref{covDirac}), and (\ref{RRfs}), it is possible to couple the pure DFT action~(\ref{pureDFT}) to various forms of matter,
\be
S_{\DFT-\Matter}=\int_{\Sigma_{D}}~e^{-2d}\So\,+\,\cL_{\Matter}(\Psi,\cD\Psi;d,V,\brV)\,,
\label{SDFTM}
\ee
where $\cL_{\Matter}(\Psi,\cD\Psi;d,V,\brV)$ represents a Lagrangian density of unit weight for generic matter fields $\Psi$ which are minimally coupled to the gravitational sector $\{d,V_{Ap},\brV_{A\brp}\}$.

Generalising the case of  pure DFT~(\ref{deltaEH}), while ignoring any surface integral and assuming that $\cL_{\Matter}$ is local Lorentz invariant~(\ref{deltaVVVV}), the infinitesimal variation of the full action~(\ref{SDFTM}) is given by~\cite{Angus:2018mep}:
\be
\delta 
S_{\DFT-\Matter}=\int_{\Sigma_{D}}\,
e^{-2d}\Big[4\brV_{A}{}^{\brq}\delta V^{Ap}\big(S_{p\brq}-K_{p\brq}\big)-2\delta d\big(\So-\To\big)\Big]+\delta\Psi\frac{\delta \cL_{\Matter}}{\delta\Psi}\,.
\ee
Here  $\frac{\delta \cL_{\Matter}}{\delta\Psi}$ denotes the Euler--Lagrange equations for the matter field. Naturally, we are led to define
\be
\ba{ll}
\dis{K_{p\brq}=\frac{1}{4}e^{2d}\left(V_{Ap}\frac{\delta \cL_{\Matter}}{\delta\brV_{A}{}^{\brq}}-\brV_{A\brq}\frac{\delta \cL_{\Matter}}{\delta V_{A}{}^{p}}\right)\,,}\qquad&\quad
\dis{\To=\frac{1}{2}e^{2d\,}\frac{\delta \cL_{\Matter}}{\delta d}\,,}
\ea
\ee
which constitute the on-shell conserved, $\ODD$-symmetric energy-momentum tensor in DFT:
\be
\ba{ll}
T_{AB}=4V_{[A}{}^{p}\brV_{B]}{}^{\brq}K_{p\brq}-\half\cJ_{AB}\To\,,\qquad&\quad\na_{A}T^{AB}=0\,,
\ea
\label{EMT}
\ee
where the conservation condition requires the Euler--Lagrange equations for the matter field. Notably, when matter couples directly to the generalised metric rather than the vielbeins, we simply have
\be
K_{p\brq}=-e^{2d}\frac{\delta \cL_{\Matter}}{\delta\cH_{AB}}V_{Ap}\brV_{B\brq}\,.
\ee
The unified field equation of DFT is obtained by equating the Einstein curvature~(\ref{EinsteinC})  with  the energy-momentum tensor~(\ref{EMT}): 
\be
G_{AB} = T_{AB}\,,
\label{EDFE}
\ee
which was dubbed the Einstein Double Field Equation (EDFE)~\cite{Angus:2018mep}.

On a $(0,0)$ Riemannian background,  the energy-momentum tensor can be parametrised as
\be
T_{AB}=\left(\ba{cc}
-K^{[\mu\nu]}& K^{(\mu\lambda)}g_{\lambda\sigma}
+K^{[\mu\lambda]}B_{\lambda\sigma}-\half\delta^{\mu}{}_{\sigma}\To\\
\!\!-g_{\rho\kappa}K^{(\kappa\nu)}-B_{\rho\kappa}K^{[\kappa\nu]}-\half\delta_{\rho}{}^{\nu}\To&\quad
K_{[\rho\sigma]}+B_{\rho\kappa}K^{[\kappa\lambda]}B_{\lambda\sigma}+B_{\rho}{}^{\kappa}K_{(\kappa\sigma)}+K_{(\rho\lambda)}B^{\lambda}{}_{\sigma}
\ea\right)\,,
\ee
and the EDFE yields the following three sets of equations:
\be
\ba{rll}
R_{\mu\nu}+2\trd_{\mu}(\partial_{\nu}\phi)-\quarter H_{\mu\rho\sigma}H_{\nu}{}^{\rho\sigma}
&\!=&\! K_{(\mu\nu)}\,,\\
\half e^{2\phi}\trd^{\rho}\!\left(e^{-2\phi}H_{\rho\mu\nu}\right)&\!=&\! K_{[\mu\nu]}\,,\\
R+4\Box\phi-4\partial_{\mu}\phi\partial^{\mu}\phi-\textstyle{\frac{1}{12}}H_{\lambda\mu\nu}H^{\lambda\mu\nu}&\!=&\!\To\,.
\ea
\label{EDFE00}
\ee
Each equation can be understood as the equation of motion  of the full action~$S_{\DFT-\Matter}$ for $g_{\mu\nu}$, $B_{\mu\nu}$, and $d$,  respectively (rather than $g,B,\phi$).   In view of $d=\phi-\frac{1}{2}\ln\sqrt{-g}$~(\ref{00d}), the traditional energy-momentum tensor in GR is  given by  $K_{(\mu\nu)}$ and $\To$~\cite{Lee:2023boi}:
\be
T_{\mu\nu}^{\scriptscriptstyle{\mathbf{GR}}}=e^{-2\phi}\left[K_{(\mu\nu)}-\frac{1}{2}g_{\mu\nu}\To\right]\,.
\ee

The on-shell conservation~(\ref{EMT}) reduces to
\be
\ba{ll}
\trd^{\mu}K_{(\mu\nu)}-2\partial^{\mu}\phi\,K_{(\mu\nu)}+\half H_{\nu}{}^{\lambda\mu}K_{[\lambda\mu]}-\half\partial_{\nu}\To=0\,,\quad&\qquad
\trd_{\mu}\!\left(e^{-2\phi}K^{[\mu\nu]}\right)=0\,,
\ea
\label{con00}
\ee
which are consistent with the geometric counterparts~(\ref{dBI00}).  In particular, the middle equation in (\ref{EDFE00}) and the latter in (\ref{con00}) represent the stringy two-form generalisations of Maxwell’s equations~(\ref{Maxwell2}) and the corresponding conservation laws:   
 $\trd_{\lambda}F^{\lambda\mu}=J^{\mu}$ and $\trd_{\mu}J^{\mu}=0$.  Strings, unlike point particles, couple to the two-form $B$-field rather than the one-form Maxwell vector potential.

It is worth noting that, after subtracting the third equation from the trace of the first in (\ref{EDFE00}), we can derive a Klein–Gordon-type equation~\cite{Choi:2022srv}:
\be
\Box\big(e^{-2\phi}\big)=\big(K_{\mu}{}^{\mu}-\To+\textstyle{\frac{1}{6}}H_{\lambda\mu\nu}H^{\lambda\mu\nu}\big)e^{-2\phi}\,,
\label{Chameleon}
\ee
where the quantity inside the parentheses on the right-hand side of the equality can be interpreted as the effective or ``Chameleon mass''~\cite{Khoury:2003aq, Brax:2004qh} of the dilaton scalar field, $e^{-2\phi}$. This interpretation links the dilaton dynamics to a scalar field with an effective mass that varies based on other fields and their interactions, similar to scalar-tensor theories where the scalar field's mass adjusts to its environment.

\subsubsection{$\ODD$ Tells Matter How to Couple to DFT: Equivalence Principle Holds in String Frame}
The EDFE~(\ref{EDFE})  provides an $\ODD$-symmetric extension of Wheeler's famous insight into  gravity: “\textit{Matter tells spacetime how to curve}.” The complementary notion, “\textit{Spacetime tells matter how to move},” is realised through the $\ODD$-symmetric minimal coupling of the gravitational fields $\{V_{Ap}, \brV_{B\brq}, d\}$ to matter, as demonstrated in (\ref{particleaction}), (\ref{stringaction}), (\ref{YMaction}), (\ref{cDT}),  (\ref{cDrho}), and (\ref{SDFTM}). Consequently, the coupling of the Riemannian trio $\{g_{\mu\nu}, B_{\mu\nu}, \phi\}$ to the Standard Model of particle physics is is fully dictated by the $\ODD$ symmetry principle, ensuring the emergence of a fixed structure that would not otherwise arise~\cite{Choi:2015bga}. Schematically, from (\ref{YMaction}) and (\ref{cDrho}), the action for a scalar field $\Phi$, electromagnetic fields $A_{\lambda}, F_{\mu\nu}$, and a spinorial fermion $\psi$ with diffeomorphic weight $\omega = \frac{1}{2}$ is given by:
\be
\dis{\int_{\Sigma_D}\sqrt{-g}e^{-2\phi}\left(-g^{\mu\nu}\partial_{\mu}\Phi\partial_{\nu}\Phi-{\frac{1}{4\alpha}}F_{\mu\nu}F^{\mu\nu}\right)+\bar{\psi}\gamma^{\lambda}\left( \trd_{\lambda}\psi-iA_{\lambda}\psi + {\frac{1}{24}} H_{\lambda\mu\nu} \gamma^{\mu\nu}\psi \right)}\,.
\label{PhiFpsi}
\ee
This indicates that bosons couple to the string dilaton, while fermions couple to the $H$-flux. As a result, they contribute distinct components to the energy-momentum tensor: bosons generate $\To$,  and  fermions generate  $K_{[\mu\nu]}$,  alongside the shared components $K_{(\mu\nu)}$.  With additional components, the gravitational physics in DFT is inherently richer than in General Relativity:  $D^2+1$ \textit{\,vs.\,} $\frac{1}{2}D(D+1)$ (off-shell) degrees of freedom.

However,  after integrating out the auxiliary potential~$\fa^{A}$, the $\ODD$-symmetric doubled particle action~(\ref{particleaction}) reduces,  on a Riemannian background~(\ref{00H}),    to the standard (undoubled) relativistic point-particle action  minimally coupled solely to the string frame metric~$g_{\mu\nu}$. Since the $\ODD$ singlet dilaton~$d$, or $e^{-2d}$, carries a nontrivial diffeomorphism weight, it cannot couple to the diffeomorphism-invariant doubled particle action~(\ref{particleaction}). Consequently, the free-falling motion of particles along geodesics---free from any fifth force---preserves the equivalence principle, not in  Einstein frame but in string frame~\cite{Ko:2016dxa}. This result aligns with string theory's foundational premise that matter consists of vibrating tiny strings, which naturally couple to the string frame metric.

Importantly, in the string frame, the kinetic term of the string dilaton has an opposite sign, with the square of its time derivative carrying a negative coefficient~(\ref{DFT00}). This property enables the formation of wormholes and drives the accelerated expansion of the Universe, as discussed below.

\summary{The equivalence principle holds in string frame: bosons couple to the dilaton,  while fermions couple  to the $H$-flux, producing dynamics beyond GR.}


\section{Solutions}
With the theoretical foundations established, we now turn to specific solutions within the DFT framework.

Any known ${D=10}$ IIA or IIB supergravity solution consistently satisfies the equations of motion, including the   Einstein Double Field Equation (EDFE), of type II supersymmetric DFT~\cite{Jeon:2012hp}. However, here we focus on more elementary ${D=4}$ solutions.

In General Relativity, the Schwarzschild geometry and de Sitter space are fundamental solutions. The Schwarzschild solution describes the spacetime around a black hole or the exterior geometry of a spherical object, while de Sitter space serves as a model for the large-scale structure of the Universe.  Below, we present their counterparts within the framework of Double Field Theory (DFT).

\subsection{Spherically Symmetric Solution: Generalisation of  the Schwarzschild Geometry\label{SphericalSOL}}
\aim{To solve the EDFE in four dimensions under spherical symmetry.}
\noindent The DFT counterpart to the Schwarzschild geometry in GR is a three-parameter family of vacuum solutions to the ${D=4}$ EDFE, \textit{i.e.}~$G_{AB}=0$ or (\ref{3beta}). These solutions can be traced back to the work of Burgess, Myers, and Quevedo in 1994~\cite{Burgess:1994kq}, who obtained them by performing $\mathbf{SL}(2,\mathbb{R})$ S-duality rotations on a dilaton–metric solution. The solutions were later re-derived~\cite{Ko:2016dxa} as the most general, spherically symmetric, static, asymptotically flat, vacuum geometry of ${D=4}$ DFT. We present the solutions with three constants $\{h,a,b\}$ in an isotropic coordinate system~\cite{Choi:2022srv}, where the metric takes the form, with $r=\sqrt{\vec{x}{\cdot\vec{x}\,}}$,
\be
\rd s^{2}= g_{tt}(r)\,\rd t^{2}+g_{rr}(r)\,\rd \vec{x}\cdot\rd \vec{x}\,.
\label{vacuum}
\ee
The two significant  components of the metric  are
\be
\ba{ll}
\scalebox{1.12}{$
g_{tt}(r)=-e^{2\phi(r)}\left(\frac{4r-\sqrt{a^{2}+b^{2}}}{4r+\sqrt{a^{2}+b^{2}}}\right)^{\frac{2a}{\sqrt{a^{2}+b^{2}}}}\,,$}~~&~~
\scalebox{1.12}{$
g_{rr}(r)=e^{2\phi(r)}\left(\frac{4r+\sqrt{a^{2}+b^{2}}}{4r-\sqrt{a^{2}+b^{2}}}\right)^{\frac{2a}{\sqrt{a^{2}+b^{2}}}}
\left(1-\frac{a^{2}+b^{2}}{16r^{2}}\right)^{\!2}\,,$}
\ea
\ee
the string dilaton is given, with $\gamma_{\pm}=\half\pm\half\sqrt{1-h^2/b^2}$,  by
\be
e^{2\phi}=\gamma_{+}\left(\frac{4r-\sqrt{a^{2}+b^{2}}}{4r+\sqrt{a^{2}+b^{2}}}\right)^{\frac{2b}{\sqrt{a^{2}+b^{2}}}}+\gamma_{-}\left(\frac{4r+\sqrt{a^{2}+b^{2}}}{4r-\sqrt{a^{2}+b^{2}}}\right)^{\frac{2b}{\sqrt{a^{2}+b^{2}}}}\,,
\ee
and the $B$-field features  electric $H$-flux,
\be
\ba{l}
H_{(3)}= h\sin\vartheta\,\rd t\wedge\rd\vartheta\wedge\rd\varphi=h\,\rd t\wedge\left(\dis{\frac{\epsilon_{ijk}x^{i\,}{\rd x^{j}\wedge\rd x^{k}}}{2r^3}}\right)\,.
\ea
\label{eH}
\ee

Several observations can be made about these solutions. 
\begin{itemize}
\item[\textit{i)}] If $b=h=0$, with $h/b\rightarrow 0$,  the solution reduces to the Schwarzschild geometry. 

\item[\textit{ii)}] If $a=0$,   the solution reduces to a wormhole geometry~\cite{Jang:2024nhm},
\be
\ba{ll}
\rd s^{2}=\dis{\frac{-\rd t^{2} +\rd y^{2}}{\cF(y)}}+\cR(y)^2\left(\rd\vartheta^{2}+\sin^{2\!}\vartheta\,\rd\varphi^{2}\right)\,,\quad&\qquad
e^{2\phi(y)}=\dis{\frac{1}{\left|\cF(y)\right|}}\,,
\ea
\ee
where 
\be
\cR(y)=\sqrt{y^{2}+\quarter h^{2}}\,,\qquad
\cF(y)=\dis{\frac{(y-b_{-})(y-b_{+})}{y^2+\frac{1}{4}h^2}}\,,\qquad b_{+}=-b\gamma_{+}\,,\qquad b_{-}=b\gamma_{-}\,.
\ee
While the wormhole throat is at $y=0$,   the points of $y=b_{\pm}$  are  curvature-wise singular within Riemannian geometry: $R\propto 1/(y-b_{\pm})$ as $y\rightarrow b_{\pm}$.   However, as a vacuum solution to the EDFE, the geometry sets both the DFT scalar and  Ricci curvature trivial, $\So=0$ and $S_{p\brq}=0$.  In fact,  by choosing the  $B$-field of the electric $H$-flux~(\ref{eH})  appropriately  to include a term that is pure gauge,
\be
B_{(2)}=h\cos\vartheta\,\rd t\wedge\rd\varphi \,+\, \frac{\rd t\wedge\rd y}{\cF(y)}\,,
\ee
the DFT metric~(\ref{00H}) and dilaton~(\ref{00d}) can both be made entirely non-singular, as can be seen from
\be
\!\!\ba{l}
g^{-1}=\!\left(\!\ba{cccc}
~-\cF~&~0~&~0~&~0~\\
0&\cF&0&0\\
0&0&\frac{1}{\cR^{2}}&0\\
0&0&0&\frac{1}{\cR^{2}\sin^2\vartheta}
\ea\!\right)\,,\quad
Bg^{-1}=\big({-g^{-1}B}\big)^{T}=\!\left(\!\ba{clcc}
~0~&1~~~~&0~&\frac{h\cos\vartheta}{\cR^{2}\sin^2\vartheta}\\
1&0&0&0\\
0&0&0&0\\
h\cos\vartheta\cF&0&0&0\ea\!\right)\,,
\ea
\label{gB}
\ee
and
\be
\ba{l}
g-Bg^{-1}B=\left(\ba{cccc}
\frac{h^2\cos^2\vartheta}{\cR^{2}\sin^2\vartheta}&0&0&0\\
0&0&0&-h\cos\vartheta\\
0&0&\cR^2&0\\
0&-h\cos\vartheta&0&\quad\scalebox{0.9}{$\cR^2\sin^2\vartheta-h^2\cF\cos^2\vartheta$}
\ea\right)\,,
\quad\quad e^{-2d}=\cR^2\sin\vartheta\,.
\ea
\label{gBgBd}
\ee
Only positive powers of $\cF(y)$ appear in (\ref{gB}) and (\ref{gBgBd}) after cancellation of the  negative powers.  This {sufficiently}  implies that, as the $B$-field gauge transformation is a part of doubled-yet-gauged diffeomorphisms, the curvature singularity as characterised within Riemannian geometry is to be identified as a coordinate singularity within  DFT~\cite{Morand:2021xeq}. When  $\cF(y)$ vanishes at ${y=b_{\pm}}$,  the upper-left block of the generalised metric, corresponding to $g^{-1}$, becomes degenerate,  no invertible Riemannian metric is defined, and thus the geometry is non-Riemannian. From the perspective of DFT, the geometry is regular everywhere~\cite{Morand:2021xeq}: it is  non-Riemannian at $y=b_{\pm}$ and  Riemannian elsewhere.     In fact,    it is  the same type of non-Riemannian  geometry, $(n,\brn)=(1,1)$, as    non-relativistic string theory~\cite{Gomis:2000bd}, as can be seen from (\ref{gB}) and (\ref{gBgBd})~\cite{Ko:2015rha,Morand:2017fnv}.\vspace{3pt}

\item[\textit{iii)}] Matching with  the  parametrised Post-Newtonian (PPN) formalism~\cite{Will:1972zz,Will:2014kxa},
\be
\rd s^{2}=-\left(1-\frac{2MG}{r}+\frac{2\betappn (MG)^2}{r^{2}}+\cdots\right)\!\rd t^{2}+\left(1+\frac{2\gammappn MG}{r}+\cdots\right)\rd x^{i}\rd x^{j}\delta_{ij}\,,
\label{PPN}
\ee
we can identify the Newtonian mass and the  Eddington--Robertson--Schiff parameters~\cite{Choi:2022srv},
\be
\ba{lll}
2MG=a+b\sqrt{1-h^{2}/b^{2}}\,,\quad&~
\betappn=1+\left(\frac{h}{a+b\sqrt{1-h^{2}/b^{2}}}\right)^2\,,\quad&~
\gammappn=1-\frac{2b\sqrt{1-h^{2}/b^{2}}}{a+b\sqrt{1-h^{2}/b^{2}}}\,.
\ea
\label{Mbg}
\ee

\item[\textit{iv)}] The vacuum solution~(\ref{vacuum}) represents the external geometry of a compact spherical object, such as a star. The full EDFE~(\ref{EDFE00}) then gives an integral formula for the Newtonian mass~\cite{Choi:2022srv},
\be
MG
=\frac{1}{4\pi}\int\rd^{3}x~e^{-2d}\left( -K_{t}{}^{t}-\frac{1}{2}H_{t\vartheta\varphi}H^{t\vartheta\varphi}\right)\,,
\label{MGint}
\ee
where the integration occurs within the star's interior. Assuming the star's geometry is regular, the integral of the squared electric $H$-flux term in (\ref{MGint}) diverges unless ${h = 0}$. To ensure physical consistency, we impose a weak energy condition, $-K_{t}{}^{t} \geq 0$, and assume a finite Newtonian mass, \textit{i.e.~}$MG \ll \infty$. Under these conditions, we deduce that the electric $H$-flux must be trivial: ${h = 0}$. Consequently, from (\ref{Mbg}), this leads to ${\betappn = 1}$, which aligns with the result in GR, or correspondingly, the  Schwarzschild geometry.  The constants $a$ and $b$ are then determined as follows:
\be
\ba{rll}
a&=&\dis{\frac{1}{4\pi}\int\rd^3 x~e^{-2d}
\left( K_{\mu}{}^{\mu}-2K_{t}{}^{t}- \To+ H_{r\vartheta\varphi}H^{r\vartheta\varphi}\right)\,,}\\
b&=&\dis{\frac{1}{4\pi}\int\rd^3 x~e^{-2d}
\left(-K_{\mu}{}^{\mu}+\To-H_{r\vartheta\varphi}H^{r\vartheta\varphi}\right)\,,}
\ea
\label{ab}
\ee
with their square sum satisfying a nontrivial  relation:  given the star's radius $r_\star$,
\be
\frac{a^{2}+b^{2}}{16}=\int_{0}^{r_{\star}}\rd r\, r\int_{r}^{r_{\star}}
\frac{\rd r^{\prime}}{r^{\prime}}
\left(\frac{e^{-2d}}{\sin\vartheta}\right)\!
\left(K_{r}{}^{r}+K_{\vartheta}{}^{\vartheta}-\To+\frac{1}{2} H_{r\vartheta\varphi}H^{r\vartheta\varphi}\right)\,.
\ee
From (\ref{Mbg}), $\gammappn$ is thereby determined as:
\be
\gammappn=1+\frac{\dis{\int}\rd^{3}x~e^{-2d}\big(K_{\mu}{}^{\mu}-\To+H_{r\vartheta\varphi}H^{r\vartheta\varphi}\big)}{\dis{\int}\rd^{3}x~e^{-2d}\big(-K_{t}{}^{t}\big)}\,,
\label{gammappn}
\ee 
where the numerator corresponds to the Chameleon mass of the string dilaton (\ref{Chameleon}) and the denominator to the Newtonian mass of the star (\ref{MGint}).

\item[\textit{v)}] The current stringent observational bounds on gravity in the solar system  are given by $\gammappn=1+(2.1\pm 2.3)\times 10^{-5}$,  as determined from the Shapiro time-delay measurements by the Cassini spacecraft~\cite{Will:2014kxa, Bertotti:2003rm}. Consequently, DFT can successfully pass solar system tests, provided the dilatonic Chameleon mass is at least $10^{-5}$ times smaller than the Sun's Newtonian mass.  
\end{itemize}
\summary{The  three-parameter family of spherically symmetric solutions to the EDFE in ${D=4}$  includes  Schwarzschild, wormhole, and dilatonic solutions. Post-Newtonian analysis shows consistency with solar-system tests provided the dilaton's  Chameleon mass   is suppressed.}

\subsection{Cosmological Solution of Open Universe: Alternative to de Sitter Universe}
\aim{To obtain a cosmological solution of DFT consistent with real data.}
\noindent The de Sitter Universe, while serving as a natural cosmological solution in GR and providing a simple model for the Universe's accelerating expansion,  encounters a fundamental limitation in DFT: the cosmological term $\sqrt{-g}\Lambda$ lacks $\ODD$ symmetry, as originally  pointed out by Gasperini and Veneziano in 1991~\cite{Gasperini:1991ak}. Consequently, the de Sitter Universe is incompatible with the EDFE~(\ref{EDFE}) and the implied $\ODD$-symmetric extension of the Friedmann equations~\cite{Angus:2019bqs}. This incompatibility aligns with the de Sitter swampland conjectures~\cite{Danielsson:2018ztv, Obied:2018sgi, Agrawal:2018own, Andriot:2018wzk}.

We introduce an alternative cosmological model: an open Universe characterised by negative spatial curvature ($k < 0$) that serves as a vacuum solution to the EDFE,  providing a compelling alternative to the de Sitter Universe. This solution  traces  back to the work of Copeland, Lahiri, and Wands in 1994~\cite{Copeland:1994vi}, who derived homogeneous and isotropic solutions to the three beta-function equations on the string worldsheet~(\ref{3beta}). Building upon this foundation, the model was further refined in~\cite{Lee:2023boi}, incorporating essential physical parameters: the Hubble constant~$H_{0}$, the spatial curvature scale~$l=1/\sqrt{-k}$, the magnetic $H$-flux~$\hh$, as well as additional redundant parameters $a_{0}$ and $\phi_{0}$, defined at the (present) conformal time $\eta = \eta_{0}$.

The vacuum geometry of the open Universe is characterized by the following triplet: the string dilaton,
\be
e^{2\phi(\eta)-2\phi_{0}}=\scalebox{0.96}{$\frac{1}{2}\!
\left[1+\sigma\sqrt{1-\frac{(\hh l\sinh\zeta)^2}{12a_{0}^4}}\right]\!\left[
\frac{\tanh\left(\frac{\eta-\eta_{0}}{l}+\frac{\zeta}{2}\right)}{\tanh\frac{\zeta}{2}}\right]^{\sqrt{3}}\vspace{7pt}+\frac{1}{2}\!
\left[1-\sigma\sqrt{1-\frac{(\hh l\sinh\zeta)^2}{12a_{0}^4}}\right]\!\left[
\frac{\tanh\left(\frac{\eta-\eta_{0}}{l}+\frac{\zeta}{2}\right)}{\tanh\frac{\zeta}{2}}\right]^{\!-\sqrt{3}\,}\,,$}
\label{analyticphi}
\ee
the (homogeneous \& isotropic)   magnetic $H$-flux,
\be
H_{(3)}=\frac{\hh\, r^2 \sin\vartheta\,\rd r\wedge\rd\vartheta\wedge\rd\varphi}{\sqrt{1+r^2/l^2}}\,,
\label{mHflux}
\ee
and the Friedmann--Lema\^{i}tre--Robertson--Walker metric with $k=-1/l^{2}<0\,$, 
\be
\rd s^2=a^2(\eta)\left[-\rd\eta^2+\frac{\rd r^2}{1+ r^2/l^2}+r^2\rd\vartheta^2+r^2{\sin^2\!\vartheta\,}\rd\varphi^2\right]\,,
\label{FRWconformal}
\ee
of which the scale factor  is given by the following expression:
\be
a^{2}(\eta)=a_{0}^2\,e^{2\phi(\eta)-2\phi_{0}}\,\frac{\sinh\left({2(\eta-\eta_{0})}/{l}+\zeta\right)}{\sinh\zeta}\,.
\label{SF}
\ee
Here, $\zeta$ is a constant and $\sigma$ is a sign factor (${\sigma^2=1}$), both determined by the parameters $\{H_{0}, l, \hh\}$. The Hubble parameter and two density parameters are  defined at arbitrary conformal time~$\eta$ as
\be
\ba{lll}
\dis{
H=\frac{1}{a^2}\frac{\rd a}{\rd\eta}\,,}\qquad&\qquad
\dis{\Omega_{k}=\frac{1}{l^2a^2H^2}\,,}\qquad&\qquad
\dis{\Omega_{\hh}=\frac{\hh^2}{12a^6H^2}\,.}
\ea
\label{HOO}
\ee
At the specific conformal time $\eta = \eta_0$, the analytic solution for the scale factor~(\ref{SF}) yields the Hubble constant, $H_0 = H(\eta_0)$,  as:
\be
H_{0}=\frac{1}{2a_{0}l\sinh\zeta}\left[2\cosh\zeta+
{\sigma\sqrt{12-a_{0}^{-4}\left(\hh l\sinh\zeta\right)^2}}
\right]\,,
\ee
which can be solved to express $\zeta$ and $\sigma$ in terms of the density parameters at $\eta = \eta_0$. These expressions are
\be
\ba{ll}\dis{
\sinh\zeta=
\sqrt{\frac{2\Omega_{0,k}}{2+\Omega_{0,k}+3\Omega_{0,\hh}-\sqrt{3+6\Omega_{0,k}+6\Omega_{0,\hh}}}}}\,,\quad&\qquad
\dis{\sigma=\frac{\,\sqrt{3}-\sqrt{1+2\Omega_{0,k}+2\Omega_{0,\hh}}\,}{\,\left|\sqrt{3}-\sqrt{1+2\Omega_{0,k}+2\Omega_{0,\hh}}\right|\,}}\,.
\ea
\label{szs}
\ee
In other words, ${\sigma = -1}\,$ if $\,\Omega_{0,k} + \Omega_{0,\hh} > 1$; otherwise, ${\sigma = +1}$.

Some comments are in order, as discussed in \cite{Lee:2023boi}.
\begin{itemize}
\item[\textit{i)}] The deceleration parameter can be expressed in terms of the density parameters as follows:
\be
q=\scalebox{1}{$-\frac{1}{H^2a}\Big(\frac{\rd~}{a\rd \eta}\Big)^{2}a
=1-\left[\frac{2\left(\frac{\,\rd\phi}{a\rd \eta}\right)^2}{3H^2}+\Omega_{k}+5\Omega_{\hh}\,\right]=-\Big[1+2\Omega_{k}+6\Omega_{\hh}-\sqrt{3(1+2\Omega_{k}+2\Omega_{\hh})}\,\Big]\,,$}
\label{decq}
\ee
which can be readily shown to assume negative values, particularly when ${\Omega_k + \Omega_\hh > 1}$. Such acceleration is only achievable in  string frame, where the string frame metric minimally couples to point particles~(\ref{particleaction}), thereby ensuring that the equivalence principle remains valid and simultaneously exhibiting a negative sign in the dilaton's kinetic term~(\ref{DFT00}). The dilaton drives the acceleration in string frame, without any need for dark energy, but not in Einstein frame.

\item[\textit{ii)}] As dictated by the $\ODD$ symmetry principle, the string dilaton is expected to couple to gauge bosons, as demonstrated in (\ref{PhiFpsi}), implying that the  fine-structure constant is  effectively proportional to the  exponential of the string  dilaton:  
\be
\alpha_{\rm{eff.}}\!(\eta)=\alpha\, e^{2\phi(\eta)}\,.
\label{alphaeff}
\ee
However, observational constraints from the absorption spectra of quasars impose stringent limits on the temporal variation of the fine-structure constant~\cite{King:2012id,Wilczynska:2015una,Martins:2017qxd,Wilczynska:2020rxx}, and consequently on the evolution of the string dilaton~$\phi$. The analytic formula~(\ref{analyticphi}) shows that the string dilaton converges at future infinity, $\eta \rightarrow \infty$. Replacing $l$ and $\zeta$ with imaginary numbers, $-il$ and $i\zeta$, yields the exact geometry of a closed Universe ($k > 0$) (see \cite{Copeland:1994vi} for the explicit expression). However, this substitution transforms the converging hyperbolic tangent functions in (\ref{analyticphi}) into diverging tangent functions, preventing the dilaton $\phi$ from stabilising during cosmic evolution, which conflicts with the observational constraints from  quasars. Similarly, when $k = 0$, the hyperbolic tangent functions reduce to linear behaviour in time, which is equally inconsistent with the observational requirements. These results underscore the necessity of an open Universe ($k < 0$) to ensure that the dilaton $\phi$ evolves slowly and remains convergent throughout the cosmic evolution.

\item[\textit{iii)}] The vacuum geometry of the open Universe given in  (\ref{analyticphi}) and (\ref{SF}), in particular the case with trivial $H$-flux (${\hh=0}$ in (\ref{mHflux})), demonstrates remarkable agreement with late-time cosmological observations, including type Ia supernovae data~\cite{Scolnic:2021amr,Riess:2021jrx} and quasar absorption spectra~\cite{King:2012id,Wilczynska:2015una,Martins:2017qxd,Wilczynska:2020rxx}, as depicted in Figure~\ref{FIGcosmo}, reproduced from \cite{Lee:2023boi}. Such observations probe the evolution of the Hubble parameter and the potentially varying fine-structure constant up to redshift $z=a^{-1}-1\approx 7$. Through an analysis of  Bayesian inference, the Hubble constant and the curvature density parameter are  estimated  to be  
\be
\ba{ll}
H_0 \simeq 71.29    \pm   0.12\,\mathrm{km/s/Mpc}\,,\quad&\qquad
\Omega_{0,k}\simeq {1+(6\pm 2)\times 10^{-7}}\,.
\ea
\ee 
These imply a curvature scale of $\,l=1/\sqrt{-k}\simeq 4.2\,\mathrm{Gpc}\,$ and select  the negative sign factor,  ${\sigma=-1}$,  in  (\ref{szs}).  Such results are consistent with the analytic limiting behaviours of the scale factor~(\ref{SF}):
\be
\ba{llll}
\dis{
\lim_{\eta\rightarrow\infty}a=\infty\,,}\quad&\qquad\dis{
\lim_{\eta\rightarrow\infty}H=0\,,}\quad&\qquad
\dis{\lim_{\eta\rightarrow\infty}\Omega_{k}=1\,,}\quad&\qquad
\dis{\lim_{\eta\rightarrow\infty}\Omega_{\hh}=0\,.}
\ea
\label{HOO}
\ee\vspace{-3pt}
In other words, there is no coincidence problem.

\item[\textit{iv)}] Extending to higher redshifts, a bounce is expected to have occurred approximately 13.7 gigayears ago. Coincidently, this is comparable to the ``age'' of the flat Universe in $\Lambda$CDM.
\begin{figure}[H]
\qquad\qquad\begin{subfigure}{.38\textwidth}
  \centering
  \includegraphics[width=\linewidth,height=0.8\textwidth]{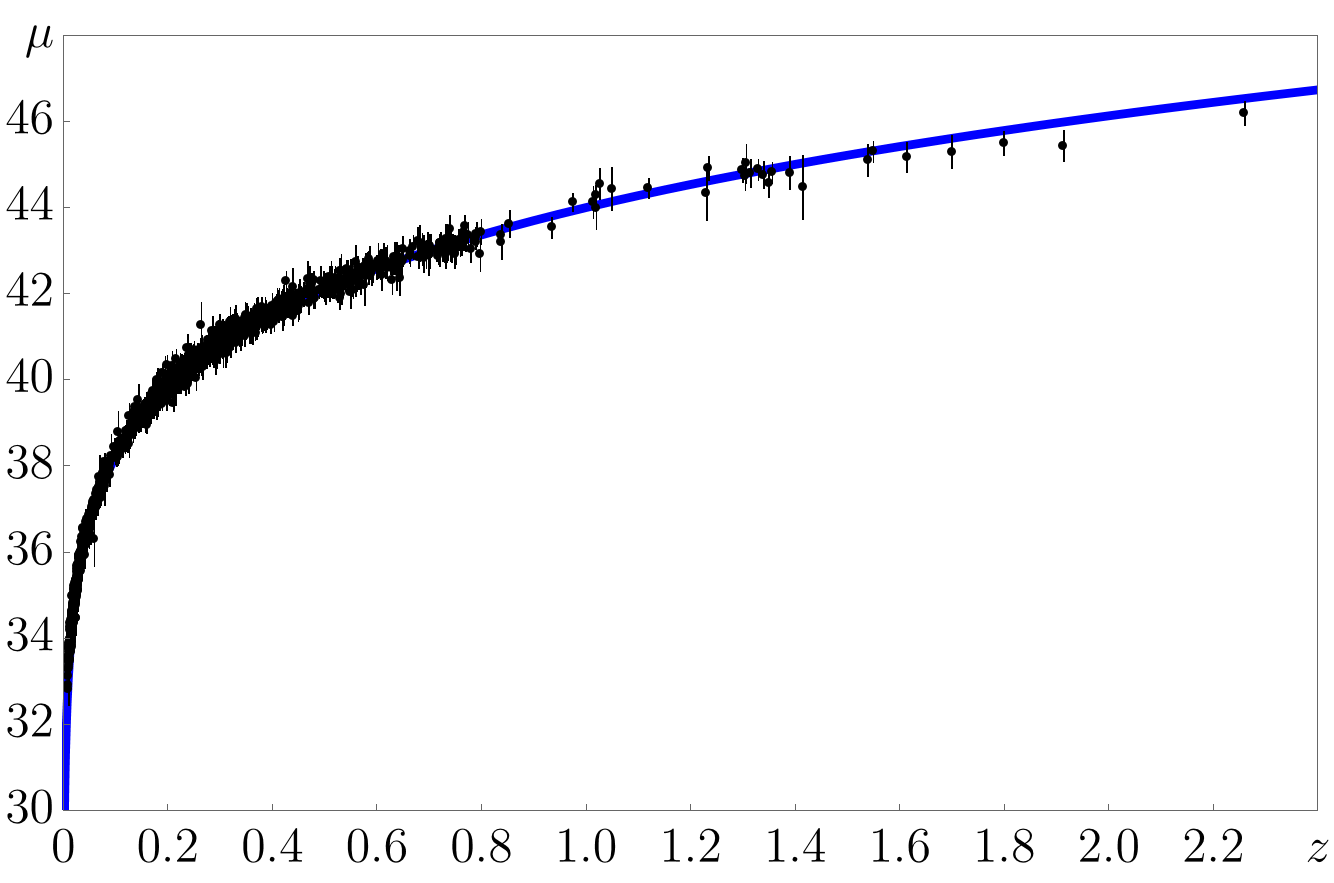}
  \caption{Type Ia Supernovae~\cite{Scolnic:2021amr,Riess:2021jrx}}
\end{subfigure}~
\quad\begin{subfigure}{0.38\textwidth}
  \centering
  \includegraphics[width=\linewidth,height=0.8\textwidth]{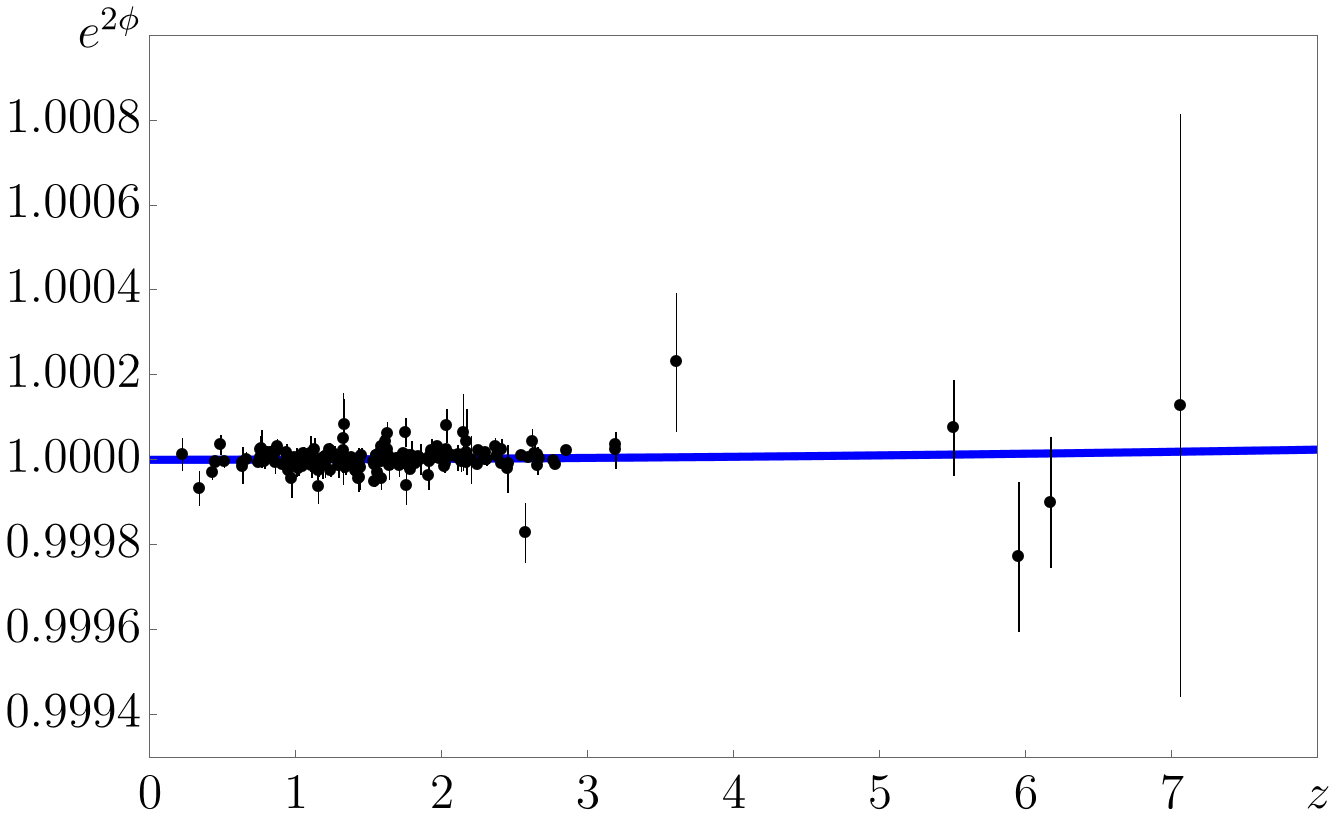}
  \caption{Quasar Absorption Spectra~\cite{King:2012id,Wilczynska:2015una,Martins:2017qxd,Wilczynska:2020rxx}}
\end{subfigure}\\~\\~\\\vspace{2pt}
\qquad\qquad\begin{subfigure}{0.38\textwidth}
  \centering
  \includegraphics[width=\linewidth,height=0.9\textwidth]{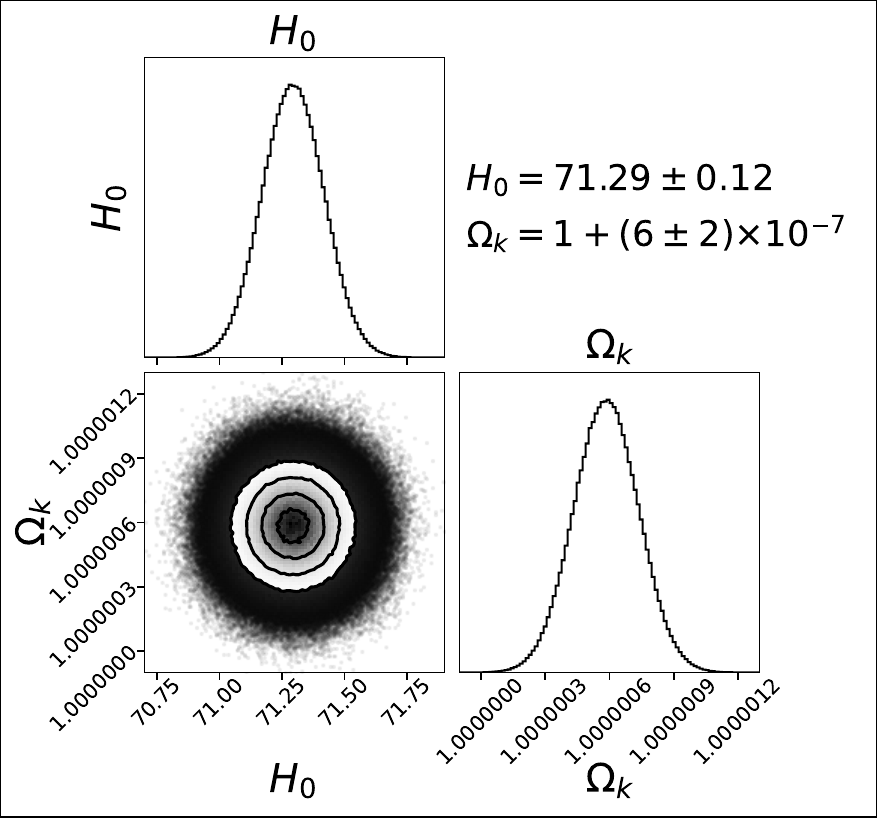}
  \caption{Bayesian Inference with Two Parameters}
\end{subfigure}~
\quad\begin{subfigure}{0.38\textwidth}
  \centering
  \includegraphics[width=\linewidth,height=0.8\textwidth]{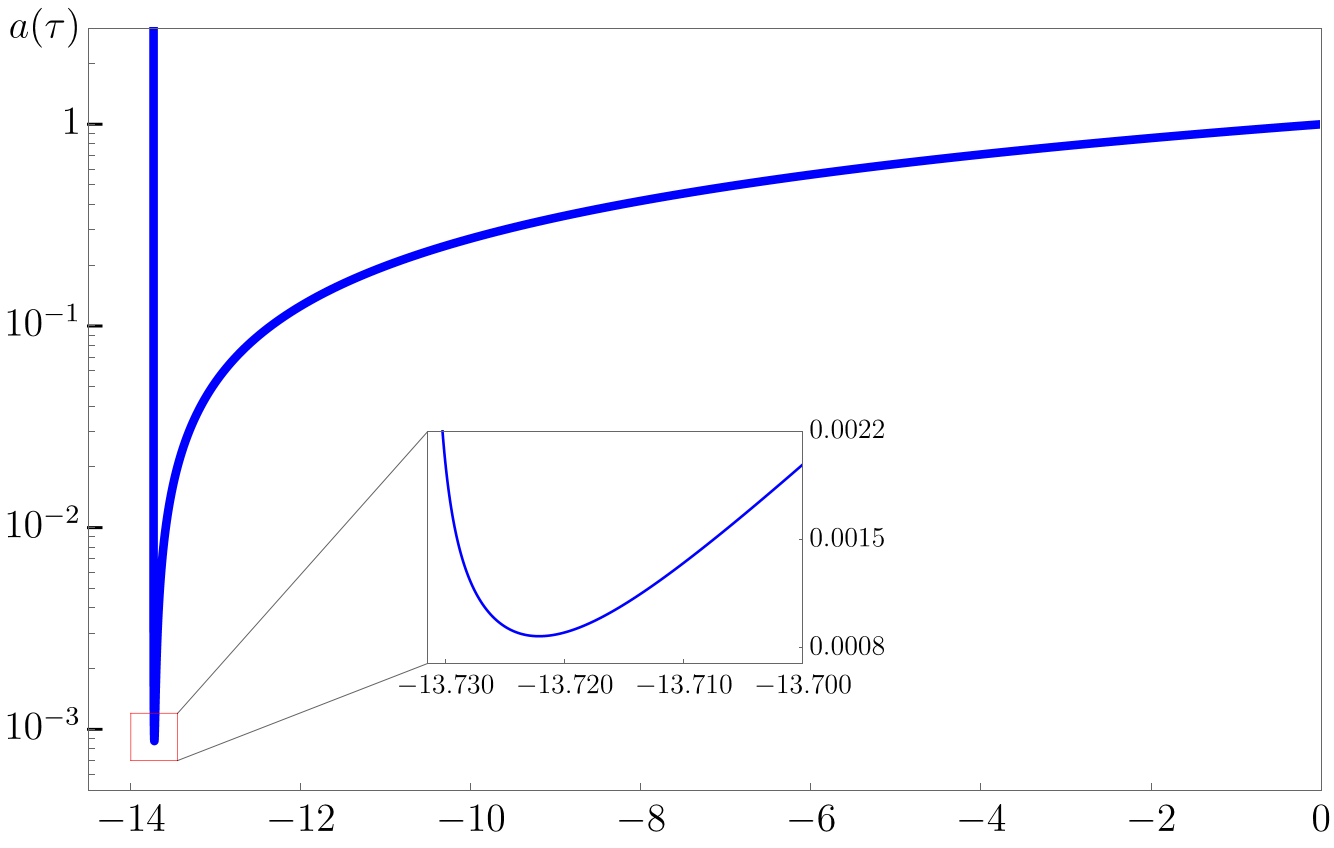}
  \caption{Bouncing: $13.7$ Gigayears ago}
\end{subfigure}
\caption{Agreement of the vacuum geometry of DFT~(\ref{analyticphi}), (\ref{SF}) with late-time cosmological data of type Ia supernovae~(a) and quasars~(b).  For the latter, the string dilaton $\phi$  already appears to be  in the convergent phase.    Bayesian inference~(c) shows that, in sharp contrast to $\Lambda$CDM, at present time the curvature density overwhelmingly  dominates,  $\Omega_{0,k}\simeq {1+(6\pm 2)\times 10^{-7}}$, implying an open Universe.  A bounce~(d) is  extrapolated to occur   $13.7$ gigayears ago.   Figures  reproduced from \cite{Lee:2023boi}.}
\label{FIGcosmo}
\end{figure}
\end{itemize}
\summary{An open Universe with negative curvature replaces de Sitter space. The dilaton drives late-time acceleration in string frame without dark energy, the solution fits supernova and quasar observations, and it predicts a cosmological bounce about $13.7$ Gyr ago.}

\section{Concluding Remarks \& Outlook}
Double Field Theory (DFT) has developed into a self-sustaining autonomous theory of gravity, offering a compelling alternative to General Relativity (GR). Governed by the $\ODD$ symmetry principle rooted in string theory, DFT uniquely fixes the self-interaction of the gravitational sector, $\{\cH_{AB}, d\}$ or the Riemannian trio $\{g, B, \phi\}$, as well as its minimal coupling to other sectors or additional matter, placing it in a distinct position among various modified theories of gravity.   
The rigorously defined structure of DFT makes it inherently falsifiable, a critical feature for endowing theoretical physics with value.

\indent The spherically symmetric solution~(\ref{vacuum}) has been evaluated against solar system experiments using the Parametrised Post-Newtonian (PPN) formalism. Unlike General Relativity (GR), where Birkhoff's theorem holds and a single component of the energy-momentum tensor determines the geometry, Double Field Theory  incorporates multiple components~(\ref{ab}), enabling the exploration of more complex phenomena. For example, studying the geometry outside compact objects or stars may offer insights into their generalised equation of state~(\ref{gammappn}) or the generalised gravitational form factor within hadrons, expressed in terms of $K_{\mu\nu}$ and $\To$ rather than  the traditional $T_{\mu\nu}$.  Notably, recent experimental findings, such as the discovery of immense pressure inside protons~\cite{Burkert:2018bqq}, challenge traditional assumptions of matter being ``cold" and may underscore the relevance of DFT's multi-component energy-momentum tensors. While these possibilities are intriguing, solar system tests of DFT remain inconclusive, necessitating further investigations, including insights from nuclear physics.

\indent On cosmological scales, the open-Universe solution given through (\ref{analyticphi}---\ref{SF})  captures accelerating expansion and demonstrates strong consistency with late-time cosmological observations, including type Ia supernovae and quasars.  However, the theory's applicability to the early Universe remains uncertain. The non-convergent phase of the string dilaton during the early Universe suggests a nontrivial coupling to electromagnetism~(\ref{PhiFpsi}), (\ref{alphaeff}), which may influence light propagation or null geodesics, thereby affecting  measurements of luminosity distance. Further corroboration, incorporating diverse data sources such as the cosmic microwave background~\cite{Planck:2018vyg}, is essential to clarify DFT's role in explaining the early Universe.

\indent In terms of symmetry, DFT distinguishes itself by being approximately twice as symmetrical as General Relativity. This profound structure invites speculation about whether Nature and the Universe might harness such symmetry. As a falsifiable prediction of string theory, DFT stands as a promising candidate awaiting experimental verification.

\indent Beyond its phenomenological significance, DFT, as the gravitational framework of string theory, faces the critical challenge of addressing extra dimensions and higher-derivative corrections. A compelling and still largely unexplored possibility regarding the former is whether these extra dimensions could take a non-Riemannian form~\cite{Morand:2017fnv, Cho:2018alk}. Regarding the latter, some pioneering contributions include~\cite{Hohm:2013jaa,Bedoya:2014pma,Hohm:2014xsa,Marques:2015vua}.

\indent Lastly, the quantisation of Double Field Theory in an $\ODD$-symmetric framework remains a vital and promising avenue for further investigation.

\indent While the aforementioned topics represent potential future works within the limitations of the author's imagination, the over 100-year legacy of GR suggests that DFT could unlock far broader and richer gravitational-related research directions. The development of DFT is an ongoing process, with immense potential to deepen our understanding of gravity.

\subsection*{Acknowledgements}  
The author extends sincere gratitude to  Manu Paranjape,  Nobuyoshi Ohta, and an anonymous peer  in Kyoto for their  encouraging suggestions regarding the writing of these lecture notes, and to Stephen Angus for his meticulous proofreading and invaluable feedback. This  work is supported by the National Research Foundation of Korea (NRF)  through  the Grants RS-2023-NR077094 and  RS-2020-NR049598   (Center for Quantum Spacetime: CQUeST).





\begin{thebibliography}{99}
\bibitem{Hardy}
G.~.H.~Hardy, 
``A Mathematician's Apology,'' 
Cambridge University Press, Cambridge: 1940.

\bibitem{Siegel:1993xq}
  W.~Siegel,
  ``Two vierbein formalism for string inspired axionic gravity,''
  Phys.\ Rev.\ D {\bf 47} (1993) 5453
  [hep-th/9302036].


\bibitem{Siegel:1993th}
  W.~Siegel,
  ``Superspace duality in low-energy superstrings,''
  Phys.\ Rev.\ D {\bf 48} (1993) 2826
  [hep-th/9305073].


\bibitem{Hull:2009mi}
  C.~Hull and B.~Zwiebach,
  ``Double Field Theory,''
  JHEP {\bf 0909} (2009) 099
  [arXiv:0904.4664 [hep-th]].


\bibitem{Hull:2009zb}
  C.~Hull and B.~Zwiebach,
  ``The Gauge algebra of double field theory and Courant brackets,''
  JHEP {\bf 0909} (2009) 090
  [arXiv:0908.1792 [hep-th]].


\bibitem{Hohm:2010jy}
  O.~Hohm, C.~Hull and B.~Zwiebach,
  ``Background independent action for double field theory,''
  JHEP {\bf 1007} (2010) 016
  [arXiv:1003.5027 [hep-th]].


\bibitem{Hohm:2010pp}
  O.~Hohm, C.~Hull and B.~Zwiebach,
  ``Generalized metric formulation of double field theory,''
  JHEP {\bf 1008} (2010) 008
  [arXiv:1006.4823 [hep-th]].

\bibitem{Duff:1989tf}
M.~J.~Duff,
``Duality Rotations in String Theory,''
Nucl. Phys. B \textbf{335} (1990), 610
doi:10.1016/0550-3213(90)90520-N

\bibitem{Aldazabal:2013sca}
G.~Aldazabal, D.~Marques and C.~Nunez,
``Double Field Theory: A Pedagogical Review,''
Class. Quant. Grav. \textbf{30} (2013), 163001
doi:10.1088/0264-9381/30/16/163001
[arXiv:1305.1907 [hep-th]].

\bibitem{Berman:2013eva}
D.~S.~Berman and D.~C.~Thompson,
``Duality Symmetric String and M-Theory,''
Phys. Rept. \textbf{566} (2014), 1-60
doi:10.1016/j.physrep.2014.11.007
[arXiv:1306.2643 [hep-th]].

\bibitem{Hohm:2013bwa}
O.~Hohm, D.~L\"ust and B.~Zwiebach,
``The Spacetime of Double Field Theory: Review, Remarks, and Outlook,''
Fortsch. Phys. \textbf{61} (2013), 926-966
doi:10.1002/prop.201300024
[arXiv:1309.2977 [hep-th]].


\bibitem{Brandenberger:2018xwl}
R.~Brandenberger, R.~Costa, G.~Franzmann and A.~Weltman,
``Dual spacetime and nonsingular string cosmology,''
Phys. Rev. D \textbf{98} (2018) no.6, 063521
doi:10.1103/PhysRevD.98.063521
[arXiv:1805.06321 [hep-th]].

\bibitem{Brandenberger:2018bdc}
R.~Brandenberger, R.~Costa, G.~Franzmann and A.~Weltman,
``T-dual cosmological solutions in double field theory,''
Phys. Rev. D \textbf{99} (2019) no.2, 023531
doi:10.1103/PhysRevD.99.023531
[arXiv:1809.03482 [hep-th]].


\bibitem{Kang:2019owv}
S.~Kang and D.~h.~Yeom,
``Causal structures and dynamics of black-hole-like solutions in string theory,''
Eur. Phys. J. C \textbf{79} (2019) no.11, 927
doi:10.1140/epjc/s10052-019-7435-7
[arXiv:1901.06857 [gr-qc]].

\bibitem{Lescano:2021nju}
E.~Lescano and N.~Mir\'on-Granese,
``Double field theory with matter and its cosmological application,''
Phys. Rev. D \textbf{107} (2023) no.4, 046016
doi:10.1103/PhysRevD.107.046016
[arXiv:2111.03682 [hep-th]].


\bibitem{Liu:2021xfs}
Y.~Liu,
``Hawking temperature and the bound on greybody factors in $D=4$ double field theory,''
Eur. Phys. J. C \textbf{82} (2022) no.11, 1054
doi:10.1140/epjc/s10052-022-11022-4
[arXiv:2201.01279 [hep-th]].

\bibitem{Liu:2021gjm}
Y.~Liu,
``Dilatonic effect in double field theory cosmology,''
Gen. Rel. Grav. \textbf{54} (2022) no.2, 18
doi:10.1007/s10714-022-02898-4
[arXiv:2112.15082 [hep-th]].


\bibitem{Lescano:2022nrb}
E.~Lescano and N.~Mir\'on-Granese,
``Double field theory with matter and the generalized Bergshoeff\textendash{}de Roo identification,''
Phys. Rev. D \textbf{107} (2023) no.8, 086008
doi:10.1103/PhysRevD.107.086008
[arXiv:2207.04041 [hep-th]].

\bibitem{Brandenberger:2023ver}
R.~Brandenberger,
``Superstring cosmology \textemdash{} a complementary review,''
JCAP \textbf{11} (2023), 019
doi:10.1088/1475-7516/2023/11/019
[arXiv:2306.12458 [hep-th]].

\bibitem{Lescano:2023gge}
E.~Lescano, N.~Mir\'on-Granese and Y.~Sakatani,
``O(D,D)-covariant formulation of perfect and imperfect fluids in the double geometry,''
Phys. Rev. D \textbf{109} (2024) no.8, 086006
doi:10.1103/PhysRevD.109.086006
[arXiv:2312.03610 [hep-th]].

\bibitem{Li:2023mhz}
S.~Li and Y.~Liu,
``Observational tests of 4D double field theory,''
Eur. Phys. J. C \textbf{84} (2024) no.5, 496
doi:10.1140/epjc/s10052-024-12858-8
[arXiv:2312.07794 [hep-th]].


\bibitem{Arapoglu:2024umz}
A.~S.~Arapo\u{g}lu, S.~\c{C}a\u{g}an and A.~\c{C}atal-\"Ozer,
``Stability analysis of the cosmological dynamics of O(D,~D)-complete stringy gravity,''
Eur. Phys. J. C \textbf{84} (2024) no.8, 848
doi:10.1140/epjc/s10052-024-13213-7
[arXiv:2405.07825 [gr-qc]].


\bibitem{Ying:2024jjr}
S.~Ying,
``Black hole solutions in double field theory,''
Eur. Phys. J. C \textbf{84} (2024) no.11, 1185
doi:10.1140/epjc/s10052-024-13537-4
[arXiv:2407.11403 [hep-th]].

\bibitem{Angus:2024owg}
S.~Angus and S.~Mukohyama,
``Perturbations in $\textbf{O}(D,D)$ string cosmology from double field theory,''
Eur. Phys. J. C \textbf{85} (2025) no.2, 173
doi:10.1140/epjc/s10052-025-13841-7
[arXiv:2408.13032 [hep-th]].

\bibitem{Angus:2018mep}
S.~Angus, K.~Cho and J.~H.~Park,
``Einstein Double Field Equations,''
Eur. Phys. J. C \textbf{78} (2018) no.6, 500
doi:10.1140/epjc/s10052-018-5982-y
[arXiv:1804.00964 [hep-th]].


\bibitem{YITP2025}
``\href{https://indico.yukawa.kyoto-u.ac.jp/event/52/}{\textit{Quantum Gravity and Information in Expanding Universe}},''    
Yukawa Institute for Theoretical Physics, Kyoto University, Japan, February  17-21, 2025.

\bibitem{Bangkok2025}
``\href{https://www.thaihep.phys.sc.chula.ac.th/BKK2025HEPTH/}{\textit{12th Bangkok workshop on High-Energy Theory}},''  
Chulalongkorn University, Thailand, January 20-24, 2025.



\bibitem{Munchen2010}
 ``\href{https://www.theorie.physik.uni-muenchen.de/activities/schools/archiv/sfp10/index.html}{ \textit{International School on Strings and Fundamental Physics}},''  M\"{u}nchen/Garching,  from 25th July to 6th August, 2010.





\bibitem{Jeon:2010rw}
  I.~Jeon, K.~Lee and J.~H.~Park,
  ``Differential geometry with a projection: Application to double field theory,''
  JHEP {\bf 1104} (2011) 014
  [arXiv:1011.1324 [hep-th]].
  
  
\bibitem{Peeters:2007wn}
K.~Peeters,
``Introducing Cadabra: A Symbolic computer algebra system for field theory problems,''
[arXiv:hep-th/0701238 [hep-th]].



\bibitem{Park:2015bza}
  J.~H.~Park, S.~J.~Rey, W.~Rim and Y.~Sakatani,
  ``$\ODD$ Covariant Noether Currents and Global Charges in Double Field Theory,''
  arXiv:1507.07545 [hep-th].
  
  
\bibitem{Duff:1986ne}
M.~J.~Duff,
``HIDDEN STRING SYMMETRIES?,''
Phys. Lett. B \textbf{173} (1986), 289-296
doi:10.1016/0370-2693(86)90519-8


\bibitem{Jeon:2011sq}
  I.~Jeon, K.~Lee and J.~H.~Park,
  ``Supersymmetric Double Field Theory: Stringy Reformulation of Supergravity,''
  Phys.\ Rev.\ D {\bf 85} (2012) 081501
   [Phys.\ Rev.\ D {\bf 86} (2012) 089903]
  [arXiv:1112.0069 [hep-th]].
  
  
\bibitem{Jeon:2012hp}
  I.~Jeon, K.~Lee, J.~H.~Park and Y.~Suh,
  ``Stringy Unification of Type IIA and IIB Supergravities under N=2 D=10 Supersymmetric Double Field Theory,''
  Phys.\ Lett.\ B {\bf 723} (2013) 245
  [arXiv:1210.5078 [hep-th]].
  
  

\bibitem{Choi:2015bga}
K.~S.~Choi and J.~H.~Park,
``Standard Model as a Double Field Theory,''
Phys. Rev. Lett. \textbf{115} (2015) no.17, 171603
doi:10.1103/PhysRevLett.115.171603
[arXiv:1506.05277 [hep-th]].




\bibitem{Berman:2010is}
D.~S.~Berman and M.~J.~Perry,
``Generalized Geometry and M theory,''
JHEP \textbf{06} (2011), 074
doi:10.1007/JHEP06(2011)074
[arXiv:1008.1763 [hep-th]].


\bibitem{Buscher:1987qj}
T.~H.~Buscher,
``Path Integral Derivation of Quantum Duality in Nonlinear Sigma Models,''
Phys. Lett. B \textbf{201} (1988), 466-472
doi:10.1016/0370-2693(88)90602-8

\bibitem{Buscher:1987sk}
T.~H.~Buscher,
``A Symmetry of the String Background Field Equations,''
Phys. Lett. B \textbf{194} (1987), 59-62
doi:10.1016/0370-2693(87)90769-6

\bibitem{Park:2013mpa}
  J.~H.~Park,
  ``Comments on double field theory and diffeomorphisms,''
  JHEP {\bf 1306} (2013) 098
  [arXiv:1304.5946 [hep-th]].



\bibitem{Lee:2013hma}
  K.~Lee and J.~H.~Park,
  ``Covariant action for a string in "doubled yet gauged" spacetime,''
  Nucl.\ Phys.\ B {\bf 880} (2014) 134
  [arXiv:1307.8377 [hep-th]].
  
  
  

\bibitem{Park:2016sbw}
J.~H.~Park,
``Green-Schwarz superstring on doubled-yet-gauged spacetime,''
JHEP \textbf{11} (2016), 005
doi:10.1007/JHEP11(2016)005
[arXiv:1609.04265 [hep-th]].

\bibitem{Ko:2016dxa}
S.~M.~Ko, J.~H.~Park and M.~Suh,
``The rotation curve of a point particle in stringy gravity,''
JCAP \textbf{06} (2017), 002
doi:10.1088/1475-7516/2017/06/002
[arXiv:1606.09307 [hep-th]].

\bibitem{Basile:2019pic}
T.~Basile, E.~Joung and J.~H.~Park,
``A note on Faddeev--Popov action for doubled-yet-gauged particle and graded Poisson geometry,''
JHEP \textbf{02} (2020), 022
doi:10.1007/JHEP02(2020)022
[arXiv:1910.13120 [hep-th]].
 


\bibitem{Morand:2017fnv}
K.~Morand and J.~H.~Park,
``Classification of non-Riemannian doubled-yet-gauged spacetime,''
Eur. Phys. J. C \textbf{77} (2017) no.10, 685
[erratum: Eur. Phys. J. C \textbf{78} (2018) no.11, 901]
doi:10.1140/epjc/s10052-017-5257-z
[arXiv:1707.03713 [hep-th]].



\bibitem{Hull:2006va}
  C.~M.~Hull,
  ``Doubled Geometry and T-Folds,''
  JHEP {\bf 0707} (2007) 080
  [hep-th/0605149].
  
  
  


  
  
  

\bibitem{Giveon:1988tt}
  A.~Giveon, E.~Rabinovici and G.~Veneziano,
  ``Duality in String Background Space,''
  Nucl.\ Phys.\ B {\bf 322} (1989) 167.

  
  
  
\bibitem{Blair:2020gng}
C.~D.~A.~Blair, G.~Oling and J.~H.~Park,
``Non-Riemannian isometries from double field theory,''
JHEP \textbf{04} (2021), 072
doi:10.1007/JHEP04(2021)072
[arXiv:2012.07766 [hep-th]].



\bibitem{Park:2020ixf}
J.~H.~Park and S.~Sugimoto,
``String Theory and non-Riemannian Geometry,''
Phys. Rev. Lett. \textbf{125} (2020) no.21, 211601
doi:10.1103/PhysRevLett.125.211601
[arXiv:2008.03084 [hep-th]].



\bibitem{Berman:2019izh}
D.~S.~Berman, C.~D.~A.~Blair and R.~Otsuki,
``Non-Riemannian geometry of M-theory,''
JHEP \textbf{07} (2019), 175
doi:10.1007/JHEP07(2019)175
[arXiv:1902.01867 [hep-th]].


\bibitem{Ko:2015rha}
S.~M.~Ko, C.~Melby-Thompson, R.~Meyer and J.~H.~Park,
``Dynamics of Perturbations in Double Field Theory \& Non-Relativistic String Theory,''
JHEP \textbf{12} (2015), 144
doi:10.1007/JHEP12(2015)144
[arXiv:1508.01121 [hep-th]].




\bibitem{Gomis:2000bd}
  J.~Gomis and H.~Ooguri,
  ``Nonrelativistic closed string theory,''
  J.\ Math.\ Phys.\  {\bf 42} (2001) 3127
  [hep-th/0009181].



\bibitem{Gomis:2005pg}
J.~Gomis, J.~Gomis and K.~Kamimura,
``Non-relativistic superstrings: A New soluble sector of $AdS(5) \times S^5$,''
JHEP \textbf{12} (2005), 024
doi:10.1088/1126-6708/2005/12/024
[arXiv:hep-th/0507036 [hep-th]].



\bibitem{Danielsson:2000gi}
U.~H.~Danielsson, A.~Guijosa and M.~Kruczenski,
``IIA/B, wound and wrapped,''
JHEP \textbf{10} (2000), 020
doi:10.1088/1126-6708/2000/10/020
[arXiv:hep-th/0009182 [hep-th]].




\bibitem{Christensen:2013lma}
M.~H.~Christensen, J.~Hartong, N.~A.~Obers and B.~Rollier,
``Torsional Newton-Cartan Geometry and Lifshitz Holography,''
Phys. Rev. D \textbf{89} (2014), 061901(R)
doi:10.1103/PhysRevD.89.061901
[arXiv:1311.4794 [hep-th]].




\bibitem{Hartong:2015zia}
J.~Hartong and N.~A.~Obers,
``Ho\v{r}ava-Lifshitz gravity from dynamical Newton-Cartan geometry,''
JHEP \textbf{07} (2015), 155
doi:10.1007/JHEP07(2015)155
[arXiv:1504.07461 [hep-th]].

\bibitem{Harmark:2017rpg}
T.~Harmark, J.~Hartong and N.~A.~Obers,
``Nonrelativistic strings and limits of the AdS/CFT correspondence,''
Phys. Rev. D \textbf{96} (2017) no.8, 086019
doi:10.1103/PhysRevD.96.086019
[arXiv:1705.03535 [hep-th]].


\bibitem{Harmark:2018cdl}
T.~Harmark, J.~Hartong, L.~Menculini, N.~A.~Obers and Z.~Yan,
``Strings with Non-Relativistic Conformal Symmetry and Limits of the AdS/CFT Correspondence,''
JHEP \textbf{11} (2018), 190
doi:10.1007/JHEP11(2018)190
[arXiv:1810.05560 [hep-th]].

\bibitem{Bergshoeff:2018yvt}
E.~Bergshoeff, J.~Gomis and Z.~Yan,
``Nonrelativistic String Theory and T-Duality,''
JHEP \textbf{11} (2018), 133
doi:10.1007/JHEP11(2018)133
[arXiv:1806.06071 [hep-th]].

\bibitem{Bergshoeff:2019pij}
E.~A.~Bergshoeff, J.~Gomis, J.~Rosseel, C.~\c{S}im\c{s}ek and Z.~Yan,
``String Theory and String Newton-Cartan Geometry,''
J. Phys. A \textbf{53} (2020) no.1, 014001
doi:10.1088/1751-8121/ab56e9
[arXiv:1907.10668 [hep-th]].

\bibitem{Harmark:2019upf}
T.~Harmark, J.~Hartong, L.~Menculini, N.~A.~Obers and G.~Oling,
``Relating non-relativistic string theories,''
JHEP \textbf{11} (2019), 071
doi:10.1007/JHEP11(2019)071
[arXiv:1907.01663 [hep-th]].


\bibitem{Bergshoeff:2021bmc}
E.~A.~Bergshoeff, J.~Lahnsteiner, L.~Romano, J.~Rosseel and C.~\c{S}im\c{s}ek,
``A Non-Relativistic Limit of NS-NS Gravity,''
[arXiv:2102.06974 [hep-th]].

\bibitem{Oling:2022fft}
G.~Oling and Z.~Yan,
``Aspects of Nonrelativistic Strings,''
Front. in Phys. \textbf{10} (2022), 832271
doi:10.3389/fphy.2022.832271
[arXiv:2202.12698 [hep-th]].



\bibitem{Hartong:2022lsy}
J.~Hartong, N.~A.~Obers and G.~Oling,
``Review on Non-Relativistic Gravity,''
Front. in Phys. \textbf{11} (2023), 1116888
doi:10.3389/fphy.2023.1116888
[arXiv:2212.11309 [gr-qc]].






\bibitem{Morand:2021xeq}
K.~Morand, J.~H.~Park and M.~Park,
``Identifying Riemannian Singularities with Regular Non-Riemannian Geometry,''
Phys. Rev. Lett. \textbf{128} (2022) no.4, 041602
doi:10.1103/PhysRevLett.128.041602
[arXiv:2106.01758 [hep-th]].









\bibitem{Jeon:2011cn}
  I.~Jeon, K.~Lee and J.~H.~Park,
  ``Stringy differential geometry, beyond Riemann,''
  Phys.\ Rev.\ D {\bf 84} (2011) 044022
  [arXiv:1105.6294 [hep-th]].

\bibitem{Hohm:2010xe}
O.~Hohm and S.~K.~Kwak,
``Frame-like Geometry of Double Field Theory,''
J. Phys. A \textbf{44} (2011), 085404
doi:10.1088/1751-8113/44/8/085404
[arXiv:1011.4101 [hep-th]].



\bibitem{Jeon:2011vx}
I.~Jeon, K.~Lee and J.~H.~Park,
``Incorporation of fermions into double field theory,''
JHEP \textbf{11} (2011), 025
doi:10.1007/JHEP11(2011)025
[arXiv:1109.2035 [hep-th]].



\bibitem{Cho:2015lha}
W.~Cho, J.~J.~Fern\'andez-Melgarejo, I.~Jeon and J.~H.~Park,
``Supersymmetric gauged double field theory: systematic derivation by virtue of twist,''
JHEP \textbf{08} (2015), 084
doi:10.1007/JHEP08(2015)084
[arXiv:1505.01301 [hep-th]].





\bibitem{Jeon:2011kp}
I.~Jeon, K.~Lee and J.~H.~Park,
``Double field formulation of Yang-Mills theory,''
Phys. Lett. B \textbf{701} (2011), 260-264
doi:10.1016/j.physletb.2011.05.051
[arXiv:1102.0419 [hep-th]].




\bibitem{Angus:2021jvm}
S.~Angus, M.~Kim and J.~H.~Park,
``Fractons, non-Riemannian geometry, and double field theory,''
Phys. Rev. Res. \textbf{4} (2022) no.3, 033186
doi:10.1103/PhysRevResearch.4.033186
[arXiv:2111.07947 [hep-th]].


\bibitem{Berkovits:2001ue}
N.~Berkovits and P.~S.~Howe,
``Ten-dimensional supergravity constraints from the pure spinor formalism for the superstring,''
Nucl. Phys. B \textbf{635} (2002), 75-105
doi:10.1016/S0550-3213(02)00352-8
[arXiv:hep-th/0112160 [hep-th]].




\bibitem{Rocen:2010bk}
A.~Rocen and P.~West,
``E11, generalised space-time and IIA string theory: the R-R sector,''
doi:10.1142/9789814412551\_0020
[arXiv:1012.2744 [hep-th]].




\bibitem{Hohm:2011zr}
O.~Hohm, S.~K.~Kwak and B.~Zwiebach,
``Unification of Type II Strings and T-duality,''
Phys. Rev. Lett. \textbf{107} (2011), 171603
doi:10.1103/PhysRevLett.107.171603
[arXiv:1106.5452 [hep-th]].


\bibitem{Hatsuda:2014aza}
M.~Hatsuda, K.~Kamimura and W.~Siegel,
``Ramond-Ramond gauge fields in superspace with manifest T-duality,''
JHEP \textbf{02} (2015), 134
doi:10.1007/JHEP02(2015)134
[arXiv:1411.2206 [hep-th]].


\bibitem{Cederwall:2016ukd}
M.~Cederwall,
``Double supergeometry,''
JHEP \textbf{06} (2016), 155
doi:10.1007/JHEP06(2016)155
[arXiv:1603.04684 [hep-th]].


\bibitem{Butter:2022gbc}
D.~Butter,
``Type II double field theory in superspace,''
JHEP \textbf{02} (2023), 187
doi:10.1007/JHEP02(2023)187
[arXiv:2209.07296 [hep-th]].



\bibitem{Butter:2022sfh}
D.~Butter,
``Notes on Ramond-Ramond spinors and bispinors in double field theory,''
JHEP \textbf{05} (2023), 039
doi:10.1007/JHEP05(2023)039
[arXiv:2208.11162 [hep-th]].



\bibitem{Jeon:2012kd}
I.~Jeon, K.~Lee and J.~H.~Park,
``Ramond-Ramond Cohomology and O(D,D) T-duality,''
JHEP \textbf{09} (2012), 079
doi:10.1007/JHEP09(2012)079
[arXiv:1206.3478 [hep-th]].



\bibitem{Aldazabal:2011nj}
G.~Aldazabal, W.~Baron, D.~Marques and C.~Nunez,
``The effective action of Double Field Theory,''
JHEP \textbf{11} (2011), 052
doi:10.1007/JHEP11(2011)052
[arXiv:1109.0290 [hep-th]].

\bibitem{Grana:2012rr}
M.~Grana and D.~Marques,
``Gauged Double Field Theory,''
JHEP \textbf{04} (2012), 020
doi:10.1007/JHEP04(2012)020
[arXiv:1201.2924 [hep-th]].



\bibitem{Geissbuhler:2013uka}
D.~Geissbuhler, D.~Marques, C.~Nunez and V.~Penas,
``Exploring Double Field Theory,''
JHEP \textbf{06} (2013), 101
doi:10.1007/JHEP06(2013)101
[arXiv:1304.1472 [hep-th]].




\bibitem{Kosmann}
Y. Kosmann, ``D\'{e}riv\'{e}es de Lie des spineurs,'' 
Annali Mat.Pura Appl. 91 (1971) 317.


\bibitem{Kim:2024ewt}
T.~Kim and P.~Yi,
``Lie, Noether, Kosmann, and Diffeomorphism Anomalies Redux,''
[arXiv:2412.03667 [hep-th]].



\bibitem{Coimbra:2011nw}
  A.~Coimbra, C.~Strickland-Constable and D.~Waldram,
  ``Supergravity as Generalised Geometry I: Type II Theories,''
  JHEP {\bf 1111} (2011) 091
  [arXiv:1107.1733 [hep-th]].






\bibitem{Cho:2019ofr}
K.~Cho and J.~H.~Park,
``Remarks on the non-Riemannian sector in Double Field Theory,''
Eur. Phys. J. C \textbf{80} (2020) no.2, 101
doi:10.1140/epjc/s10052-020-7648-9
[arXiv:1909.10711 [hep-th]].




\bibitem{Hohm:2011si}
  O.~Hohm and B.~Zwiebach,
  ``On the Riemann Tensor in Double Field Theory,''
  JHEP {\bf 1205} (2012) 126
  [arXiv:1112.5296 [hep-th]].







\bibitem{Blair:2015eba}
C.~D.~A.~Blair,
``Conserved Currents of Double Field Theory,''
JHEP \textbf{04} (2016), 180
doi:10.1007/JHEP04(2016)180
[arXiv:1507.07541 [hep-th]].



\bibitem{Dyer:2008hb}
E.~Dyer and K.~Hinterbichler,
``Boundary Terms, Variational Principles and Higher Derivative Modified Gravity,''
Phys. Rev. D \textbf{79} (2009), 024028
doi:10.1103/PhysRevD.79.024028
[arXiv:0809.4033 [gr-qc]].

\bibitem{Berman:2011kg}
D.~S.~Berman, E.~T.~Musaev and M.~J.~Perry,
``Boundary Terms in Generalized Geometry and doubled field theory,''
Phys. Lett. B \textbf{706} (2011), 228-231
doi:10.1016/j.physletb.2011.11.019
[arXiv:1110.3097 [hep-th]].


\bibitem{Lee:2025fme}
Kawon~Lee and J.~H.~Park,
``$\mathbf{O}(D,D)$-Symmetric Box Operator and $\alpha^{\prime}$-Corrections with Riemann Curvature,''
[arXiv:2506.21143 [hep-th]].



\bibitem{Hohm:2015ugy}
O.~Hohm and D.~Marques,
``Perturbative Double Field Theory on General Backgrounds,''
Phys. Rev. D \textbf{93} (2016) no.2, 025032
doi:10.1103/PhysRevD.93.025032
[arXiv:1512.02658 [hep-th]].

\bibitem{Cho:2019npq}
K.~Cho, K.~Morand and J.~H.~Park,
``Stringy Newton Gravity with $H$-flux,''
Phys. Rev. D \textbf{101} (2020) no.6, 064020
doi:10.1103/PhysRevD.101.064020
[arXiv:1912.13220 [hep-th]].

\bibitem{Lee:2023boi}
H.~Lee, J.~H.~Park, L.~Velasco-Sevilla and L.~Yin,
``Late-time cosmology without dark sector but with closed string massless sector,''
Phys. Lett. B \textbf{860} (2025), 139215
doi:10.1016/j.physletb.2024.139215
[arXiv:2308.07149 [hep-th]].



\bibitem{Choi:2022srv}
K.~S.~Choi and J.~H.~Park,
``Post-Newtonian Feasibility of the Closed String Massless Sector,''
Phys. Rev. Lett. \textbf{129} (2022) no.6, 061603
doi:10.1103/PhysRevLett.129.061603
[arXiv:2202.07413 [hep-th]].



\bibitem{Khoury:2003aq}
J.~Khoury and A.~Weltman,
``Chameleon fields: Awaiting surprises for tests of gravity in space,''
Phys. Rev. Lett. \textbf{93} (2004), 171104
doi:10.1103/PhysRevLett.93.171104
[arXiv:astro-ph/0309300 [astro-ph]].

\bibitem{Brax:2004qh}
P.~Brax, C.~van de Bruck, A.~C.~Davis, J.~Khoury and A.~Weltman,
``Detecting dark energy in orbit: The cosmological chameleon,''
Phys. Rev. D \textbf{70} (2004), 123518
doi:10.1103/PhysRevD.70.123518
[arXiv:astro-ph/0408415 [astro-ph]].




\bibitem{Burgess:1994kq}
C.~P.~Burgess, R.~C.~Myers and F.~Quevedo,
``On spherically symmetric string solutions in four-dimensions,''
Nucl. Phys. B \textbf{442} (1995), 75-96
doi:10.1016/S0550-3213(95)00090-9
[arXiv:hep-th/9410142 [hep-th]].



\bibitem{Jang:2024nhm}
H.~Jang, M.~Kim, H.~Lee and J.~H.~Park,
``Traversable wormhole for string, but not for particle,''
[arXiv:2412.04128 [hep-th]].






\bibitem{Will:1972zz}
C.~M.~Will and K.~Nordtvedt, Jr.,
``Conservation Laws and Preferred Frames in Relativistic Gravity. I. Preferred-Frame Theories and an Extended PPN Formalism,''
Astrophys. J. \textbf{177} (1972), 757
doi:10.1086/151754






\bibitem{Will:2014kxa}
C.~M.~Will,
``The Confrontation between General Relativity and Experiment,''
Living Rev. Rel. \textbf{17} (2014), 4
doi:10.12942/lrr-2014-4
[arXiv:1403.7377 [gr-qc]].

\bibitem{Bertotti:2003rm}
B.~Bertotti, L.~Iess and P.~Tortora,
``A test of general relativity using radio links with the Cassini spacecraft,''
Nature \textbf{425} (2003), 374-376
doi:10.1038/nature01997


\bibitem{Gasperini:1991ak}
M.~Gasperini and G.~Veneziano,
``O(d,d) covariant string cosmology,''
Phys. Lett. B \textbf{277} (1992), 256-264
doi:10.1016/0370-2693(92)90744-O
[arXiv:hep-th/9112044 [hep-th]].




\bibitem{Angus:2019bqs}
S.~Angus, K.~Cho, G.~Franzmann, S.~Mukohyama and J.~H.~Park,
``$\mathbf {O}(D,D)$ completion of the Friedmann equations,''
Eur. Phys. J. C \textbf{80} (2020) no.9, 830
doi:10.1140/epjc/s10052-020-8379-7
[arXiv:1905.03620 [hep-th]].




\bibitem{Danielsson:2018ztv}
U.~H.~Danielsson and T.~Van Riet,  
``What if string theory has no de Sitter vacua?,''
Int. J. Mod. Phys. D \textbf{27} (2018) no.12, 1830007
doi:10.1142/S0218271818300070
[arXiv:1804.01120 [hep-th]].



\bibitem{Obied:2018sgi} 
  G.~Obied, H.~Ooguri, L.~Spodyneiko and C.~Vafa,
  ``De Sitter Space and the Swampland,''
  arXiv:1806.08362 [hep-th].




\bibitem{Agrawal:2018own} 
  P.~Agrawal, G.~Obied, P.~J.~Steinhardt and C.~Vafa,
  ``On the Cosmological Implications of the String Swampland,''
  Phys.\ Lett.\ B {\bf 784}, 271 (2018)
  doi:10.1016/j.physletb.2018.07.040
  [arXiv:1806.09718 [hep-th]].


\bibitem{Andriot:2018wzk} 
  D.~Andriot,
  ``On the de Sitter swampland criterion,''
  Phys.\ Lett.\ B {\bf 785}, 570 (2018)
  doi:10.1016/j.physletb.2018.09.022
  [arXiv:1806.10999 [hep-th]].



\bibitem{Copeland:1994vi}
E.~J.~Copeland, A.~Lahiri and D.~Wands,
``Low-energy effective string cosmology,''
Phys. Rev. D \textbf{50} (1994), 4868-4880
doi:10.1103/PhysRevD.50.4868
[arXiv:hep-th/9406216 [hep-th]].







\bibitem{King:2012id}
J.~A.~King, J.~K.~Webb, M.~T.~Murphy, V.~V.~Flambaum, R.~F.~Carswell, M.~B.~Bainbridge, M.~R.~Wilczynska and F.~E.~Koch,
``Spatial variation in the fine-structure constant -- new results from VLT/UVES,''
Mon. Not. Roy. Astron. Soc. \textbf{422} (2012), 3370-3413
doi:10.1111/j.1365-2966.2012.20852.x
[arXiv:1202.4758 [astro-ph.CO]].

\bibitem{Wilczynska:2015una}
M.~R.~Wilczynska, J.~K.~Webb, J.~A.~King, M.~T.~Murphy, M.~B.~Bainbridge and V.~V.~Flambaum,
``A new analysis of fine-structure constant measurements and modelling errors from quasar absorption lines,''
Mon. Not. Roy. Astron. Soc. \textbf{454} (2015) no.3, 3082-3093
doi:10.1093/mnras/stv2148
[arXiv:1510.02536 [astro-ph.CO]].


\bibitem{Martins:2017qxd}
C.~J.~A.~P.~Martins and A.~M.~M.~Pinho,
``Stability of fundamental couplings: a global analysis,''
Phys. Rev. D \textbf{95} (2017) no.2, 023008
doi:10.1103/PhysRevD.95.023008
[arXiv:1701.08724 [astro-ph.CO]].

\bibitem{Wilczynska:2020rxx}
M.~R.~Wilczynska, J.~K.~Webb, M.~Bainbridge, S.~E.~I.~Bosman, J.~D.~Barrow, R.~F.~Carswell, M.~P.~Dk{a}browski, V.~Dumont, A.~C.~Leite and C.~C.~Lee, \textit{et al.}
``Four direct measurements of the fine-structure constant 13 billion years ago,''
Sci. Adv. \textbf{6} (2020) no.17, eaay9672
doi:10.1126/sciadv.aay9672
[arXiv:2003.07627 [astro-ph.CO]].












\bibitem{Scolnic:2021amr}
D.~Scolnic, D.~Brout, A.~Carr, A.~G.~Riess, T.~M.~Davis, A.~Dwomoh, D.~O.~Jones, N.~Ali, P.~Charvu and R.~Chen, \textit{et al.} ``The Pantheon+ Analysis: The Full Data Set and Light-curve Release,''
Astrophys. J. \textbf{938}, no.2, 113 (2022)
doi:10.3847/1538-4357/ac8b7a
[arXiv:2112.03863 [astro-ph.CO]].

\bibitem{Riess:2021jrx}
A.~G.~Riess, W.~Yuan, L.~M.~Macri, D.~Scolnic, D.~Brout, S.~Casertano, D.~O.~Jones, Y.~Murakami, L.~Breuval and T.~G.~Brink, \textit{et al.}
``A Comprehensive Measurement of the Local Value of the Hubble Constant with 1 $\mathrm{km s^{-1}Mpc^{-1}}$ Uncertainty from the Hubble Space Telescope and the SH0ES Team,''
Astrophys. J. Lett. \textbf{934} (2022) no.1, L7
doi:10.3847/2041-8213/ac5c5b
[arXiv:2112.04510 [astro-ph.CO]].



\bibitem{Burkert:2018bqq}
V.~D.~Burkert, L.~Elouadrhiri and F.~X.~Girod,
``The pressure distribution inside the proton,''
Nature \textbf{557} (2018) no.7705, 396-399
doi:10.1038/s41586-018-0060-z


\bibitem{Planck:2018vyg}
N.~Aghanim \textit{et al.} [Planck],
``Planck 2018 results. VI. Cosmological parameters,''
Astron. Astrophys. \textbf{641} (2020), A6
[erratum: Astron. Astrophys. \textbf{652} (2021), C4]
doi:10.1051/0004-6361/201833910
[arXiv:1807.06209 [astro-ph.CO]].

\bibitem{Cho:2018alk}
K.~Cho, K.~Morand and J.~H.~Park,
``Kaluza\textendash{}Klein reduction on a maximally non-Riemannian space is moduli-free,''
Phys. Lett. B \textbf{793} (2019), 65-69
doi:10.1016/j.physletb.2019.04.042
[arXiv:1808.10605 [hep-th]].



\bibitem{Hohm:2013jaa}
O.~Hohm, W.~Siegel and B.~Zwiebach,
``Doubled $\alpha'$-geometry,''
JHEP \textbf{02} (2014), 065
doi:10.1007/JHEP02(2014)065
[arXiv:1306.2970 [hep-th]].


\bibitem{Bedoya:2014pma}
O.~A.~Bedoya, D.~Marques and C.~Nunez,
``Heterotic $\alpha$'-corrections in Double Field Theory,''
JHEP \textbf{12} (2014), 074
doi:10.1007/JHEP12(2014)074
[arXiv:1407.0365 [hep-th]].

\bibitem{Hohm:2014xsa}
O.~Hohm and B.~Zwiebach,
``Double field theory at order $\alpha'$,''
JHEP \textbf{11} (2014), 075
doi:10.1007/JHEP11(2014)075
[arXiv:1407.3803 [hep-th]].


\bibitem{Marques:2015vua}
D.~Marques and C.~A.~Nunez,
``T-duality and \ensuremath{\alpha}'-corrections,''
JHEP \textbf{10} (2015), 084
doi:10.1007/JHEP10(2015)084
[arXiv:1507.00652 [hep-th]].


\end{thebibliography}
\end{document}